\documentclass[useAMS,usenatbib,aas_macros]{mn2e}
\usepackage{aas_macros}
\usepackage{amsmath}
\usepackage{subfigure}
\usepackage{graphicx}
\usepackage{epsfig,times}
\usepackage{float}

\usepackage{color}

\newcommand{\equ}[1]{eq.~(\ref{eq:#1})}

\newcommand{\se}[1]{\S\ref{sec:#1}}
\newcommand{\fig}[1]{Fig.~\ref{fig:#1}}
\newcommand{\figs}[1]{Figs.~\ref{fig:#1}}
\newcommand{\Fig}[1]{Figure~\ref{fig:#1}}
\newcommand{\Figs}[1]{Figures~\ref{fig:#1}}
\newcommand{\tab}[1]{Table~\ref{tab:#1}}
\newcommand{\be}{\begin{equation}}
\newcommand{\ee}{\end{equation}}
\newcommand{\bea}{\begin{eqnarray}}
\newcommand{\eea}{\end{eqnarray}}
\def\ra{\rangle}
\def\la{\langle}

\newcommand{\msun}{M_\odot}

\newcommand{\sy}{\,M_\odot\, {\rm yr}^{-1}}

\newcommand{\ifm}[1]{\relax\ifmmode#1\else$\mathsurround=0pt #1$\fi}
\newcommand{\kms}{\ifmmode\,{\rm km}\,{\rm s}^{-1}\else km$\,$s$^{-1}$\fi}
\newcommand{\hmpc}{\,\ifm{h^{-1}}{\rm Mpc}}

\newcommand{\Mpc}{\,{\rm Mpc}}

\newcommand{\kpc}{\,{\rm kpc}}
\newcommand{\pc}{\,{\rm pc}}
\newcommand{\Gyr}{\,{\rm Gyr}}

\newcommand{\Myr}{\,{\rm Myr}}

\newcommand{\cmc}{\,{\rm cm}^{-3}}
\newcommand{\cms}{\,{\rm cm}^{-2}}

\newcommand{\ltsima}{$\; \buildrel < \over \sim \;$}
\newcommand{\lsim}{\lower.5ex\hbox{\ltsima}}
\newcommand{\gtsima}{$\; \buildrel > \over \sim \;$}
\newcommand{\gsim}{\lower.5ex\hbox{\gtsima}}
\newcommand{\prop}{\propto}
\newcommand{\dd}{\rm d}

\def\omm{\Omega_{\rm m}}
\def\omb{\Omega_{\rm b}}
\def\oml{\Omega_{\Lambda}}

\def\Mv{M_{\rm v}}

\def\Rv{R_{\rm v}}

\def\Mg{M_{\rm g}}
\def\Mgdot{\dot{M}_{\rm g}}
\def\Ms{M_{\rm s}}

\def\Mdot{\dot{M}}
\def\Mt{M_{\rm tot}}

\def\fc{f_{\rm c}}

\def\Mcold{M_{\rm cold}}

\def\Jneg{J_{-}}
\def\Jd{J_{\rm d}}
\def\td{t_{\rm d}}
\def\torb{t_{\rm orb}}

\def\tinf{t_{\rm inf}}
\def\tsfr{t_{\rm sfr}}

\def\Re{R_{\rm e}}

\def\md{m_{\rm d}}

\def\Sig1{\Sigma_{1\kpc}}
\def\slos{\sigma_{\rm los}}
\newcommand{\Se}{\Sigma_{\rm e}}
\newcommand{\Sem}{\Sigma_{\rm e,max}}
\newcommand{\Skpc}{\Sigma_{1}}
\newcommand{\Skpcm}{\Sigma_{{\rm 1,max}}}
\newcommand{\Sm}{\Sigma_{\rm max}}

\def\M12{M_{\rm v,12}}
\def\R100{R_{\rm v,100}}
\def\z3{(1+z)_3}

\def\ga{\tau_0^{-1}}

\def\rf{\par\smallskip\noindent\hangindent 9pt}

\def\bul{$\bullet\ $}

\title[Compaction and Quenching]
{Compaction and quenching of high-z galaxies 
in cosmological simulations: blue and red nuggets}

\author[Zolotov et al.]
{\parbox[t]{\textwidth} 
{
Adi Zolotov$^{1,2}$\thanks{E-mail: zolotov.1@osu.edu},
Avishai Dekel$^{1}$\thanks{E-mail: dekel@huji.ac.il},
Nir Mandelker$^{1}$,
Dylan Tweed$^{1,3}$,
Shigeki Inoue$^{1}$,
\\
Colin DeGraf$^{1}$, 
Daniel Ceverino$^{1,4}$,
Joel R. Primack$^{5}$,
Guillermo Barro$^{6}$,
\\
Sandra M. Faber$^{6}$
}
\\
\\
$^1$Center for Astrophysics and Planetary Science,
Racah Institute of Physics, The Hebrew University, Jerusalem 91904, Israel\\
$^2$Center for Cosmology and Astroparticle Physics, Department of Physics, The
Ohio State University, OH 43210, USA\\
$^3$Center for Astronomy and Astrophysics, Shanghai Jiao Tong University,
Shanghai 200240, China\\
$^4$Departamento de Fisica Teorica, Universidad Autonoma de Madrid, 28049
Madrid, Spain\\
$^5$Department of Physics, University of California, Santa Cruz, CA 95064,
USA\\
$^6$UCO/Lick Observatory, Department of Astronomy and Astyrophysics, 
University of California, Santa Cruz, CA 95064, USA
}
\begin{document}

\large
%\Large
%\LARGE

\pagerange{\pageref{firstpage}--\pageref{lastpage}} \pubyear{2002}

\maketitle

\label{firstpage}

\begin{abstract}
We use cosmological simulations to study a characteristic evolution 
pattern of high redshift galaxies. Early, stream-fed, highly perturbed, 
gas-rich discs undergo phases of dissipative contraction into compact, 
star-forming systems (``blue" nuggets) at $z\!\sim\!4-2$.  The peak of gas 
compaction marks the onset of central gas depletion and inside-out quenching 
into compact ellipticals (red nuggets) by $z\!\sim\!2$.  These are sometimes 
surrounded by gas rings or grow extended dry stellar envelopes. The 
compaction occurs at a roughly constant specific star-formation rate (SFR), 
and the quenching occurs at a constant stellar surface density within the 
inner kpc ($\Skpc$). Massive galaxies quench earlier, faster, and at a higher 
$\Skpc$ than lower-mass galaxies, which compactify and attempt to quench more 
than once.  This evolution pattern is consistent with the way galaxies populate 
the SFR-size-mass space, and with gradients and scatter across the main 
sequence.
The compaction is triggered by an intense inflow episode, involving 
(mostly minor) mergers, 
counter-rotating streams or recycled gas, and is commonly associated with 
violent disc instability.  The contraction is dissipative, with the inflow 
rate $>$SFR, and the maximum $\Skpc$ anti-correlated with the initial spin 
parameter \citep{db14}. The central quenching is 
triggered by the high SFR and stellar/supernova feedback (maybe also AGN 
feedback) due to the high central gas density, while the central inflow weakens
as the disc vanishes.  Suppression of fresh gas supply by a hot halo allows the 
long-term maintenance of quenching once above a threshold halo mass, inducing 
the quenching downsizing. 
%self-gravitating. 
\end{abstract}


%
%Using linear regression to fit a line for the data with log S1 < 9
%
%I. For alpha =0.14 beta =2.5
%The RMS error about the horizontal branch (log S1 < 9) for:
%--  raw data, with slope = -0.38 (best fit from linear regression), RMS = 0.31
%--  SFR-scaled data, with slope = 0.02, RMS = 0.23
%--  SFR scaled, S1 shifted data and cut <9.3, slope = 0.016, RMS = 0.23
%-- SFR scaled, S1 shifted data and cut <9.0, slope = 0.008, RMS = 0.23

%II. For alpha =0.0 beta =2.0
%The RMS error about the horizontal branch (log S1 < 9) for:
%--  raw data, with slope = -0.38 (best fit from linear regression), RMS = 0.31
%--  SFR-scaled data, with slope = 0.04, RMS = 0.23
%--  SFR scaled, S1 shifted data and cut <9.3, slope = 0.038, RMS = 0.23
%-- SFR scaled, S1 shifted data and cut <9.0, slope = 0.038, RMS = 0.24

%So the slopes are always ~0 for the scaled data, and the RMS is ~0.23 in all
%cases.

%%%%%%%%%%%%%%%%%%%%%1
\section{Introduction}
\label{sec:intro}

% RN
Observations indicate that a significant fraction of the massive galaxies
at redshifts $z=2-3$ are compact ellipticals with suppressed star formation
rates (SFR), for which we adopt the nickname ``red nuggets"\footnote{The term
``nugget" does not imply a small mass; the nuggets are massive but small 
in size, namely compact.}
\citep{daddi05,trujillo06a,trujillo06b,dokkum08,damjanov09,newman10,
dokkum10,damjanov11,whitaker12,bruce12,dokkum14}. 
While the massive star-forming discs of stellar mass $\sim 10^{11}\msun$ 
extend to effective radii of several kpc
\citep{elmegreen07,genzel06,genzel08},
the quenched spheroids of a similar mass have effective radii of order 1 kpc
\citep{carollo13,vanderwel14b}.

% wet
\smallskip
The sizes of extended discs are roughly consistent with the theoretical 
expectations based on gas infall through dark-matter haloes into
rotating discs while conserving angular-momentum 
\citep[][Danovich et al. 2014]{fe80,mmw98,bullock01_j,fall13}.
Since stellar systems tend to conserve energy and angular momentum,
further contraction to form the compact nuggets would require
further loss of energy and angular momentum, which cannot be easily
achieved by stellar systems. Thus, the formation of nuggets is likely to be
a dissipative process, 
namely associated with gas inflow into the central regions of the galaxies.
We refer to this as a ``wet" process. 
Gas inflow is naturally expected at high redshift, since the gas fraction in 
discs is high at these early times \citep{daddi10,tacconi10,tacconi13}.

% BN
\smallskip
Indeed, there are indicative observational identifications of the progenitors 
of the red nuggets in the form of ``blue nuggets", which are compact, 
star-forming galaxies.\footnote{The ``blue" nuggets could actually be quite red
due to dust, so we sometimes refer to them as ``blue", with quotation marks.}
Their masses, kinematics and abundances are consistent with those of the red
nuggets 
\citep{barro13,williams14a,williams14b,bruce14,nelson14,barro14a,barro14b}.
The observed abundances of blue nuggets depend on their lifetimes, 
which could be rather short, depending on the galaxy properties and on the 
actual quenching mechanism (see below).

% pattern of phases
\smallskip
We thus envision a generic pattern of evolution for high-redshift galaxies
through several characteristic phases.
First, the early formation of a gas-rich, star-forming, highly perturbed disc,
subject to intense inflows involving multiple (mostly minor) wet mergers,
and developing violent disc instability (VDI).
Second, the dissipative, quick {\it compaction} of the gas disc into a compact, 
star-forming blue nugget.
Third, immediately following the compaction, is the rather fast 
{\it quenching} of star formation into a compact red nugget.
Finally, the gradual growth and expansion of the elliptical galaxy
by dry mergers, and/or the development of a new gas disc or ring 
surrounding the red nugget.
% goal
The origins of the compaction and the subsequent quenching are the
theoretical challenges addressed in this paper. 

% inflow and early disc
\smallskip
In the early disc phase, streams from the cosmic web, consisting of smooth gas 
and merging galaxies, continuously feed galactic discs
\citep{bd03,keres05,db06,ocvirk08,keres09,dekel09,danovich12,danovich14}.
The detailed thermal history of the streams in the inner halo
\citep{cdb10,nelson13}
does not make a difference, as long as the discs are fed with cold gas
at the levels consistent with the observed high SFR and gas fraction. 
The high gas fraction and the high density of the Universe
at these high redshifts, combined with constant triggering by minor
mergers, induce and maintain VDI, which is characterized by 
turbulence and perturbations in the form of large transient features and 
giant clumps
\citep{noguchi98,immeli04_a,immeli04_b,genzel06,bournaud07c,genzel08,
dsc09,agertz09,cdb10,ceverino12,mandelker14}. 

% origin of wet inflow 6
\smallskip
The {\it onset of wet compaction} is the first open issue we wish to
address here.
We can identify several potential reasons for the compaction.
Gas-rich mergers tend to drive gas into the galaxy centre
\citep[e.g.][]{barnes91,mihos96,hopkins06}.
Counter-rotating streams, low-angular-momentum recycled gas, and tidal
compression could also generate shrinkage (see \se{onset_comp}).
Finally, intense gas inflow within the disc is naturally driven by VDI
\citep{noguchi99,gammie01,dsc09,krum_burkert10,burkert10,bournaud11,
forbes12,cdg12,elmegreen12,dekel13,forbes14a}.
The timescale for VDI-driven inflow has been estimated in several different ways
\citep[e.g.,][]{dsc09,dekel13} to be 
$\tinf \sim \delta^{-2} \td$,
where $\delta$ is the fraction of ``cold" mass (mostly gas and young stars)
within the disc radius
with respect to the total mass (including the bulge and dark matter),
and $\td$ is the typical dynamical crossing time of the disc. 
It appears that at high redshift more gas is driven 
into the bulge by processes that do not involve major mergers (of a
stellar mass ratio larger than 1:3), 
both based on observations 
\citep[e.g.][]{genzel06,genzel08,bournaud08,kaviraj13b} 
and on theory including simulations 
\citep[e.g.][]{neistein08b,dekel09,bournaud09,cattaneo13,dekel13}.
Semi-analytic models that try to include VDI-driven inflows confirm that it
is a major source of spheroid growth \citep{porter14a,porter14b}.
In fact, our developing understanding is that
minor mergers, counter-rotation, recycling and tidal compression
stimulate the VDI, and they actually work in concert
(Inoue et al., in preparation.).

% db14
\citet[][hereafter DB14]{db14} 
addressed the formation of blue nuggets by wet compaction
of VDI discs, based on the requirement that 
for the inflow to be intense and dissipative 
the characteristic timescale for star formation, $\tsfr$, should be longer
than the timescale for inflow, $\tinf$. Otherwise, most of the disc mass will
turn into stars before it reaches the bulge, the inflow rate will be
suppressed, and the galaxy will become an extended stellar system.
DB14 thus defined a ``wetness" parameter
$w \equiv \tsfr/\tinf$
that should be greater than unity for a wet inflow. 
They showed that 
\be
w \equiv \frac{\tsfr}{\tinf} \sim \frac{\delta^2}{\epsilon} \, ,
\label{eq:w}
\ee
where $\epsilon$ is the efficiency of SFR per dynamical time,
and as said above $\delta$ is the fraction of cold mass 
with respect to the total mass within the disc radius.
Given that one expects at high redshift $\delta \geq 0.2$ \citep{dsc09}
and $\epsilon \leq 0.02$ \citep{kdm12,dm14}, 
the value of $w$ for typical galaxies is expected to be larger than unity,
and especially so for galaxies for which the initial spin parameter 
$\lambda$ is smaller than average. 
DB14 showed that, at a given mass and redshift, the distribution of $w$ values,
and the subsequent variation in evolution path, 
may be dominated by variations in contraction factor from the virial
radius to the disc radius, which can be traced back to variations in
spin parameter of the baryons that make the disc. 
The threshold $w=1$ can be translated to a critical value of spin parameter,
$\lambda_{w=1} \sim 0.04 (\fc/0.5)(\md/0.04)$, where 
$\fc$ is the fraction of cold mass with respect to the baryonic disc mass,
and $\md$ is the disc to halo mass ratio. 
The fraction of star-forming galaxies that will become blue
nuggets then depends on the value of $\lambda_{w=1}$ with respect to the
average value of the spin distribution, $\la \lambda \ra \sim 0.04$ 
\citep{bullock01_j},
and thus a larger fraction of blue nuggets is expected at high redshifts, where
$\fc$ is large.
Once $w>1$, DB14 argued that compaction should continue till the 
system becomes dispersion dominated and the VDI phases out.
They therefore predicted that star-forming galaxies should show a 
bimodality in central density, separating the extended and compact galaxies.
While DB14 specifically addressed the compaction of a VDI disc, the 
ideas concerning wet compaction are also valid in the general case, where
the compaction is triggered by an external process.

% quenching
\smallskip
The way in which {\it compaction leads to quenching} is the second open 
question to be addressed in this paper.
As discussed in DB14,
the internal, {\it bulge quenching}, associated with the compaction, 
may involve gas starvation by rapid consumption into stars,
the associated gas loss via outflows driven by stellar feedback 
\citep[e.g.][]{ds86,murray05} 
or AGN feedback \citep[e.g.][for a review]{ciotti07,cattaneo09}, 
and possibly a slowdown of gas supply to the centre \citep{feldmann14}.
A massive stellar bulge could also suppress star formation by shutting off disc
instability (by increasing the Toomre Q parameter to above unity), 
either by the gravitational effect on the rotation curve
\citep[morphological quenching,][]{martig09,martig13,genzel14a}, 
by the reduced gas surface density in the disc, or by high feedback-driven 
velocity dispersion. 
These processes typically operate in a {\it fast mode} that may be the 
dominant trigger for quenching at high redshift.

\smallskip
However, observations indicate that compactness is a necessary, but not
sufficient,  condition for quenching 
\citep{cheung12,fang13,barro13,db14,woo13,woo14}. 
Obviously, a long-term 
suppression of external gas supply is required for maintaining quenching.
This happens naturally once the halo mass grows above a threshold 
mass of order $10^{11.5-12.5}\msun$, either via virial shock heating 
\citep{bd03,keres05,db06,keres09},
by gravitational infall heating \citep{db08,khochfar08},
or by AGN feedback coupled to the hot halo gas
\citep{db06,cattaneo09,fabian12}. 
These external processes typically operate in a {\it slow, maintenance mode}
that is 
expected to be dominant at low redshift or at the late stages of quenching,
and their interplay with the internal
bulge-driven quenching is yet to be investigated.

% This paper
In this paper we address the processes of compaction and quenching
using high-resolution, zoom-in, hydro-cosmological, Adaptive Mesh Refinement 
(AMR) simulations of galaxies in
the redshift range $z=7$ to $z=1$. The suite of galaxies analyzed here
were simulated at a maximum resolution of $\sim 25 \pc$
including supernova and radiative stellar feedback.
At $z \sim 2$, the halo masses are in the range $\Mv \sim 10^{11-12}\msun$ 
and the stellar masses are in the range $\Ms \sim 10^{9.1-10.8}\msun$. 

\smallskip
This paper is organized as follows.
In \se{sims} we describe the simulations.
In \se{wet_bulge} we address the wetness of bulge formation by measuring 
the fraction of bulge stars that were formed in-situ in the bulge,
namely after dissipative gas contraction .
In \se{sample} we study the evolution of the properties of the 
whole sample of simulated galaxies, and address in particular the mass 
and time dependence of the quenching events. 
In \se{mass} we extend the analysis to the dependence of quenching 
on halo mass.
In \se{origin} we discuss the possible origins of 
the compaction and quenching processes. 
In \se{conc} we summarize our conclusions and discuss them.

%%%%%%%%%%%%%%%%%%%%%%%%%%%%%2
\section{Simulations}
\label{sec:sims}

%------------
\subsection{Simulation method and subgrid physics}

%Intro. Code. 
We use zoom-in hydro-cosmological simulations of 26 moderately massive
galaxies with an AMR maximum resolution that varies between $17.5$ and $35\pc$,
all evolved to $z=2$ and most reaching $z=1$.
They utilize the Adaptive Refinement Tree (ART) code
\citep{krav97,ceverino+k09},
which accurately follows
the evolution of a gravitating $N$-body system and the Eulerian gas dynamics
using an adaptive mesh.
Beyond gravity and hydrodynamics, the code incorporates at the subgrid level
many of the physical processes relevant for galaxy formation.
These include gas cooling by atomic hydrogen and helium as well as by
metals and molecular hydrogen, photoionization heating by the UV background
with partial self-shielding,
% short version:
star formation, stellar mass loss, metal enrichment of the ISM,
and stellar feedback.
Supernovae and stellar winds are implemented by local injection
of thermal energy as in \citet{ceverino+k09,cdb10,ceverino12}.
Radiative stellar feedback is implemented at a moderate level,
with no significant infrared trapping \citep[in the spirit of][]{dk13}, as described in \citet{ceverino14}.

% detailed version:
\smallskip
A few relevant details concerning the subgrid physics are as follows.
Cooling and heating rates are tabulated for a given gas
density, temperature, metallicity and UV background based on the CLOUDY code
\citep[version 96b4][]{ferland98}, assuming a slab of thickness 1 kpc. 
A uniform UV background
based on the redshift-dependent \citet{haardt96} model is assumed,
except at gas densities higher than $0.1\cmc$, where a substantially
suppressed UV background is used
($5.9\times 10^{26}{\rm erg}{\rm s}^{-1}{\rm cm}^{-2}{\rm Hz}^{-1}$)
in order to mimic the partial self-shielding of dense gas.
This allows the dense gas to cool down to temperatures of $\sim 300$K.
The assumed equation of state is that of an ideal mono-atomic gas.
Artificial fragmentation on the cell size is prevented by introducing
a pressure floor, which ensures that the Jeans scale is resolved by at least
7 cells \citep[see][]{cdb10}.

\smallskip
% SFR and feedback 
Star formation is assumed to occur at densities above a threshold of $1\cmc$
and at temperatures below $10^4$K. More than 90\% of the stars form at
temperatures well below $10^3$K, and more than half the stars form at 300~K
in cells where the gas density is higher than $10\cmc$.
The code implements a stochastic star-formation model that yields a
star-formation efficiency per free-fall time of $\sim 2\%$.\footnote{This is
lower by a factor of about 3 compared to the efficiency used 
in earlier simulations of somewhat lower resolution
\citep{ceverino+k09,cdb10,ceverino12,dekel13,mandelker14}.}  
At the given resolution,
this efficiency roughly mimics the empirical Kennicutt-Schmidt law
\citep{kennicutt98}.
The current version of the codes uses the stellar initial mass function of
\citet{chabrier05}.\footnote{Replacing the Miller-Scalo IMF used in earlier
simulations.}

\smallskip
% feedback
The code incorporates a thermal stellar feedback model, in which the combined
energy from stellar winds and supernova explosions is released as a constant
heating rate over $40\Myr$ following star formation, the typical age of the
lightest star that explodes as a type-II, core-collapse supernova.
The heating rate due to feedback may or may not overcome the cooling
rate, depending on the gas conditions in the star-forming regions
\citep{ds86,ceverino+k09}. We note that no artificial 
shutdown of cooling is implemented in these simulations.
On the other hand, we include the effect of runaway stars by applying a 
velocity kick of $\sim 10 \kms$ to 30\% of the newly formed stellar particles.
The code also includes the later effects of type-Ia supernova and
stellar mass loss, and it follows the metal enrichment of the ISM.
% RP new
Radiation pressure is incorporated through the addition of a non-thermal
pressure term to the total gas pressure in regions where ionizing photons
from massive stars are produced and may be trapped. This ionizing radiation
injects momentum in the cells neighboring massive star particles younger
than 5 Myr, and whose column density exceeds $10^{21}\cms$, isotropically
pressurizing the star-forming regions 
\citep[as described also in][Appendix B]{agertz13}.
More details are provided in \citet{ceverino14}.

% haloes
\smallskip
The initial conditions for the high-resolution, zoom-in, hydrodynamical
simulations that are used in this paper are based on dark-matter haloes that
were drawn from dissipationless $N$-body simulations at lower resolution
in three large comoving cosmological boxes.
The assumed cosmology is the standard $\Lambda$CDM model with the WMAP5 values
of the cosmological parameters, namely
$\omm=0.27$, $\oml=0.73$, $\omb= 0.045$, $h=0.7$ and $\sigma_8=0.82$
\citep{komatsu09}.
Each halo was selected to have a given virial mass at $z=1$.
%(or $z=0$ in a few cases).
The only other selection criterion was that they show no ongoing major merger
at $z=1$. This eliminates less than $10\%$ of the haloes which
tend to be in a dense environment at $z \sim 1$,
and it induces only a minor selection effect at higher redshifts.
The target virial masses at $z=1$ were selected to be in the range
$\Mv = 2\times 10^{11}-2\times 10^{12}\msun$, 
about a median of $4.6 \times 10^{11}\msun$. 
If left in isolation, the median mass at $z=0$ was intended to be 
$\sim 10^{12}\msun$, namely comparable to the Milky Way.
In practice, the actual mass range is broader, 
with some of the haloes merging into more massive haloes
that eventually host groups at $z=0$.

\smallskip
%zoom-in details 6
The initial conditions corresponding to each of the selected haloes
were filled with gas and refined to a much higher resolution on an adaptive
mesh within a zoom-in Lagrangian volume that encompasses the mass within
twice the virial radius at $z=0.5-1$, which is roughly a sphere of comoving
radius $1\Mpc$.
This was embedded in a comoving cosmological box of side that ranges from
$10$ to $40\hmpc$.
Each galaxy has been evolved with the full hydro ART and subgrid physics on an
adaptive comoving mesh refined in the dense regions
to cells of minimum size between $17.5$ and $35\pc$ in physical units at all
times.\footnote{This range is dictated by the fact that the refinement
level is changed discretely by a factor of two once the Hubble-expanding 
box has doubled in size.} 
This maximum resolution is valid in particular throughout the cold discs
and dense clumps, allowing cooling to $\sim 300$K and maximum gas densities of
$\sim 10^3\,{\rm cm}^{-3}$.
The force resolution is two grid cells, as required for computing the 
gradient of the gravitational potential.
The dark-matter particle mass is $8.3\times 10^4\msun$,
and the particles representing stars have a minimum mass of $10^3\msun$.
% refinement new 
Each AMR cell is split into 8 cells once it contains a
mass in stars and dark matter higher than $2.6 \times 10^{5}\msun$,
equivalent to three dark-matter particles, or once it contains
a gas mass higher than $1.5\times 10^6\msun$. This quasi-Lagrangian
strategy ends at the highest level of refinement that marks the
minimum cell size at each redshift. 

\begin{table*}
\centering
    \begin{tabular}{ccccccccccccc}
\hline
\hline
Galaxy & $\Mv$ & $\Ms$ & $\Rv$ & $\Re$
   & $\Mv$ & $\Ms$ & $\Rv$ & $\Re$ &$a_{\rm fin}$&$z_{\rm fin}$ \\
       & $10^{12}\msun$ & $10^{10}\msun$  & kpc & kpc
   & $10^{12}\msun$ & $10^{10} \msun$  & kpc  & kpc &\\
       & $(z=2)$ & $(z=2)$ &$(z=2)$ & $(z=2)$&
    $(z_{\rm fin})$ & $(z_{\rm fin})$ & $(z_{\rm fin}$) &$(z_{\rm fin}$) &&\\
\hline
V01  &  0.16  &  0.22  & 58.25  &  1.06  &  0.48  &  1.51  &123.75  &  2.18  &  0.50  &  1.00\\ 
V02  &  0.13  &  0.19  & 54.50  &  2.19  &  0.39  &  0.92  &115.25  &  2.09  &  0.50  &  1.00\\ 
V03  &  0.14  &  0.43  & 55.50  &  1.70  &  0.32  &  1.00  &108.00  &  1.91  &  0.50  &  1.00\\ 
V06  &  0.55  &  2.16  & 88.25  &  1.06  &  0.75  &  2.57  &108.75  &  1.13  &  0.37  &  1.70\\ 
V07  &  0.90  &  5.67  &104.25  &  2.78  &  1.51  &  7.06  &183.00  &  3.37  &  0.50  &  1.00\\ 
V08  &  0.28  &  0.35  & 70.50  &  0.76  &  1.20  &  3.37  &167.25  &  3.40  &  0.50  &  1.00\\ 
V09  &  0.27  &  1.06  & 70.50  &  1.82  &  0.80  &  4.18  &121.25  &  1.47  &  0.40  &  1.50\\ 
V10  &  0.13  &  0.64  & 55.25  &  0.53  &  0.73  &  2.38  &142.25  &  0.79  &  0.50  &  1.00\\ 
V11  &  0.27  &  0.91  & 69.50  &  2.98  &  0.38  &  1.55  &105.75  &  3.12  &  0.46  &  1.17\\ 
V12  &  0.27  &  2.03  & 69.50  &  1.22  &  0.28  &  2.22  & 93.00  &  1.32  &  0.44  &  1.27\\ 
V13  &  0.31  &  0.69  & 72.50  &  3.21  &  0.56  &  2.08  &108.50  &  4.25  &  0.40  &  1.50\\ 
V14  &  0.36  &  1.30  & 76.50  &  0.35  &  0.28  &  2.78  & 86.25  &  0.70  &  0.41  &  1.44\\ 
V15  &  0.12  &  0.56  & 53.25  &  1.31  &  0.35  &  1.04  &111.50  &  1.95  &  0.50  &  1.00\\ 
V20  &  0.53  &  3.70  & 87.50  &  1.81  &  1.06  &  6.87  &146.25  &  3.74  &  0.44  &  1.27\\ 
V21  &  0.62  &  4.10  & 92.25  &  1.76  &  0.86  &  5.74  &151.50  &  3.53  &  0.50  &  1.00\\ 
V22  &  0.49  &  4.45  & 85.50  &  1.32  &  0.62  &  4.51  &136.00  &  1.92  &  0.50  &  1.00\\ 
V23  &  0.15  &  0.83  & 57.00  &  1.38  &  0.47  &  2.51  &123.00  &  1.98  &  0.50  &  1.00\\ 
V24  &  0.28  &  0.92  & 70.25  &  1.79  &  0.36  &  2.15  &108.25  &  1.73  &  0.48  &  1.08\\ 
V25  &  0.22  &  0.73  & 65.00  &  0.82  &  0.32  &  1.39  &108.00  &  1.11  &  0.50  &  1.00\\ 
V26  &  0.36  &  1.60  & 76.75  &  0.76  &  0.42  &  2.14  &120.00  &  1.97  &  0.50  &  1.00\\ 
V27  &  0.33  &  0.80  & 75.50  &  2.45  &  0.35  &  1.86  &114.50  &  3.99  &  0.50  &  1.00\\ 
V29  &  0.52  &  2.34  & 89.25  &  1.96  &  0.90  &  3.33  &152.50  &  2.78  &  0.50  &  1.00\\ 
V30  &  0.31  &  1.66  & 73.25  &  1.56  &  0.32  &  1.67  & 76.25  &  1.64  &  0.34  &  1.94\\ 
V32  &  0.59  &  2.68  & 90.50  &  2.60  &  0.59  &  2.68  & 90.50  &  2.60  &  0.33  &  2.03\\ 
V33  &  0.83  &  4.80  &101.25  &  1.22  &  1.46  &  8.91  &143.75  &  1.63  &  0.39  &  1.56\\ 
V34  &  0.52  &  1.61  & 86.50  &  1.90  &  0.62  &  1.90  & 97.00  &  2.06  &  0.35  &  1.86\\ 
\hline
\hline
\end{tabular}
\caption{The suite of 26 simulated galaxies.
The galaxy name Vxx is short for VELA\_V2\_xx.
Quoted are the
total mass, $\Mv$, the stellar mass, $\Ms$, the virial radius $\Rv$ and the
effective stellar (half-mass) radius $\Re$
both at $z=2$ and at the final simulation snapshot,
$a_{\rm fin}=(1+z_{\rm fin})^{-1}$.
%\adr{Nir says that for V11-14 afin should be smaller by about 0.04.
%I think it doesn't matter, and willing to use the larger values.}
}
\label{tab:sim_table}
\end{table*}

%------------------
\subsection{The sample of galaxies: physical quantities}

\smallskip % a_fin 
We start the analysis at the cosmological time corresponding to expansion
factor $a=0.125$ (redshift $z=7$). At earlier times, the fixed resolution scale
typically corresponds to a larger fraction of the galaxy size, 
which may bias some of the quantities that we wish to study here.
Most galaxies reach $a=0.50$ ($z=1$). 
The output of each simulation is analyzed at output times separated by a
constant interval in $a$, $\Delta a =0.01$ (which at $z=2$ corresponds to about
$100\Myr$).

\smallskip % table 1  Mv, Rv, afin
Global properties of the galaxies in our sample are listed in \tab{sim_table}.
This includes the total virial mass $\Mv$ and virial radius $\Rv$, 
the galaxy stellar mass $\Ms$, and the effective, half-mass radius $\Re$,
both at $z=2$ and at the last available time for each simulation. 
The latest time of analysis for each galaxy in terms of the expansion factor,
$a_{\rm fin}$, and redshift, $z_{\rm fin}$, is provided.
The virial mass $\Mv$ is the total mass within a sphere of radius $\Rv$
that encompasses an
overdensity of $\Delta(z) = (18\pi^2-82\oml(z)-39\oml(z)^2)/\omm(z)$,
where $\oml(z)$ and $\omm(z)$ are the cosmological parameters at $z$.
\citep[][Appendix A1]{bryan98,db06}.

\smallskip

The stellar mass of the galaxy, $\Ms$, is the instantaneous mass in stars
(after the appropriate mass loss),
measured within a sphere of radius $10\kpc$ about the galaxy center.
The effective radius $\Re$ is the three-dimensional half-mass radius 
corresponding to this $\Ms$.
Compactness is measured in terms of the 
stellar surface density within the effective radius or within the inner
$1\kpc$, $\Se$ and $\Skpc$ respectively.
In practice, for either $r=\Re$ or $r=1\kpc$,
we measure $\Sigma_r=\Ms(r)/(\pi r^2)$, where $\Ms(r)$ is the
stellar mass within a three-dimensional sphere of radius $r$,
while observationally it is measured in two-dimensional projection along a
given line of sight, using the two-dimensional half-mass radius.
The surface density measured using the three-dimensional quantities
turns out to be only a $10-20\%$ underestimate of the surface density 
observed in two dimensions for $r\!=\!\Re$. The value of $\Skpc$ is thus
a $10-20\%$ underestimate when $\Re\!\sim\!1\kpc$, and it is a better 
approximation when $\Re$ is significantly smaller than $1\kpc$.

\smallskip % SFR, sSFR

The SFR is measured within spheres of radius $10\kpc$ or $1\kpc$,
as the initial stellar mass (before mass loss) 
in stars younger then $\Delta t$ divided by $\Delta t$,
for $\Delta t\!=\!60\Myr$\footnote{This is a crude proxy for
SFR estimates based on H$_\alpha$ measurements, while UV-based estimates
are sensitive to stars younger than $\sim\!100\Myr$.}.
In practice, in order to reduce fluctuations due to a $\sim\!5\Myr$
discreteness in stellar birth times in the simulation,
we average the SFR as deduced using different $\Delta t$ 
values equally spaced ($0.2\Myr$) in the range $\Delta t\!=\!40\!-\!80\Myr$.
The specific star-formation rate in the corresponding volume is simply 
${\rm sSFR}\!=\!{\rm SFR}/\Ms$.

%----------------
\subsection{Limitations of the current simulations}

%Limitations:
These simulations are state-of-the-art in terms of the high-resolution
AMR hydrodynamics and the treatment of key physical processes at the
subgrid level. In particular, they properly trace the cosmological streams
that feed galaxies at high redshift, including mergers and smooth flows,
and they resolve the violent disc instability that governs the high-$z$ disc
evolution and the bulge formation 
\citep{cdb10,ceverino12,ceverino14_e,mandelker14}.
AMR codes more accurately trace some of the high-resolution hydrodynamical
processes involved in galaxy formation than SPH codes that use the traditional
density formulation \citep[e.g.][]{agertz07,scannapieco12,bauer12}.
They are comparable in accuracy to codes using modern formulations of SPH
\citep{hopkins14,schaye15} or a moving unstructured grid \citep{bauer12},
but implementations of the latter in a cosmological context have not yet
reached the resolution currently achieved with AMR codes.

\smallskip
Like other simulations,
the simulations used in this paper
are not yet doing the most accurate possible job
in treating the star formation and feedback processes.
For example,
while the code now assumes a SFR efficiency per free-fall time that is more
realistic than in earlier versions,
it does not yet follow in detail the formation of molecules
and the effect of metallicity on SFR \citep{kd12}.
Furthermore, the resolution does not allow the capture of the Sedov-Taylor
adiabatic phase of supernova feedback. 
The radiative stellar feedback assumed no infrared trapping,
in the spirit of the low trapping advocated by
\citet{dk13} based on \citet{krum_thom13}. 
On the other hand, other works assume more significant trapping 
\citep{murray10,kd10,hopkins12c}, which makes the assumed
strength of the radiative stellar feedback still somewhat ad hoc.
Finally, AGN feedback, and feedback associated with cosmic rays and magnetic
fields, are not yet incorporated.
%\citep{silk98,hopkins06,booth09,cattaneo09}.
As shown in \citet{ceverino14},
the star formation rates, gas fractions, outflow rates,
and stellar to halo mass fractions are all in the ballpark of the estimates 
deduced from observations, providing a better match to observations than 
earlier versions of the ART simulations, but
this match is still only at the semi-quantitative level, with an accuracy at the level of a factor $\sim\!2$.

\smallskip
As a result of this non-perfect match, the dramatic events in the evolution of
galaxies that concern us here may occur somewhat earlier than in the 
real Universe.
In particular, we will see that compaction and onset of quenching occurs in 
some of our galaxies at very high redshifts, possibly too early. 
On the other hand, with some of the feedback mechanisms not yet incorporated
(e.g., resolved supernova feedback and AGN feedback),
full quenching to very low sSFR values is not fully materialized in many cases 
by the end of the simulation at $z \sim 1$.
We adopt the hypothesis that these inaccuracies are not of a qualitative nature,
and assume that the simulations are accurate enough for
acquiring a basic qualitative understanding of the phenomena of compaction
and quenching and the processes that drive them.

\smallskip
Additional analysis of the same suite of simulations,
especially the properties of giant clumps in VDI discs,
are discussed in \citet{moody14} and \citet{snyder14}.

%\subsection{Group Finding}
%We trace each star particle located within the main galaxy at
%the final snapshot back to higher redshift using the hierarchical 
%group-finder {\tt AdaptaHOP}
%\citep{aubert04, tweed09}. This fully topological code was modified to
%run on the stellar particles of the {\tt Hydro ART} simulations outputs.

%At each time step we identify the dark matter
%(DM) halo to which each particle belonged using {\tt AdaptaHOP}. Stars are
%considered bound to a DM halo if they are identified as being a
%member of that halo for two consecutive time steps. We follow
%the particlesÕ histories back to z = 9. This procedure results
%in four different classifications for stars origin: accreted, in situ,
%ambiguous, and other. These classifications are described in the
%following sections.
%\smallskip
%\adr{A proper description of the merger tree is necessary 
%if we wish to mark mergers in \fig{ssfr_sigma_ex}.
%}

\bigskip
%%%%%%%%%%%%%%%%%%%%%%%%3
\section{Wet Bulge Formation}
\label{sec:wet_bulge}

The in-situ star formation in the central bulge is a measure of the 
wetness of bulge formation, namely gas contraction into the bulge at a rate
faster than the SFR along the way (DB14). 
A large fraction of in-situ star formation in the bulge would therefore be
evidence for wet compaction and a blue-nugget phase.
Given the typically low SFR in galactic spheroids at moderate and late 
redshifts, 
a high fraction of in-situ SFR in high-$z$ bulges is not at all obvious. 
We next describe how we identify the bulge stars at a given redshift. 
Then we trace the birthplaces of these stars, and compute the fraction of the
stars that were born in-situ in the bulge of the main progenitor. 
The remaining stars in the bulge either formed in the disc and migrated inwards
to the bulge or formed ex-situ in other galaxies that have merged with 
the main galaxy.

%Because the main focus of this paper is the stellar component in the 
%central-most regions of hi-z galaxies, we first discuss the physical 
%processes involved in the formation of stellar bulge in our simulations.  
%We begin with a brief discussion of the kinematic decomposition and group 
%finding procedures we employ, and then continue in Section 3.2 to show that 
%at $z>3$, more than half of the stellar mass contained within the spheroid 
%formed in situ.

%-------------------
\subsection{Kinematic Decomposition}

A decomposition of the stellar component in every snapshot of each galaxy
into disc and spheroid has been carried out based on kinematics
\citep[as in][]{ceverino14_e}. 
We first define the spin axis of the galaxy using the stars 
within a face-on projected radius of $10\kpc$. 
%$0.1\Rv$ as described in \citet{mandelker14} for the gas
%but here applied to the stars.
%\adr{Adi: the above is from Dylan, but somebody else, possibly Nir, said 
%$10\kpc$. Nir now says that what he did for the gas AM was: 
%``AM within a cylinder of thickness $\pm 1\kpc$ and 
%a radius containing 85\% of the mass out to $0.15\Rv$".
%So, what was applied to the stars here?}
We then assign each stellar particle a ratio $j_z/j_{\rm max}$, where $j_z$ 
is the specific angular momentum with respect to the galaxy centre
along the spin axis, and $j_{\rm max}= r v$ is the maximum specific angular 
momentum the star particle could have with its given energy at its distance 
$r$ from the centre and with its given speed $v$.
A star on a co-rotating circular orbit has $j_z/j_{\rm max} = 1$. 
Disc stars are selected with a cut of $j_z/j_{\rm max} \geq 0.7$ while
the remaining stars are assigned to the spheroid.
The spheroid is further divided into a bulge and a halo based on the radial
distance from the galaxy centre, where the bulge radius 
is defined as the half-mass radius of the stellar component of the galaxy, 
$\Re$.
%of the stellar spheroid, xxx 
%%excluding clumps that largely represent merging galaxies. 
%%Clumps are identified by applying the hierarchical group-finder AdaptaHOP 
%%\citep{tweed09,colombi13} to the stellar particles.
%%In the current set of simulated galaxies, across the whole redshift range,
%%the clumps may account for $0-35\%$ of the extended stellar spheroid mass, 
%%but only $0-10\%$ of the stellar bulge, with their distribution peaking below
%%2\%.
%
%Clumps are defined as local topological over-densities within the galaxy. 
%Ex-situ and In-situ clumps are further distinguished from studying the
%origin of the star particles they contains. In-situ clumps are dominated by
%stars born within the clumps, while Ex-situ clumps are dominated by stars born
%ex-situ.

\begin{figure} % 1 wet_bulge 
%\vskip 5.7cm
\includegraphics[width=0.48\textwidth]{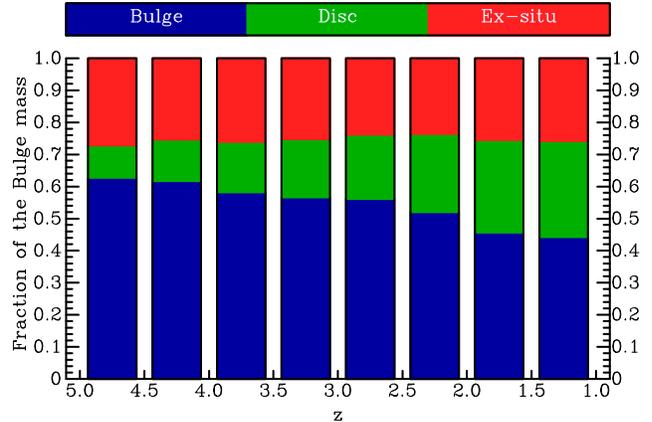}
\caption{Wet bulge formation.
Shown are the fractions of bulge stars at $z$ according to their
birth place with respect to the main-progenitor galaxy, averaged over all the
simulated galaxies and snapshots in the sample.
The bulge stars either formed in-situ in the bulge (blue), or
formed in the disc and migrated to the bulge (green), or formed ex-situ
outside the main-progenitor galaxy and joined the bulge through a merger --
major, minor or mini-minor (red).
At $z \sim 2-3$, more than half  the bulge stars have formed in-situ in the
bulge, indicating a rather wet bulge formation.
%\adr{For immediate submission use $\Re$ (or even this version). 
%We may eventually replace by $1\kpc$.}
}
\label{fig:wet_bulge}
\end{figure}

%--------------------------
\subsection{In Situ Star Formation in the Bulge}

\Fig{wet_bulge} shows the fraction of 
bulge stars that have formed in any of three different locations with
respect to the main-progenitor galaxy, namely, 
(a) in the bulge itself (blue), (b) in the disc and migrated into the bulge
(green), and (c) in external galaxies and joined the bulge via mergers (red). 
These fractions are averaged all simulated galaxies, and are shown in bins of
redshift in the range $z\!=\!5-1$.
One can see that the fraction of in-situ star formation in the bulge is high -- 
it gradually declines 
from 62\% at $z\!\sim\!5$ to 44\% at $z\!\sim\!1$.
This is clear evidence for wet compaction, preferentially at high redshifts.

\smallskip
We also see in \fig{wet_bulge} that
the fraction of stars that formed ex-situ to the galaxy and joined the bulge
by mergers is varying about 25\%.
This reflects the slow evolution of accretion rate in a growing galaxy
\citep{dekel13}, the slow growth of stellar fraction in the total 
accreted baryons \citep{oser10}, and the evolution of SFR in the
disc and bulge. 
%\adr{xxx The above is really not saying much, as I don't really understand 
%why the ex situ fraction is constant (and so low).}
We also note in passing that the fraction of bulge stars that formed
in the disc and migrated to the bulge, mostly by VDI-driven clump migration
\citep[e.g.,][]{noguchi98,bournaud07c,dsc09},
is growing systematically from $10\%$ at $z\!\sim\!5$ to 30\% at $z\!\sim\!1$.

%%%%%%%%%%%%%%%%%%%%%%%%%4
\section{Compaction and Quenching: Prototypical Cases}
\label{sec:evolution}

Before we present the results for the entire simulated sample, we first study
in this section the detailed evolution of eight individual galaxies from our
sample. We will show that these galaxies all undergo phases of dissipative
contraction followed by quenching attempts or full quenching. 
In the following section we will use the entire sample to show
that these phases are characteristic of the evolutionary pattern of high
redshift galaxies.

%----------------  4.1
\subsection{Massive Galaxies}
% high redshift, high compactness, efficient quenching}

\begin{figure*}  % 2 sSFR vs Sigma, examples of high Sigma
\centering
\vspace{-5mm}
\subfigure{\includegraphics[angle=-90,width=0.52\textwidth]
{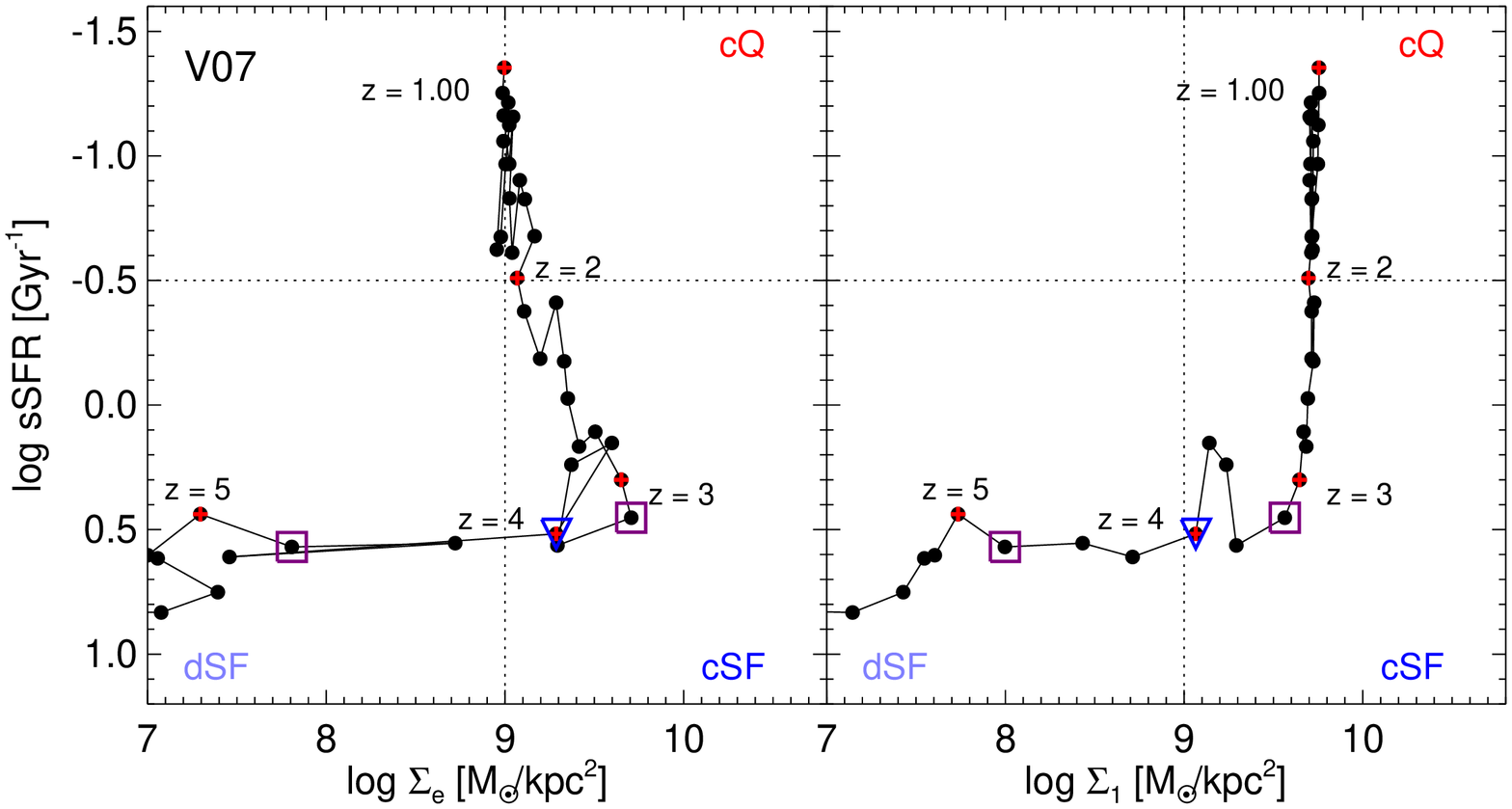}} 
\hspace{-10mm}
\subfigure{\includegraphics[angle=-90,width=0.52\textwidth]
{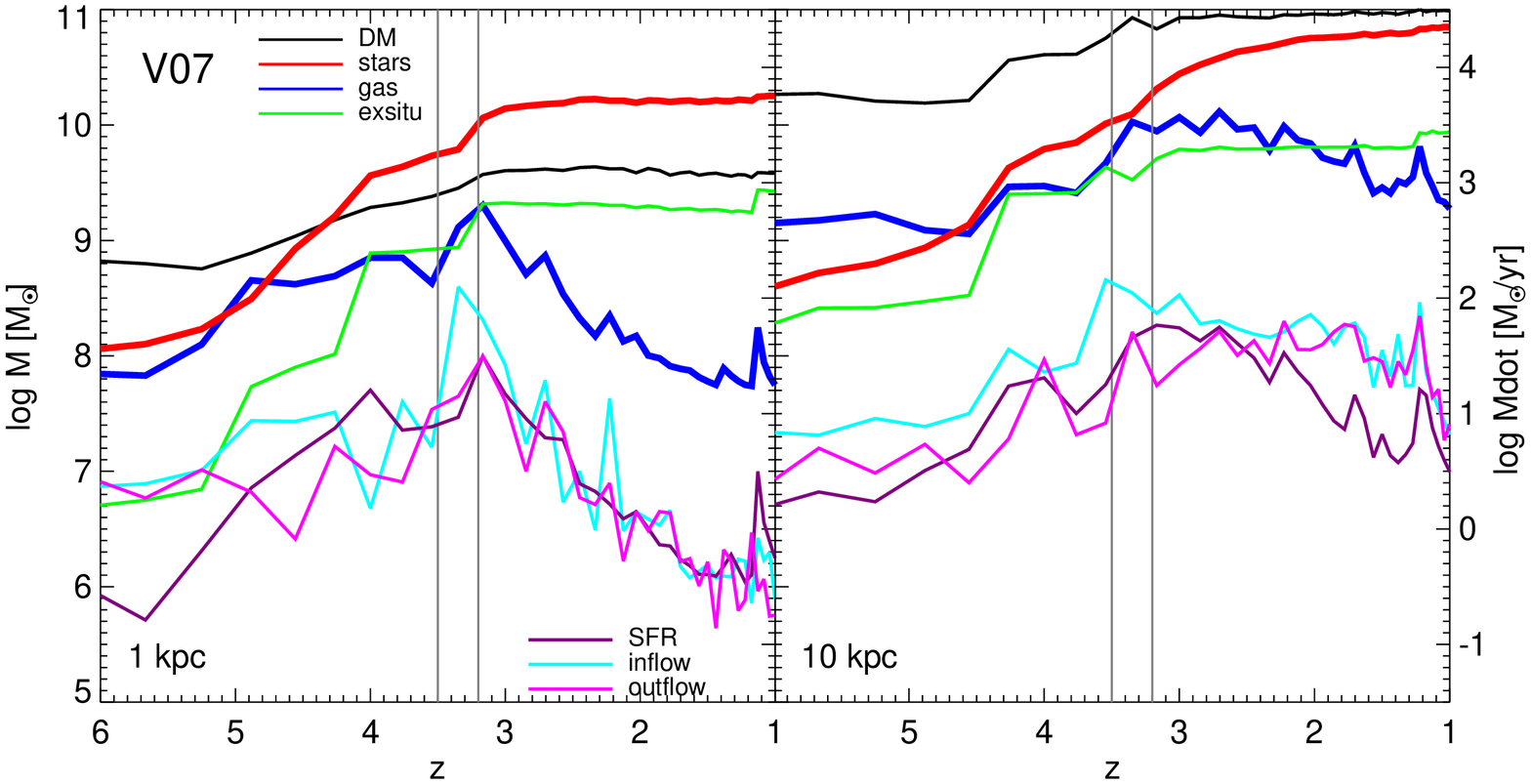}}
\vspace{-5mm}
\subfigure{\includegraphics[angle=-90,width=0.52\textwidth]
{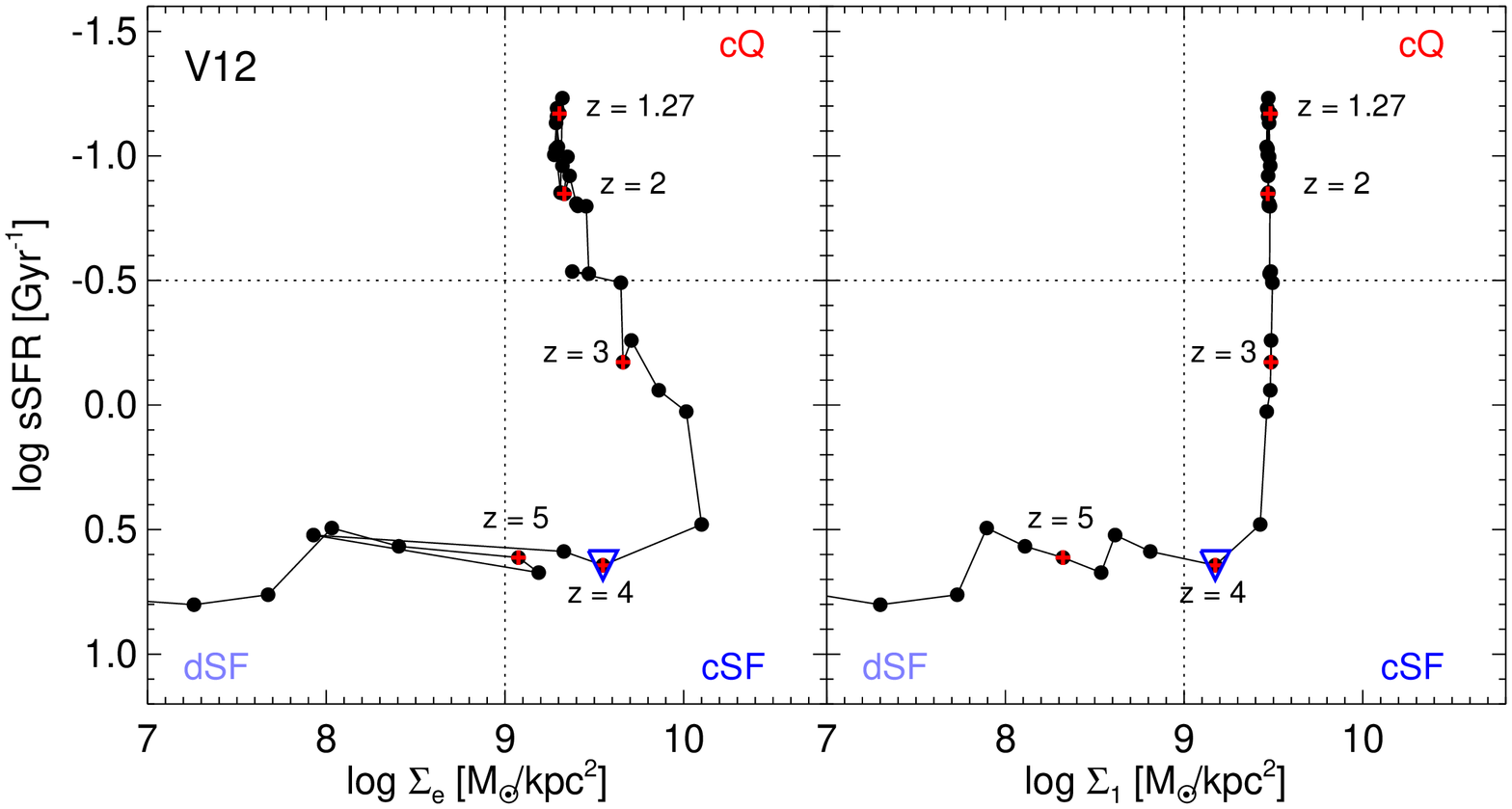}}
\hspace{-10mm}
\subfigure{\includegraphics[angle=-90,width=0.52\textwidth]
{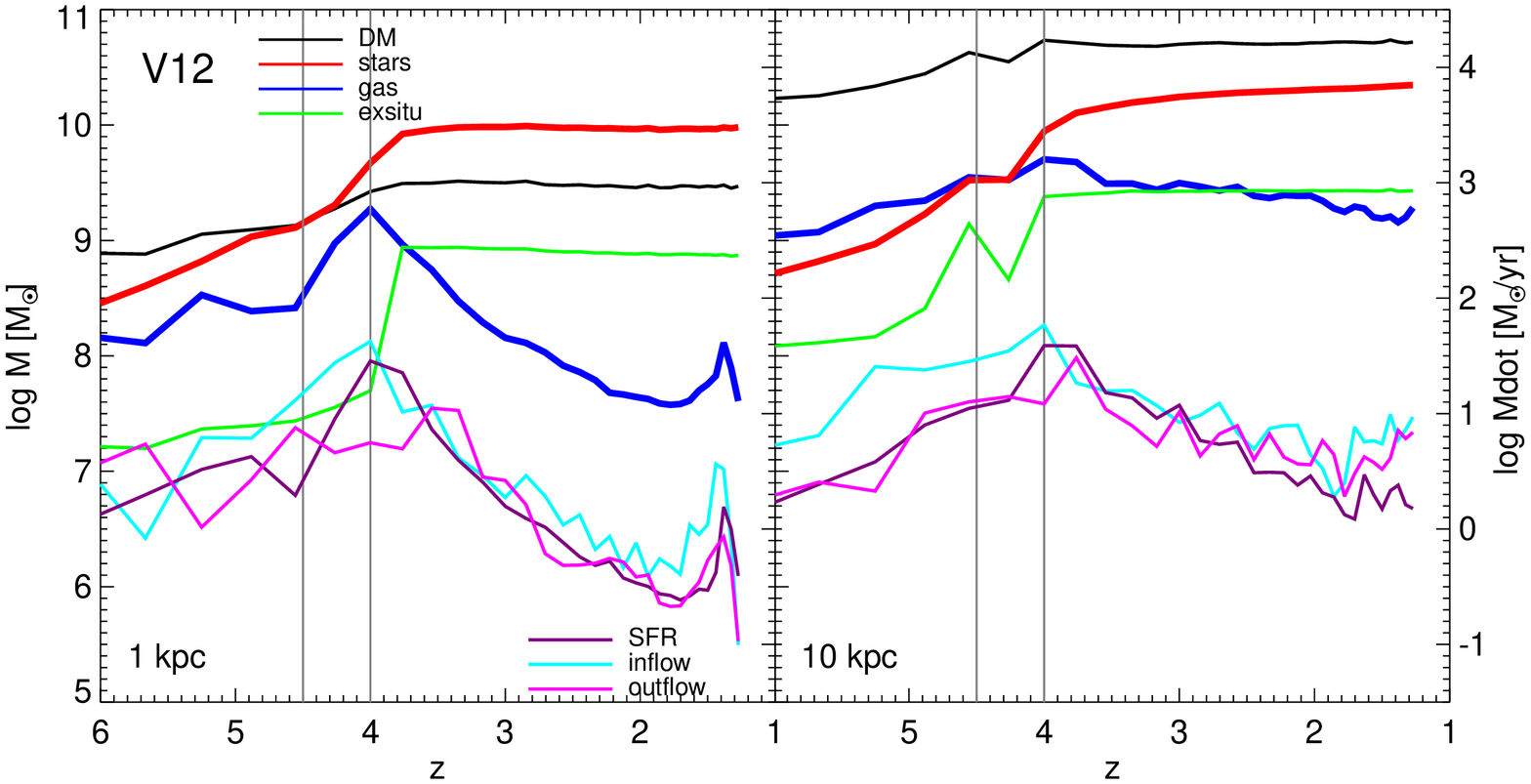}}
\vspace{-5mm}
\subfigure{\includegraphics[angle=-90,width=0.52\textwidth]
{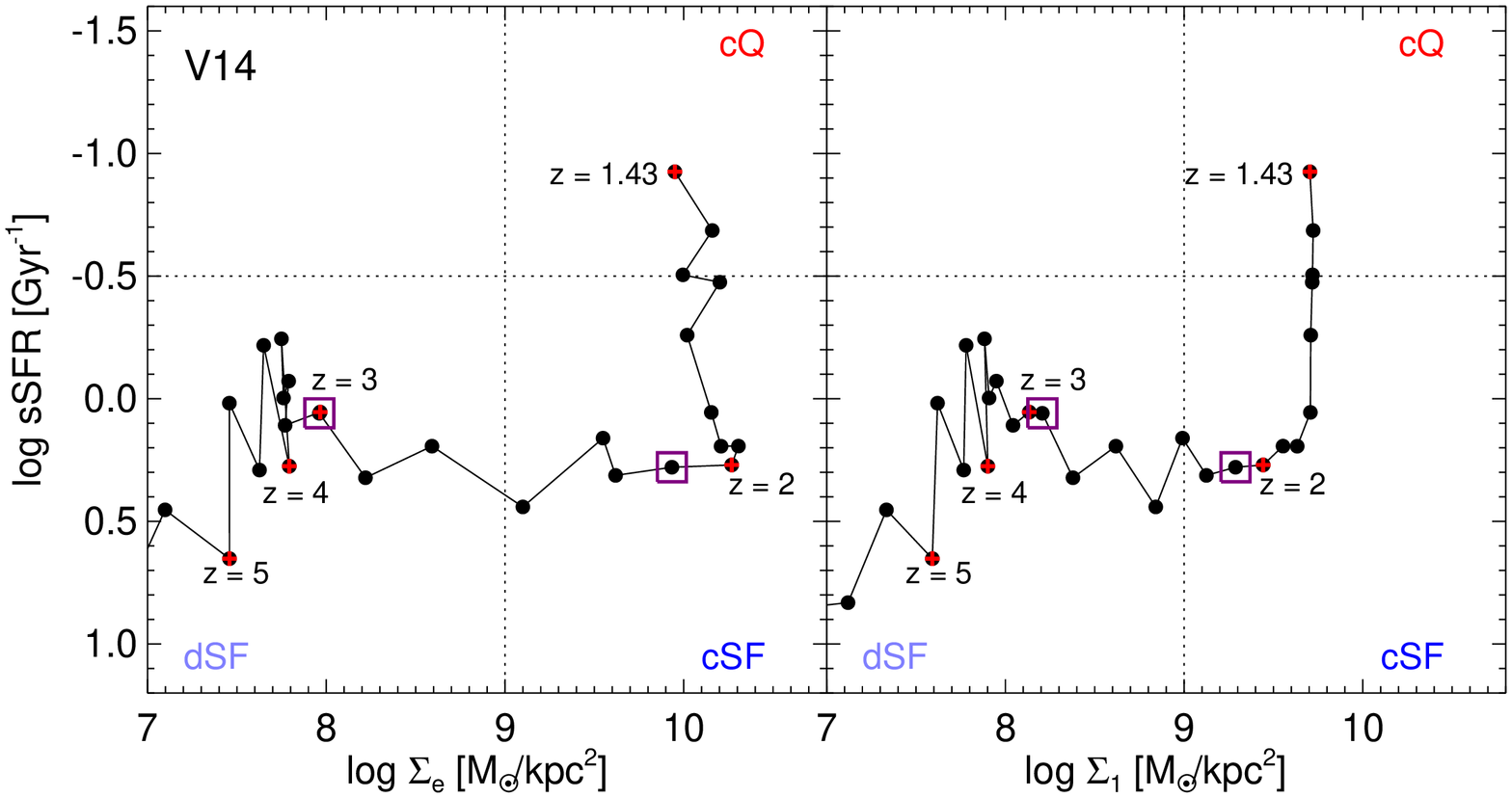}}
\hspace{-10mm}
\subfigure{\includegraphics[angle=-90,width=0.52\textwidth]
{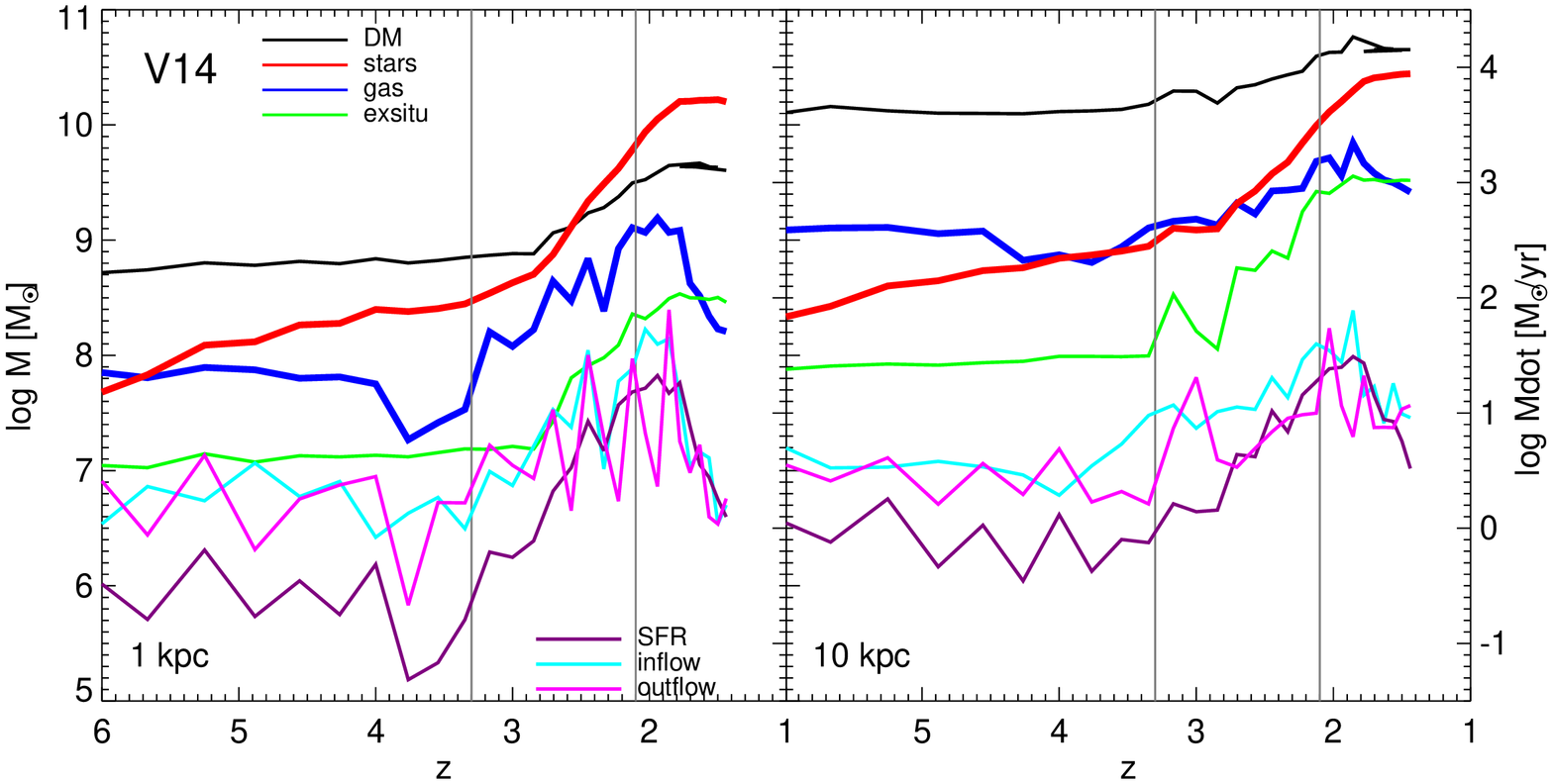}}
\subfigure{\includegraphics[angle=-90,width=0.52\textwidth]
{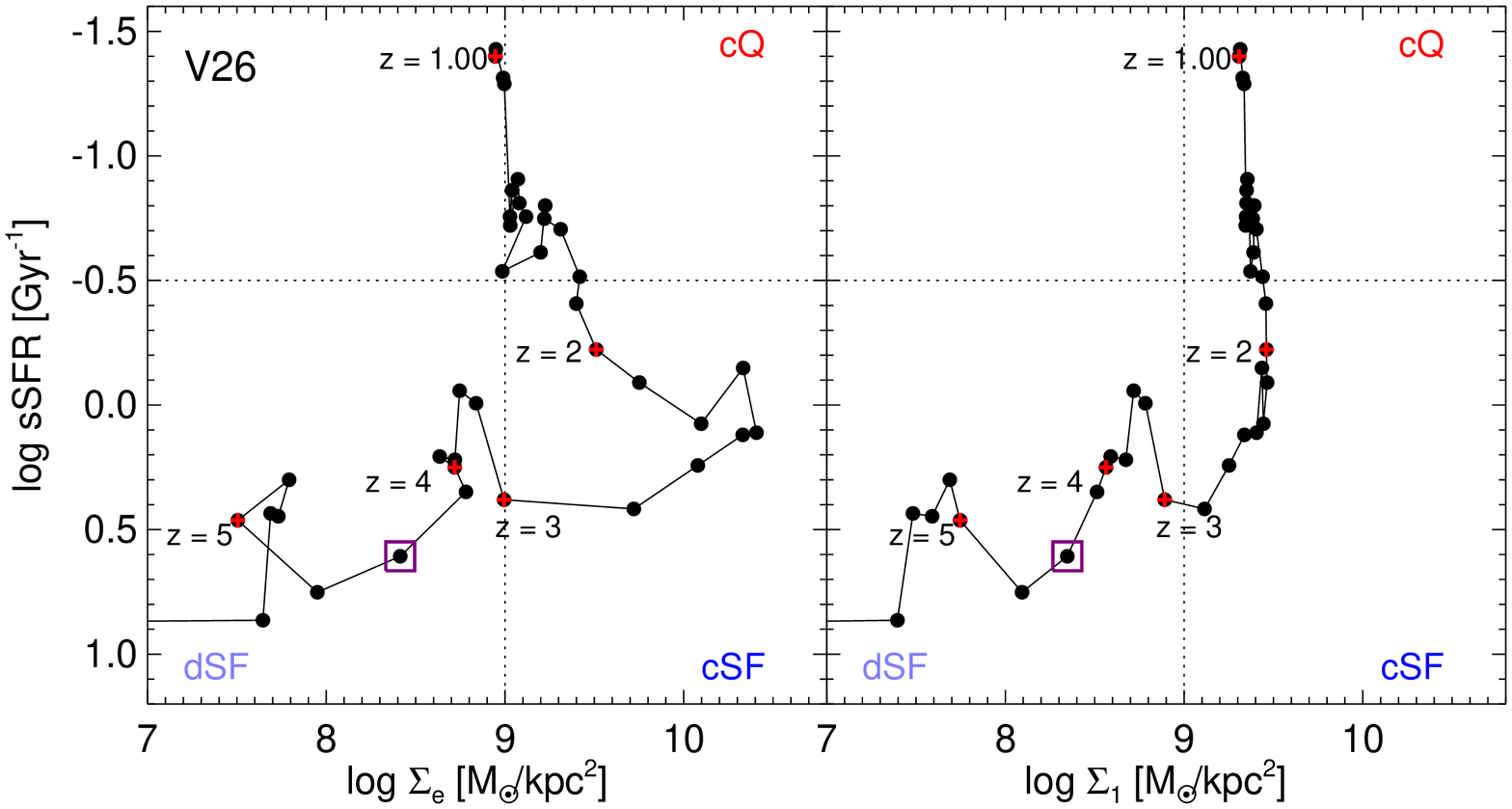}}
\hspace{-10mm}
\subfigure{\includegraphics[angle=-90,width=0.52\textwidth]
{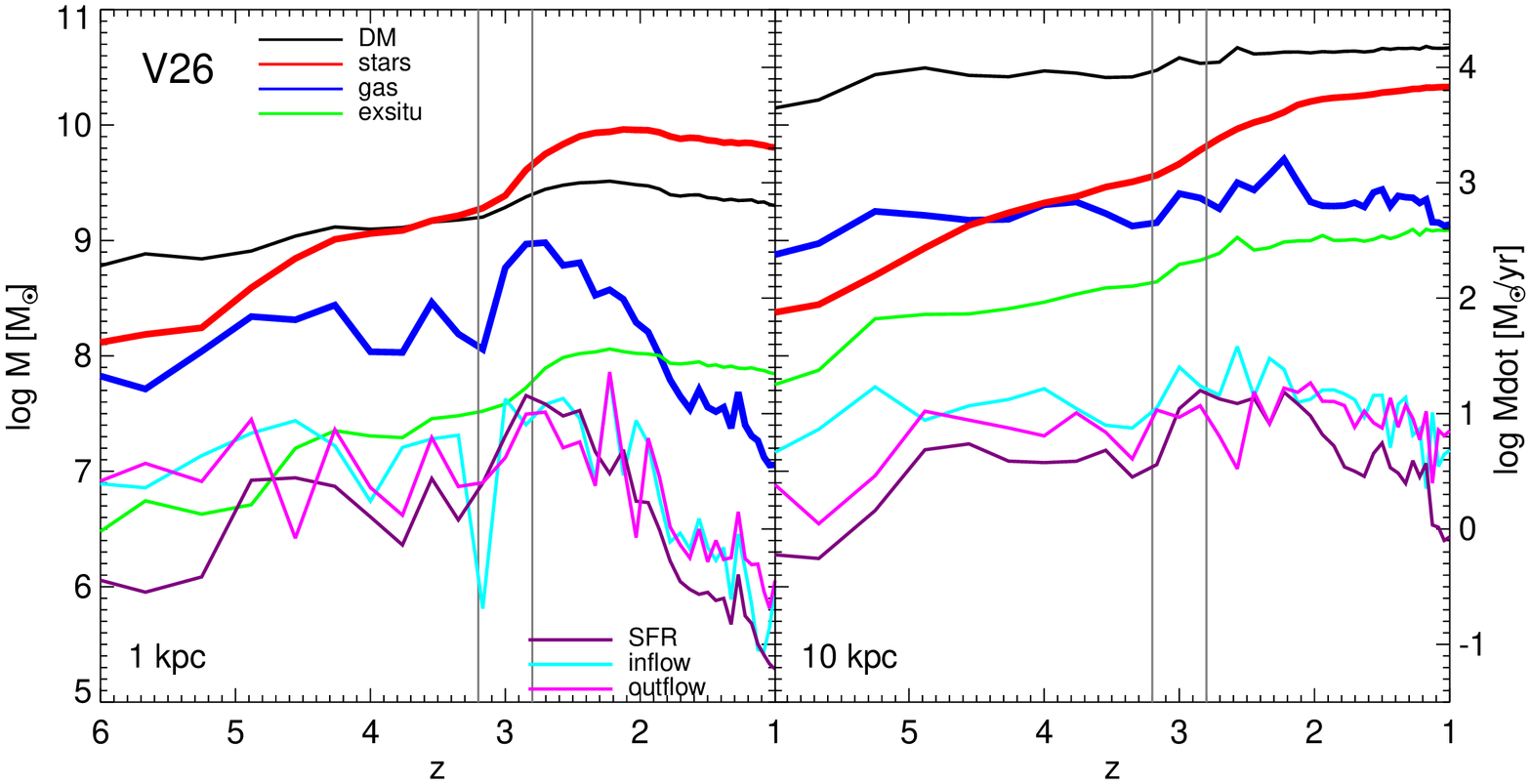}}
\caption{Evolution of four galaxies of relatively high stellar masses
that compactify at relatively high redshift to a high central
surface density and quench efficiently.
{\bf Two left panels:} 
Evolution tracks in sSFR and compactness as measured by 
$\Se$ (left) and $\Skpc$ (second from left).
The redshifts from $z=5$ to $z=1$ are marked along the tracks by red symbols.
Major mergers are marked by open blue upside-down triangles,
and minor mergers by open purple squares.
{\bf Two right Panels:}
Evolution of mass and its rate of change inside a central sphere of 
radius 1 kpc (second from right) and 10 kpc (right).
Shown at the top (scale along the left axis) are the masses 
in gas (blue), stars (red), and dark matter (black).
Also shown is the mass in ex-situ stars, as a merger indicator (green).
Shown at the bottom (scale along the right axis) are the rates of change
of gas mass
due to SFR (purple), gas inflow (cyan), and gas outflow (magenta). 
%%The right panel shows at the bottom in grey the evolution of the effective 
%%radius (scale for $\log (\Re/\!\kpc)+6$ along the left axis, 
%%and $1\kpc$ marked by a horizontal line).
Each of these galaxies shows at least one well-defined compaction phase 
that is immediately followed by gas depletion and quenching. 
The onset of gas compaction
in the central 1 kpc and the point of maximum central gas compaction 
are marked by vertical lines.
}
\label{fig:ssfr_sigma_ex}
\end{figure*}

\begin{figure*} % 3
\centering
\vspace{-5mm}
\subfigure{\includegraphics[angle=-90,width=0.52\textwidth]
{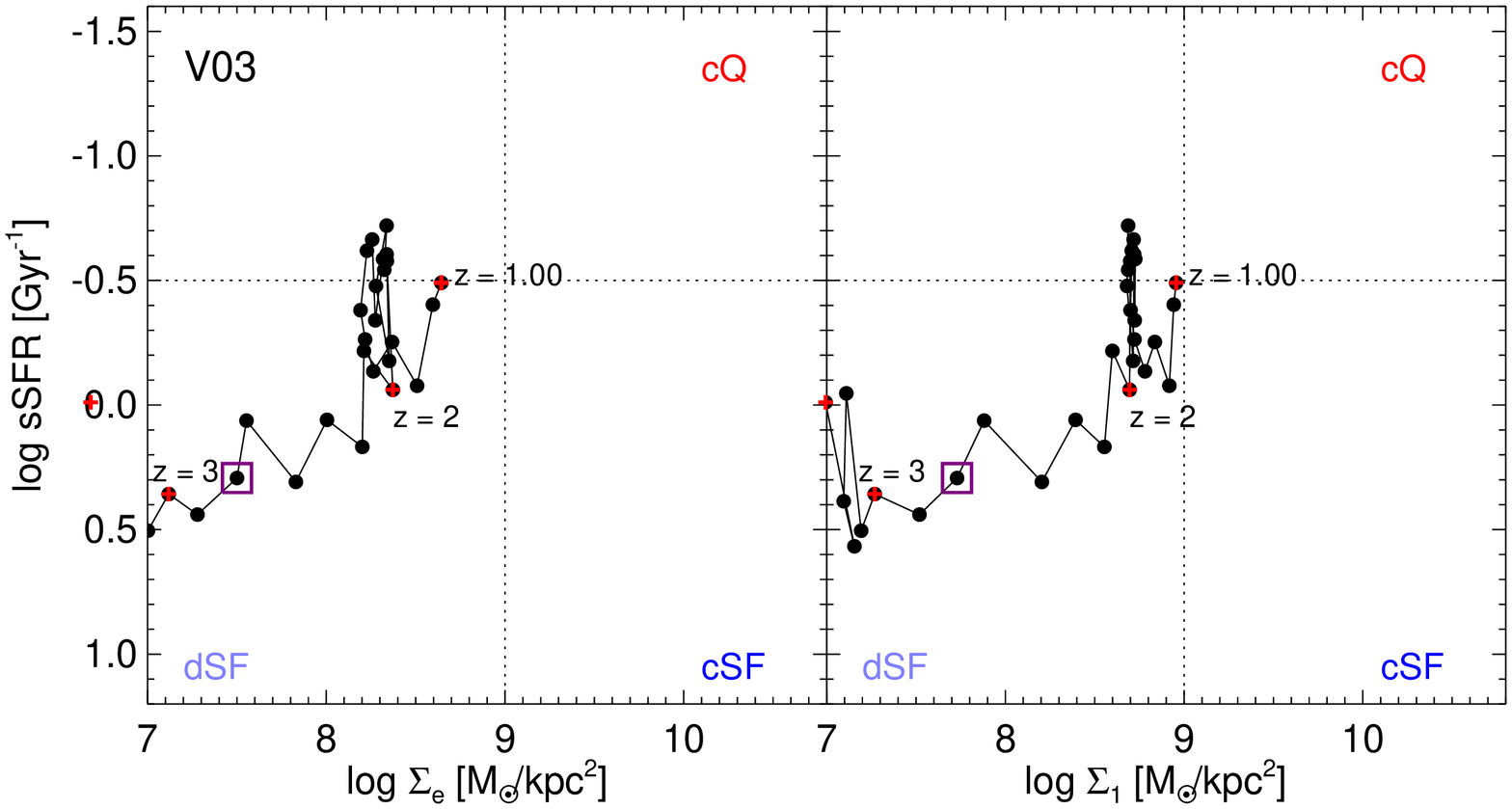}}
\hspace{-10mm}
\subfigure{\includegraphics[angle=-90,width=0.52\textwidth]
{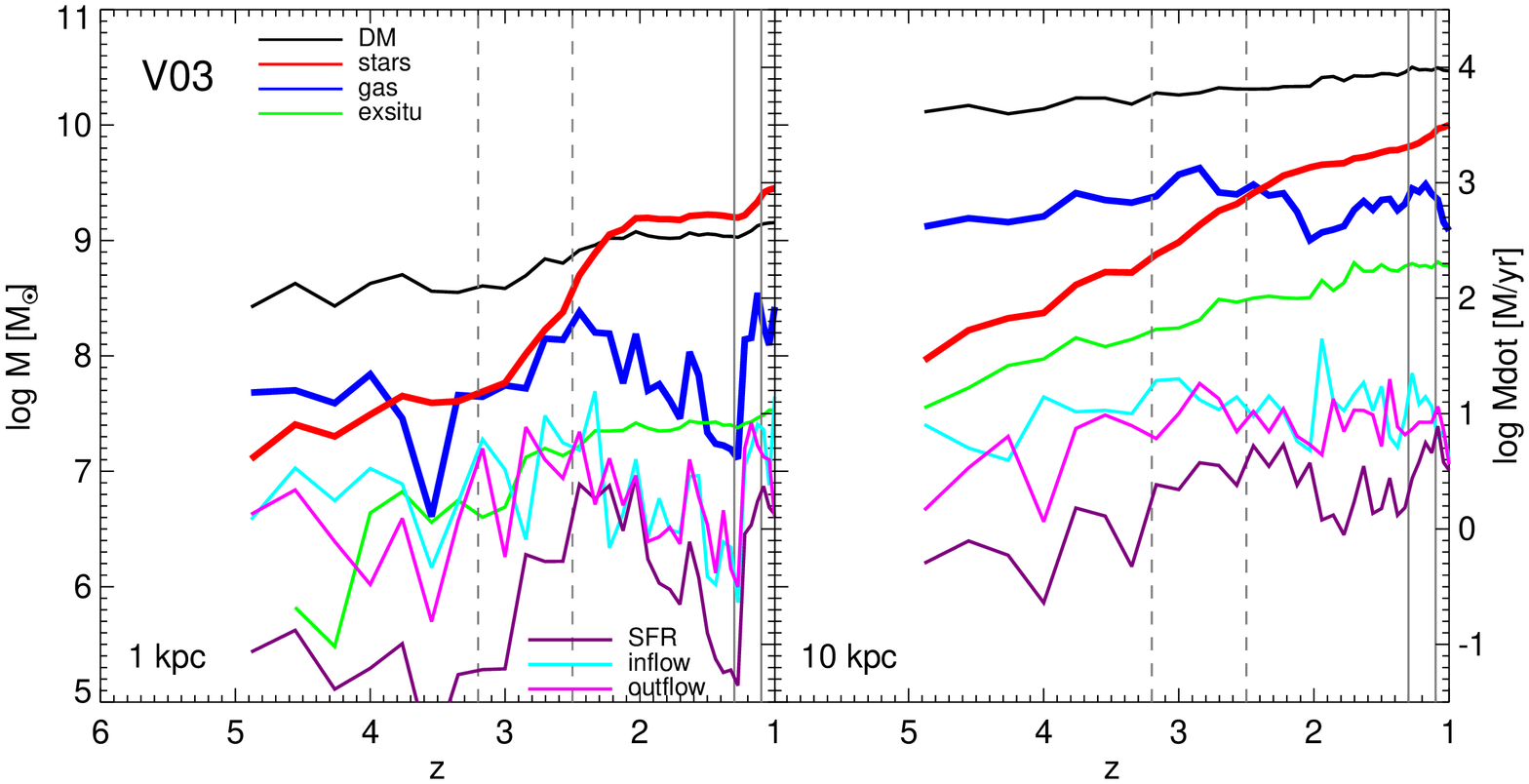}}
\vspace{-5mm}
\subfigure{\includegraphics[angle=-90,width=0.52\textwidth]
{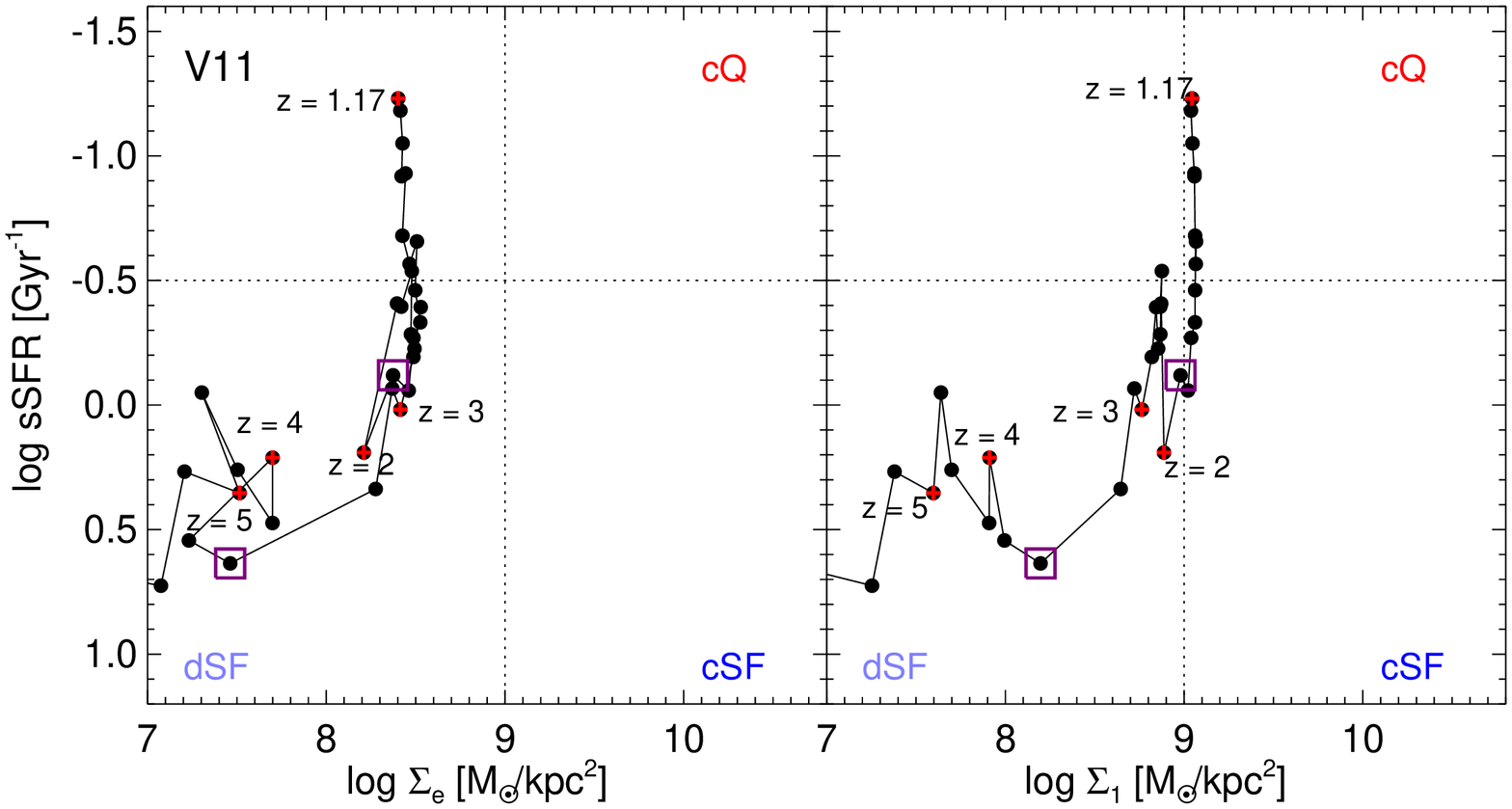}}
\hspace{-10mm}
\subfigure{\includegraphics[angle=-90,width=0.52\textwidth]
{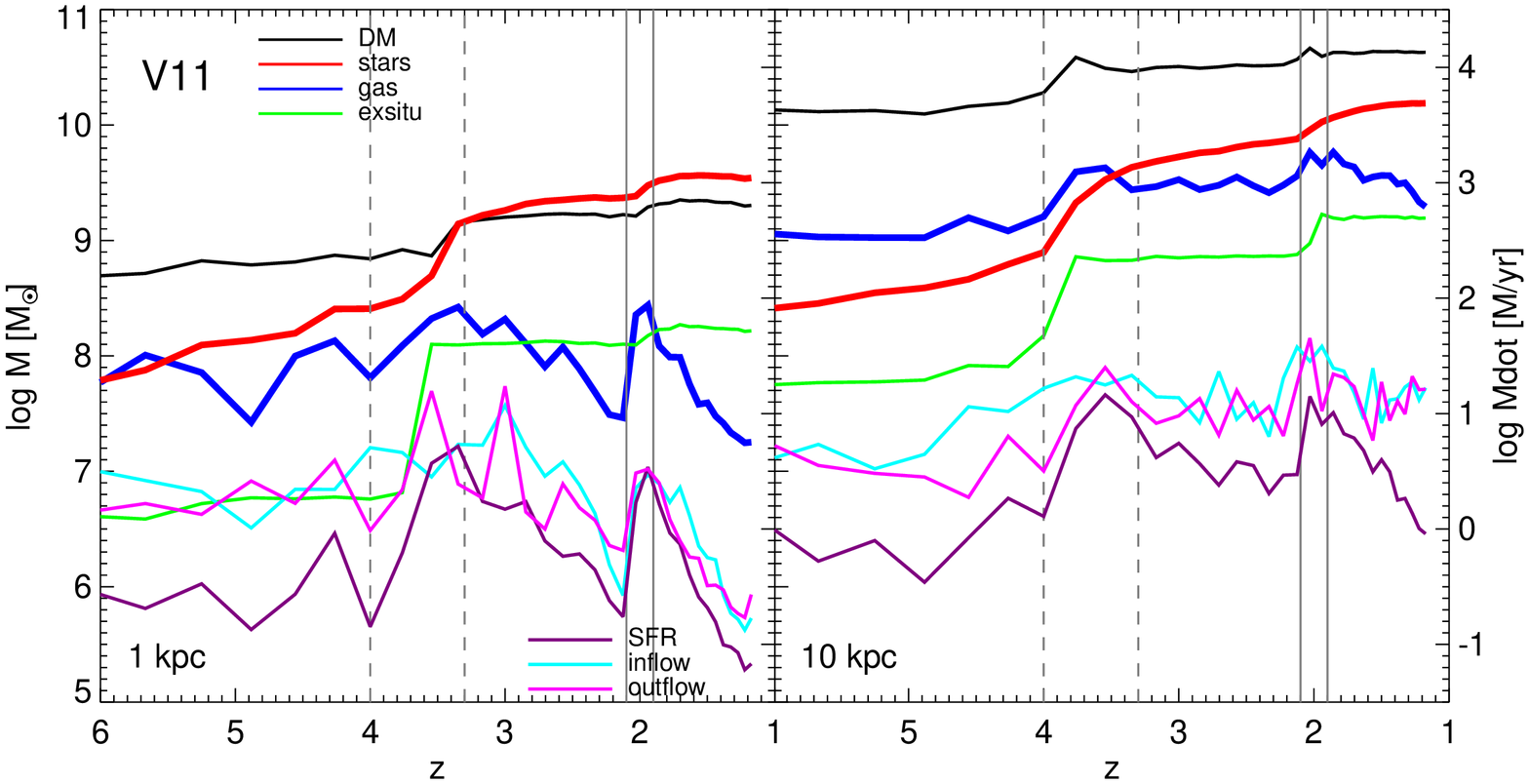}}
\vspace{-5mm}
\subfigure{\includegraphics[angle=-90,width=0.52\textwidth]
{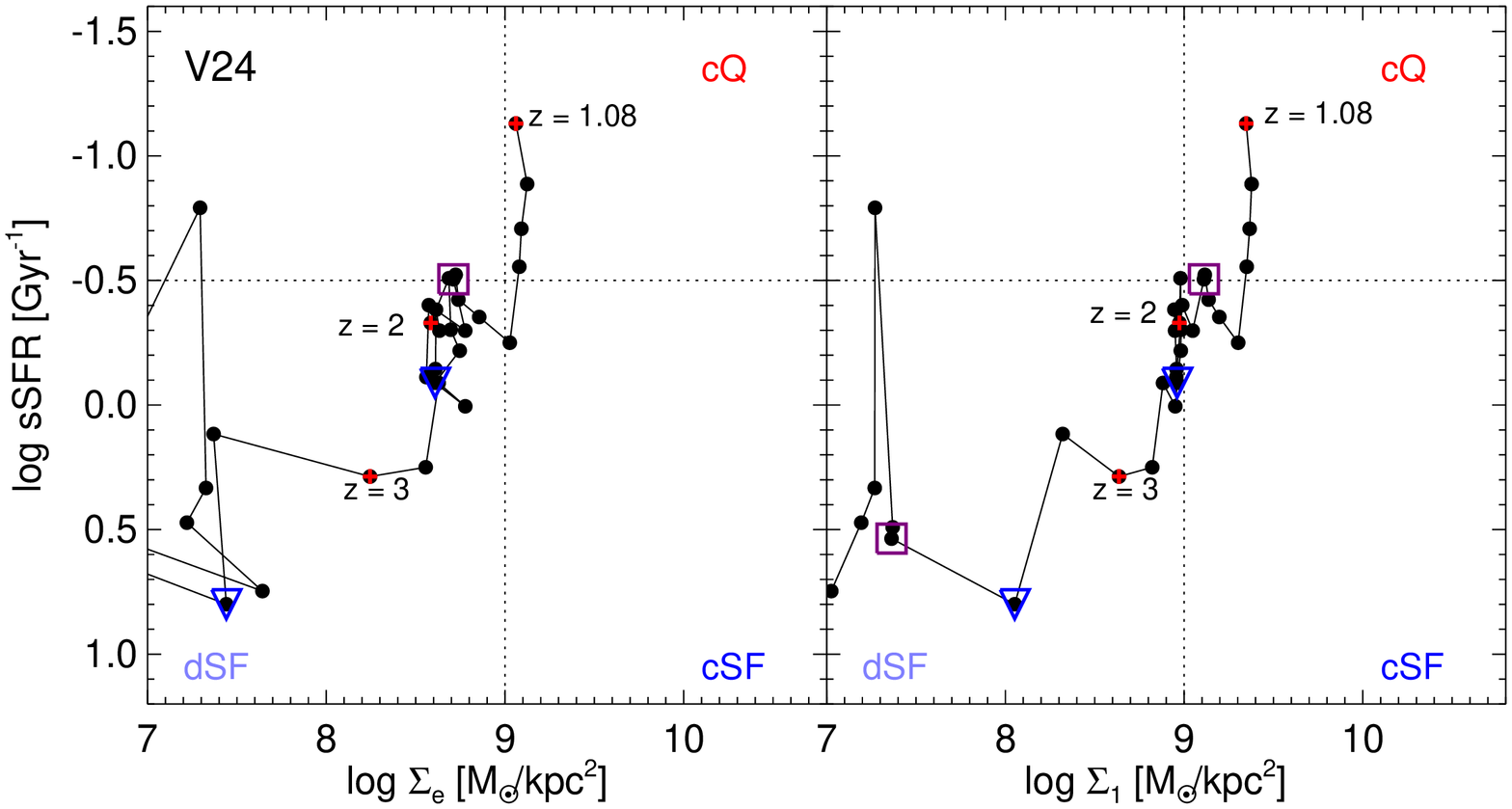}}
\hspace{-10mm}
\subfigure{\includegraphics[angle=-90,width=0.52\textwidth]
{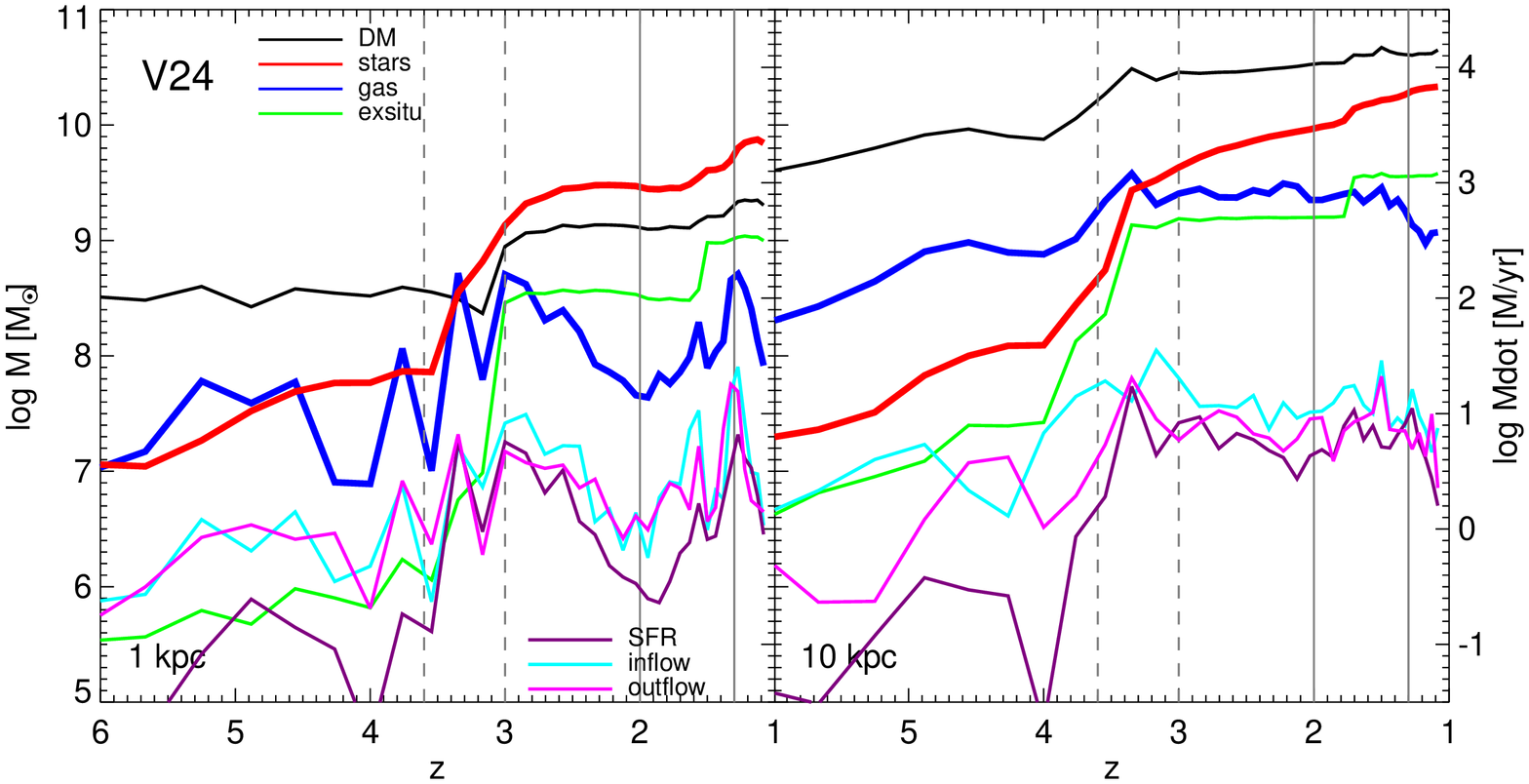}}
\subfigure{\includegraphics[angle=-90,width=0.52\textwidth]
{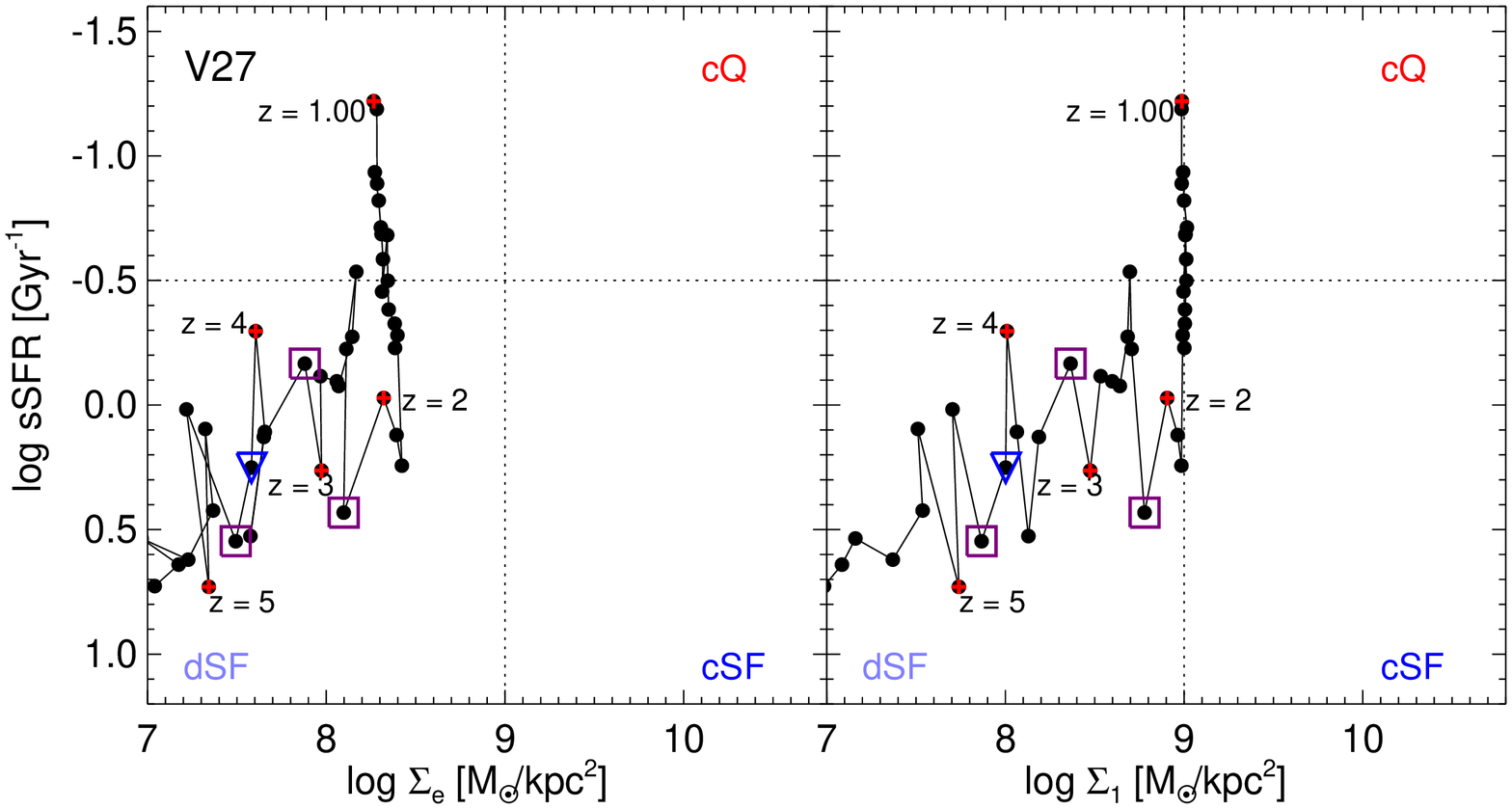}}
\hspace{-10mm}
\subfigure{\includegraphics[angle=-90,width=0.52\textwidth]
{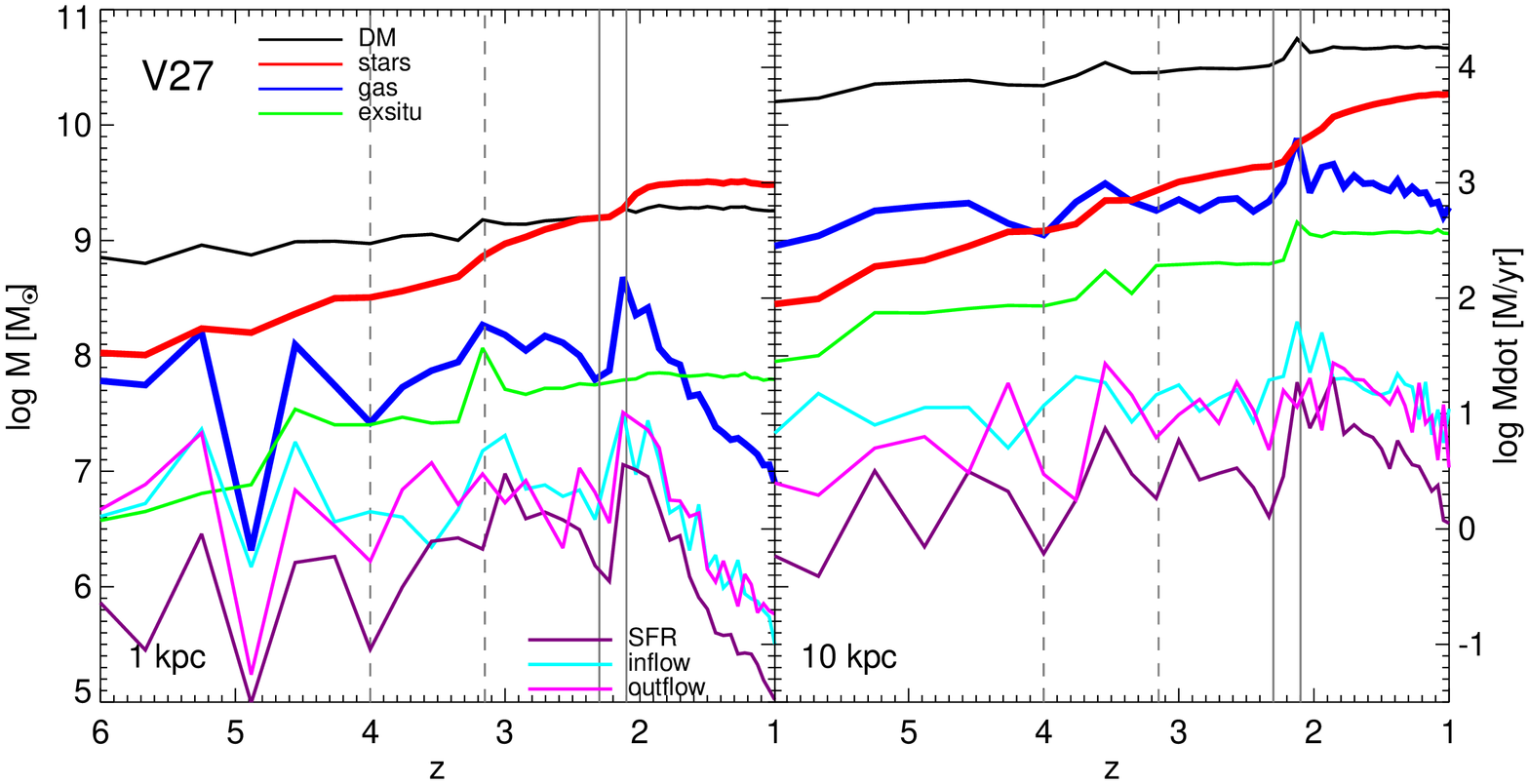}}
\caption{Same as \fig{ssfr_sigma_ex}, but for four galaxies of lower masses.
The dashed vertical lines mark the onset and peak of earlier compaction
events.
These galaxies compactify to lower central densities and make more than one 
quenching attempt.
}
\label{fig:ssfr_sigma_low}
\end{figure*}

\begin{figure*} % 4
\vskip 5.5cm
\includegraphics{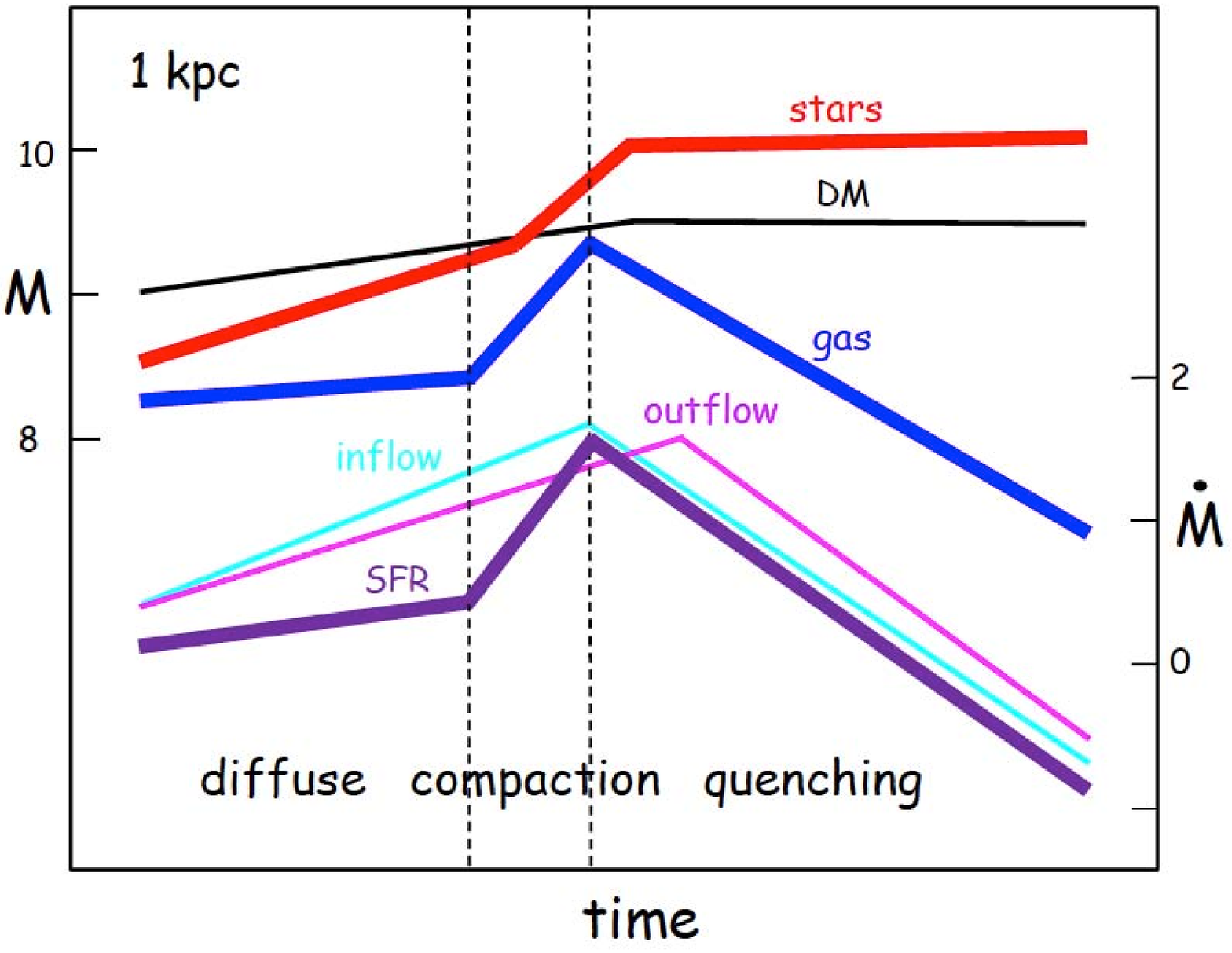}
\includegraphics{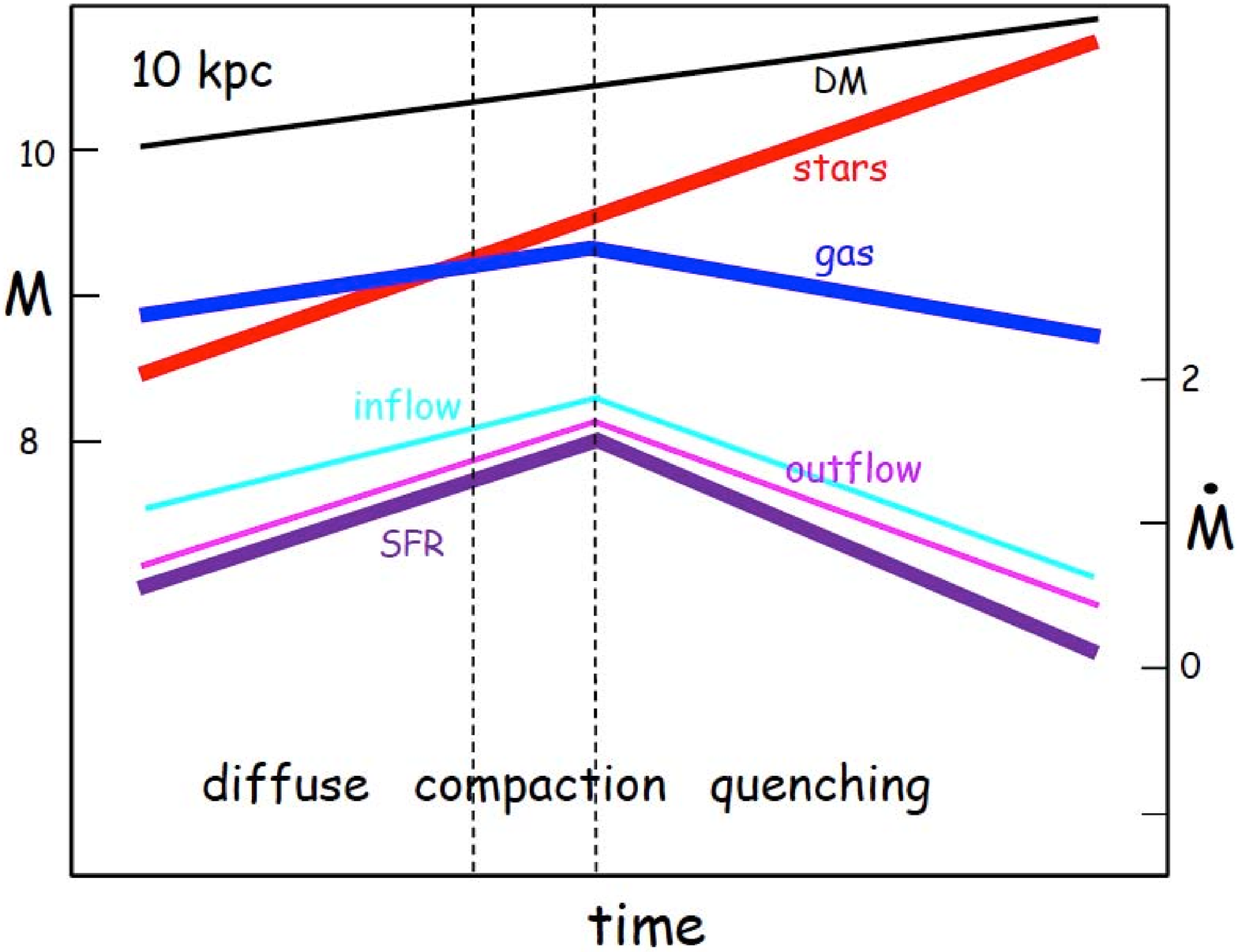}
\caption{The three successive phases of diffuse galaxies, compaction and 
quenching in a cartoon based on the examples shown in the right panels of 
\fig{ssfr_sigma_ex} for the galaxies of high stellar mass
(and to some extent also in \fig{ssfr_sigma_low} for the less massive galaxies).
Shown is the characteristic evolution of mass and its rate of change in the
central 1 kpc (left) and in the galaxy as a whole (out to 10 kpc, right).
After an early phase of gradual mass growth and star formation, there is a
well-defined, relatively short phase of wet compaction in the inner 1 kpc,
reaching a peak of central gas density and SFR (a blue nugget). In the central
kpc, this is immediately followed by a longer phase of gas depletion and 
quenching of SFR caused by a low rate of inflow to the center compared to the
sum of SFR and outflow rate. 
The result is a compact quenched galaxy (a red nugget), where the central 
stellar density remains roughly constant from the blue nugget phase and on.
The whole galaxy typically quenches in a slower pace due to residual star 
formation in an extended gas ring.
}
\label{fig:mass_z_cartoon}
\end{figure*}

\begin{figure} % 5 new 
\centering
\includegraphics[angle=90,width=0.50\textwidth]{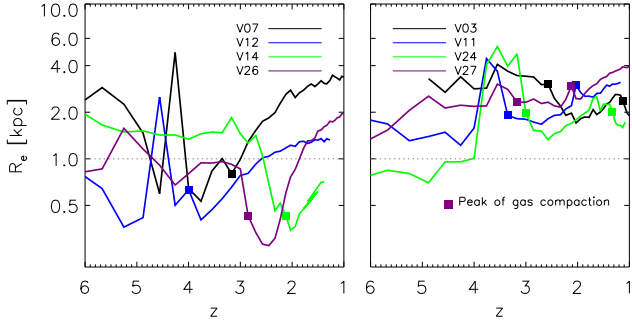}
\caption{
Evolution of the stellar effective radius $\Re$ for the eight example
galaxies of \fig{ssfr_sigma_ex} (left) and \fig {ssfr_sigma_low} (right). The
effective radius tends to systematically grow in time from below $1\kpc$ to
above it.
The peaks of gas compactness are marked (squares) -- they tend to be
associated with local minima in $\Re$.
}
\label{fig:re_z}
\end{figure}

% figures
\Figs{ssfr_sigma_ex} and \ref{fig:ssfr_sigma_low} describe the evolution 
of 8 example galaxies from our sample. The first four galaxies
shown in \fig{ssfr_sigma_ex} have relatively 
high stellar masses and they tend to compactify to higher densities and then 
quench rather efficiently, while the second four, shown in \fig{ssfr_sigma_low}
are of lower masses, lower
density at compaction, and more hesitant quenching that is commonly followed 
by a new compaction phase. 
The companion \fig{mass_z_cartoon} is a cartoon summarizing the main features 
characterizing the evolution through compaction and quenching phases.
\Fig{re_z} shows the corresponding evolution of the effective radius of these 
8 examples. Then 
\fig{re_mass_tracks} displays evolution tracks in the plane of effective radius
and stellar mass for 12 galaxies that evolve through a nugget phase 
(including the 8 default examples shown in previous figures).

\smallskip % par 1
We first focus on \fig{ssfr_sigma_ex}, showing the evolution of four massive 
galaxies that compactify to high densities and quench efficiently.
The left panels show the evolution tracks of these
galaxies in diagrams of sSFR (increasing from top to bottom)
versus central compactness. 
This diagram is a proxy for diagrams commonly used to present
observational results for samples of galaxies in given redshift bins 
\citep[e.g.][]{barro13}. 
In the left-most panel, compactness is measured by the effective  
stellar surface density $\Se$ within the effective (half-mass) radius $\Re$. 
In the second panel from the left, compactness is measured by the stellar 
surface density within the inner 1 kpc, $\Skpc$.
% re_z
One can see in \fig{re_z} that the effective radius is on the order of 1 kpc 
(to within a factor of 3)
in most of our galaxies and most of the times; 
it tends to grow systematically with cosmological time from below 1 kpc to 
above it, and it typically fluctuates during compaction events. Therefore,  
the two left panels of \fig{ssfr_sigma_ex} provide complementary information. 

%Major mergers are marked in the figure by blue upside-down open triangles, 
%while minor mergers are marked by open purple squares.
%These are based on merger trees described in  
%\citet{tweed09} and Tweed et al. (in preparation).

\smallskip % L-shape 2
In the selected examples shown in \fig{ssfr_sigma_ex} 
we see that the evolution tracks have a characteristic L shape in the 
sSFR-$\Sigma$ plane. 
At early times, the galaxy is in a diffuse phase where it
forms stars at a low surface density, $\Skpc\sim \Se\sim10^8 \msun\kpc^{-2}$.
Then there is a rather quick compaction 
to a maximum surface density at $\Skpcm\sim 10^{9.7}$
and $\Sem\sim 10^{10} \msun\kpc^{-2}$. 
At this point there is a sharp onset of quenching, followed by 
a continuous decline in sSFR by 1-2 orders of magnitude 
while $\Skpc$ remains high.
A similar behavior is seen both for $\Se$ and $\Skpc$, except that during the
quenching phase $\Skpc$ remains rather constant while $\Se$ tends to gradually
decline, reflecting the systematic growth of $\Re$.

\smallskip % incomplete quenchinga 3
We note that the quenching is not always complete in our simulations. One
reason for this is that the simulations were only run to $z\sim1$, while
complete quenching may be achieved only at a later redshift.
The other possibility is that the quenching efficiency is underestimated 
in our simulations, and may not be sufficient for complete, long-term
quenching. This could be due to the missing AGN feedback or to a possible
underestimate of the supernova or radiative stellar feedback.
We therefore consider a partial reduction in sSFR, say by an order of
magnitude from its maximum value, as an indication of potentially complete 
quenching.

In all such cases, the decline rate of sSFR is faster than the overall
decline rate of the ridge of the star-forming main sequence (MS)
as defined below in \equ{sSFR}, 
indicating that this is indeed a real quenching process.

\smallskip % master diagrams 4
The two right panels of \fig{ssfr_sigma_ex} provide another useful way 
to follow the details of the evolution of these four galaxies.
The right panels show the evolution of masses $M$ and gas mass rates of change 
$\dot M$, for the central sphere of radius 1 kpc 
(the panel second from the right)
and for the main body of the galaxy contained within the sphere of radius 
10 kpc (right-most panel).
Shown are the gas mass, stellar mass, and dark-matter mass. 
Note that the stellar mass within the inner 1 kpc is a proxy for the surface
density in the central 1-kpc region of the galaxy, $\Skpc$.
The mass rates of change shown are the 
SFR, the gas inflow rate, and the gas outflow rate. 
These rates are measured in spherical shells of radii $r=1$ and $10\kpc$ 
and of width $\Delta r =\pm 0.1r$ via 
\be
\dot{M}= \frac{1}{\Delta r} \sum_i m_i v_{r,i} \, , 
\ee
where the sum is either over all inflowing gas cells in the shell or over all
outflowing cells, $m_i$ is the gas mass in cell $i$, 
and $v_{r,i}$ is the radial velocity of the gas in that cell.
%
%\adb{\Fig{re_z} shows the corresponding evolution of the effective radius of
%each of the example galaxies.}

\smallskip % 5
In all four galaxies shown in \fig{ssfr_sigma_ex}
we identify the appearance of a characteristic pattern,
which occurs in the different galaxies at different times, 
and sometimes more than once in the spanned period of evolution. 
This pattern is schematically summarized in the cartoon shown in
\fig{mass_z_cartoon}, referring to the evolution of $M$ and $\dot{M}$ as in the
right panels of \fig{ssfr_sigma_ex}.

\smallskip % early diffuse phase 6
We focus first on the evolution of gas and stellar mass within 1 kpc.
There is an early phase where the gas mass is constant or growing very slowly,
and where the stellar mass is growing at a slow pace, reflecting continuous
star formation in the central 1 kpc (blue and red curves respectively). 
The sSFR is therefore constant or rising slowly. Take, for example, V14, 
where the gas mass in \Fig{ssfr_sigma_ex} is $\sim$ constant
until $z=3$, and the stellar mass (red) grows slowly during this time. 
One can see that the SFR (purple) is also approximately constant during 
this period. 

\smallskip % gas compaction and quenching 7
At a certain point in time, one can identify a beginning of a faster growth 
rate for the gas mass -- this is the onset of the gas compaction phase.
It is marked by a vertical grey line for each galaxy in the right panels 
of \figs{ssfr_sigma_ex} and \ref{fig:ssfr_sigma_low}, where a second line 
marks the peak of gas compactness.
%\adr{Adi: specify how the onset of compaction is measured}
The gas compaction starts at 
$z \simeq 3.5, 4.5, 3.3, 3.2$ for V07, V12, V14, V26, respectively. 
Then, the central gas mass grows quickly by an order of magnitude or more, 
reaching a peak at a certain time, after which it begins to continuously drop. 
This occurs at 
$z \simeq 3.2, 4.0, 2.1, 2.8$ in V07, V12, V14, V26, respectively. 
The steep decline in gas mass is associated with a similar decline in SFR, 
namely this peak marks the onset of the central quenching phase, where the
central region is becoming devoid of gas. 

\begin{figure*} % 6 <- 5 Images
\centering
\includegraphics[width=\linewidth]{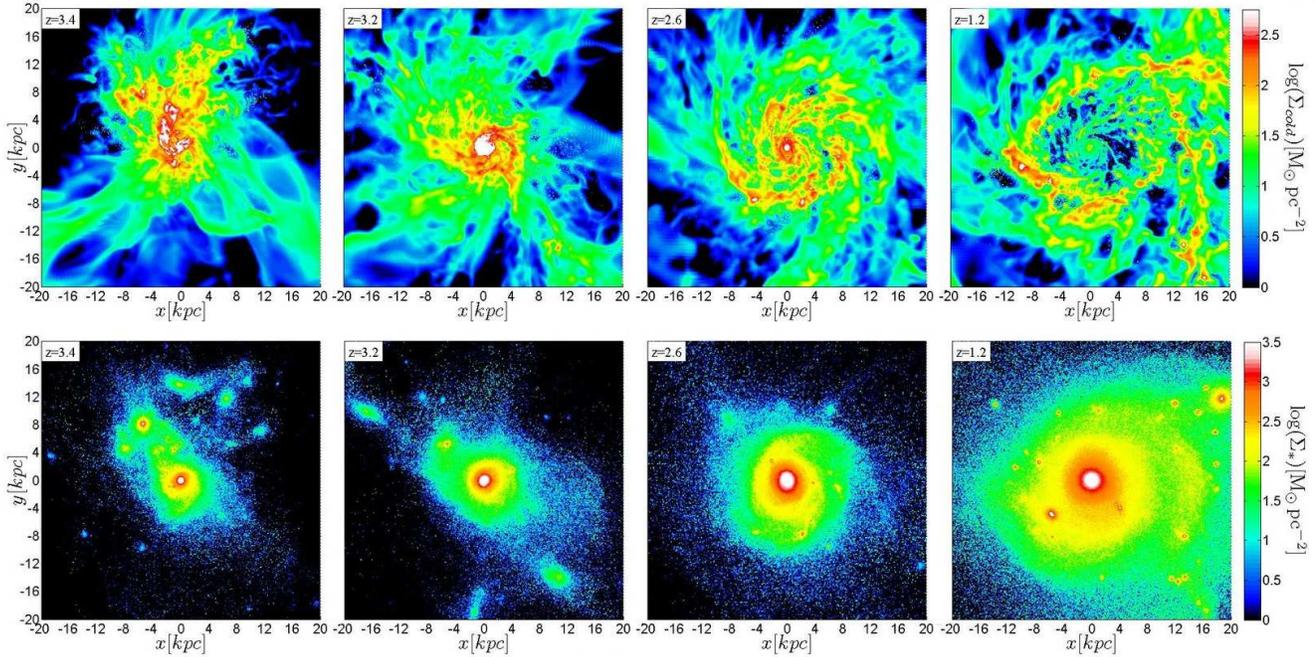}
\caption{
Compaction and quenching in V07.
Shown are images of face-on projected density of the cold component made of
gas and stars younger than $100\Myr$ (top)
and of the stellar component (bottom), in a cubic box of side 40 kpc.
The snapshots from left to right correspond to
(a) prior to or during the compaction phase,
(b) the blue nugget phase near maximum gas compaction,
(c) the ``green nugget" phase during the quenching process,
and (d) the red nugget phase after quenching.
A BN with a dense core of gas and stars develops via 
dissipative compaction. It leads to gas depletion in the core while an 
extended ring develops. The dense stellar core remains intact from the BN
to the RN phase, while in this case an extended stellar envelope develops
around the RN core.
}
\label{fig:images_07}
\end{figure*}

\begin{figure*} % 7 <- 6
\centering
\includegraphics[width=\linewidth]{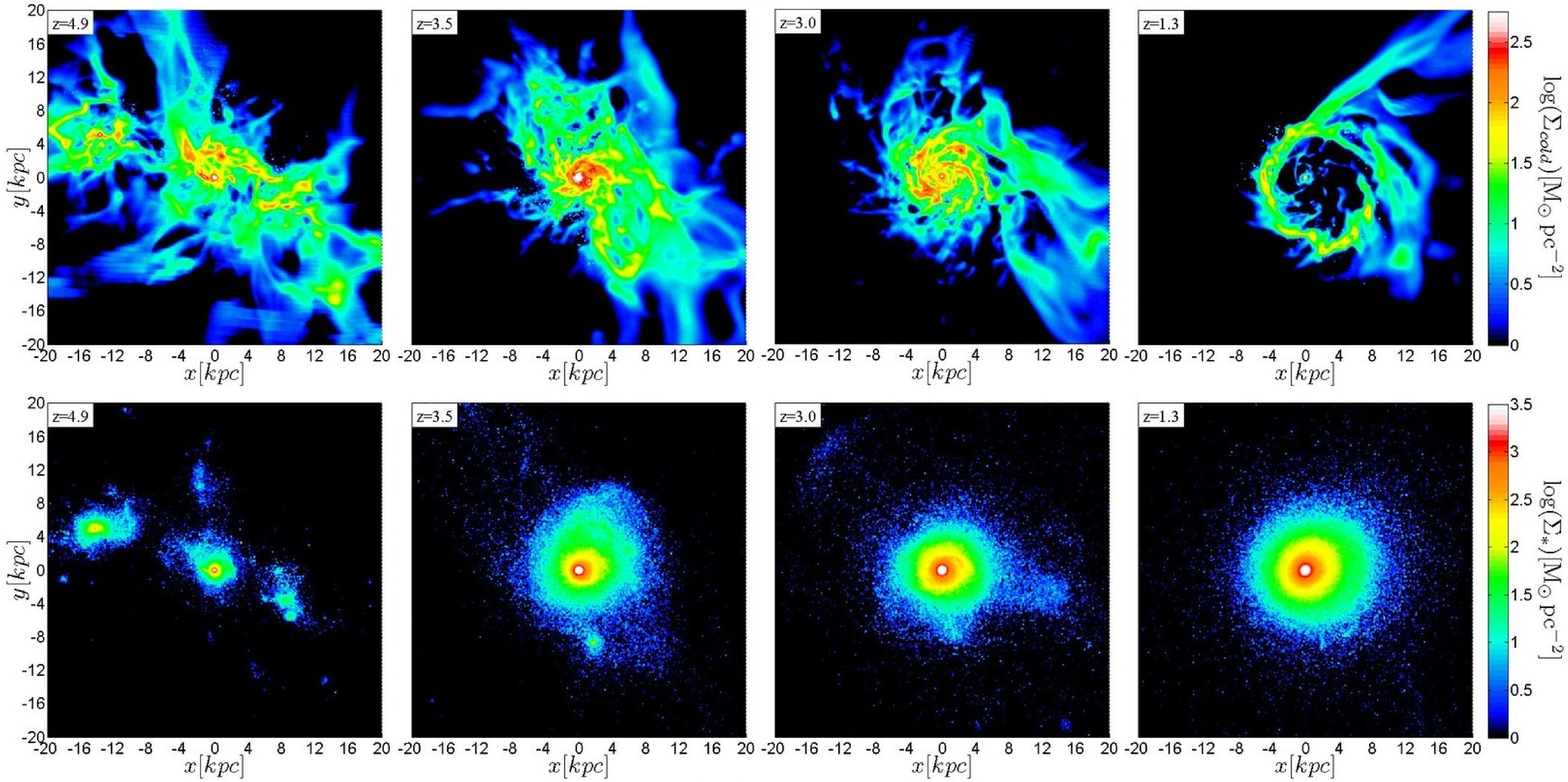}
\caption{Same as \fig{images_07}, but for V12.
In this case, a post-compaction gas ring develops,
while the RN remains naked and similar to the BN.
}
\label{fig:images_12}
\end{figure*}

\smallskip % stars
The central stellar mass typically shows analogous features that occur
following the growth of the central gas mass, with a certain time delay 
between the two. 
The stellar compaction typically starts a little later than the gas 
compaction, and the stellar mass growth during the compaction
is typically slower than the gas mass growth.
The stellar compaction reaches a maximum density at a slightly later time 
than the gas density peak, after which the central stellar mass within the
inner 1 kpc remains rather constant at an asymptotic value.
The qualitative behavior of the stellar mass compared to the gas mass is 
naturally expected if the compaction is driven by gas dissipation
and given that the gas continuously turns into stars at a high rate,
and more so as the system becomes more compact.
% Re
The evolution of effective radii shown in \fig{re_z}, 
where the peaks of gas compactness are marked, indicates a correlation between 
maxima in core gas density and minima in stellar effective radius, 
though the correspondence is not always one-to-one.

\smallskip
% wet inflow
Before and during compaction, the inflow rate (cyan) tends to be
significantly larger than the SFR (purple). 
This indicates that the compaction is wet, as expected.

\smallskip
% SFR
One can see in these figures that the overall SFR follows the total gas mass,
both in the central 1 kpc and in the whole galaxy. 
A local relation between gas density and SFR is built into the simulations, 
and we see that it translates to this global 
scaling \citep{silk97,elmegreen97,kennicutt98}.
The peak of gas compaction coincides with a peak in central SFR,
and the subsequent decline of gas mass is associated with a similar decline in
SFR. During the compaction phase, the sSFR keeps a roughly constant 
high level. Beyond the SFR peak, reflecting the constancy of the central 
stellar mass, there is a continuous decline in sSFR, namely quenching.  

\smallskip
% inflow, outflow
In the vicinity of the SFR peak, the rates of inflow and outflow from the 
central region also tend to peak, and from then onwards, throughout the
quenching phase, they are all declining and remain
comparable to each other. In particular,
near the onset of quenching and somewhat after it, there is marginal 
evidence for certain enhancement of the outflow rate compared to the 
inflow rate, but no evidence for a dramatic burst of outflow that could 
serve as the dominant driver of the quenching. 
The onset of central quenching is due to the tilt of the balance from a state 
where the inflow to the center is dominant to a state where the inflow rate is 
insufficient for balancing the sum of SFR and outflow rate,
which naturally leads to depletion.

\smallskip
% DM
We note that during compaction the central region makes a drastic
transition from being dominated by the dark matter to becoming governed by the
self-gravitating baryons. This can be seen by comparing the red and black
curves in the second-from-right panels of \figs{ssfr_sigma_ex} and
\ref{fig:ssfr_sigma_low}.
The fact that the quenching occurs in the
self-gravitating phase may be an important clue for the origin of quenching, 
to be discussed in \se{origin}.

\begin{figure*} % 8 <- 7
\centering
\includegraphics[width=\linewidth]{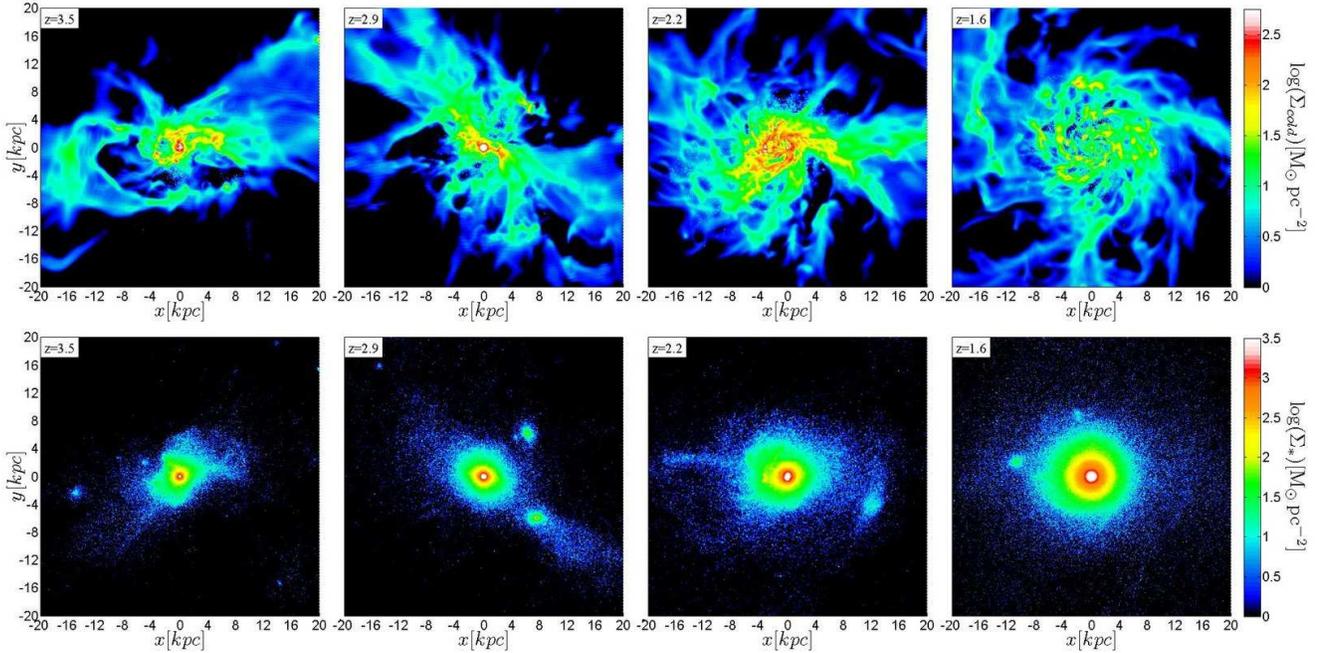}
\caption{Same as \fig{images_07}, but for V26.
Here, mergers do not play an important role. A late diffuse gas disc develops,
and the stellar density profile grows both in amplitude and in extent
from the BN to the RN phase.
}
\label{fig:images_26}
\end{figure*}

\smallskip
% galaxy
The right-most panels of \fig{ssfr_sigma_ex}, referring to the whole galaxy
within 10 kpc, show that the overall SFR also reaches a peak at the same time
as the SFR in the inner 1 kpc, and the whole galaxy is also gradually
quenching from the SFR peak onward.
% inside-out quenching
However, in this post-compaction phase, 
the overall SFR quenching rate is slower than the inner quenching,
with the gas mass declining even slower, 
reflecting the development of an extended gaseous ring forming stars
around the quenched bulge (see images in \figs{images_07}-\ref{fig:images_26}).
% obs
This implies that the quenching process in the post-compaction phase
progresses inside-out \citep[see a detailed analysis in][]{tacchella15_prof}.
This is consistent with preliminary observational indications for 
{\it inside-out}
quenching based on sSFR profiles of a sample of galaxies at $z\sim 2.2$
\citep{tacchella15_science}.

\smallskip
% mergers
In the left panels of \fig{ssfr_sigma_ex}, major mergers 
(with stellar
mass ratios
larger than $1:3$) and minor mergers ($1:10$ to $1:3$) are indicated 
by open blue upside-down triangles and open purple squares, respectively.
These are based on merger trees described in  
\citet{tweed09} and Tweed et al. (in preparation).
Another merger indicator is provided by jumps in the evolution of mass in 
ex-situ stars, those that formed outside the galaxy,
shown in green curves in the right panels, especially within the whole galaxy.
The merger times are identified by the two indicators only in a crude way. 
The role of mergers seems to be different in the different
galaxies and at different times. 
For example, 
galaxy V07 has a major merger prior to $z\sim 4$, which may or may not be
associated with the onset of compaction at $z\sim 3.5$.
It then has a minor merger prior to $z\sim 3$, which may be associated with
either the compaction or the quenching.
In V12 there is no major or minor mergers that could trigger the compaction,
but there is a major merger near $z \sim 4$, which may be associated with the
onset of quenching.
Galaxy V14 has a minor merger near $z\sim 3$ that could trigger its long-term
compaction, and another minor merger just prior to $z \sim 2$ that could be
associated with the onset of quenching. 
Finally, V26 does not show evidence for major or minor mergers associated with 
the compaction or the quenching.
It thus seems that the compaction and quenching could be triggered by one
of different mechanisms, including major mergers in a fraction of the cases, 
minor mergers in another fraction, and something else, possibly related to 
counter-rotating streams or recycled inflows and possibly associated with 
VDI in a third fraction of the cases.
We will return to the role of mergers and VDI in compaction and quenching 
in \se{origin}.

%------------------------- 4.2
\subsection{Less Massive Galaxies}

\Fig{ssfr_sigma_low} shows the evolution of four galaxies of lower stellar
mass that end their compaction and start their final quenching at lower 
central surface densities, near $\Skpc \simeq 10^9\msun\kpc^{-2}$. 
These galaxies do go through events of compaction followed by 
quenching, similar to the characteristic chain of events seen in the 
high-$\Sm$ examples shown in \fig{ssfr_sigma_ex}.
However, in the low-mass, low-$\Sm$ cases these events tend to occur at later 
redshifts, and the L-shape evolution track is more fluctuative.
In particular, the quenching is less decisive: the sSFR fluctuates
down and up several times before it eventually quenches beyond the green
valley.

\smallskip
An inspection of \fig{re_z} indicates
that while the massive galaxies 
that compactify earlier do so to effective radii smaller than 1 kpc, 
the effective radii of the low-mass galaxies that compactify
later tend not to drop to below 1 kpc.  

%---------------------------- 4.3
\subsection{Images and Compactness of Blue and Red Nuggets}

\smallskip
% images
\Figs{images_07} to \ref{fig:images_26} show two-dimensional images of the
density of the cold component of gas and stars younger than $100\Myr$
and of the 
stellar density at 4 snapshots in the history of three of the massive
galaxies, V07, V12 and V26.
The cold component of the disc refers to the mass that is directly involved in 
VDI, which is typically roughly twice the gas mass alone.
The cold mass density could serve as a crude proxy for the density measured 
from H$_\alpha$ or UV observations.
The projections are face-on in a cubic box of side 40 kpc.   
The 4 snapshots correspond to 
(a) prior to or during the early stages of compaction, 
(b) the peak blue-nugget phase near maximum compaction and beginning of
quenching,
(c) the green-nugget phase during the quenching process,
and (d) the red-nugget phase after quenching.

\smallskip
% discuss images
The pre-compaction phase is characterized by a clumpy gas appearance,
typically associated with a major merger (V12),
minor mergers (V07), and no mergers (V26).

\smallskip
The blue-nugget phase shows a high gas density associated with a high stellar
density in the inner 1 kpc, with only low-density gas left at large radii.

\smallskip
The quenching phase is characterized by a massive, centrally condensed stellar
bulge, which may gradually grow a more diffuse stellar envelope (V07). 
%In some cases, a new gas disc grows about the bulge and develops VDI. 
The gas is gradually depleted from the central regions while fresh incoming gas
may develop an extended unstable gas ring (V07, V12) or a diffuse disc (V26).
We should comment that these post-quenching phenomena might be suppressed when
stronger feedback is implemented, e.g., when AGN feedback is added.

\begin{figure} % 9 new 
\centering
\includegraphics[angle=90,trim=0mm 0mm -10mm 13mm, clip, width=0.50\textwidth,
height=0.48\textwidth]{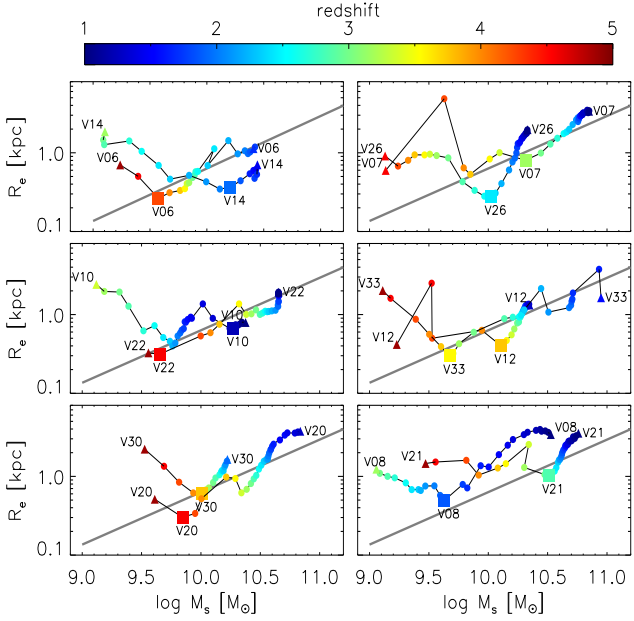}
\caption{
Evolution tracks of galaxies in the mass-radius plane, with respect to
the line adopted in \citet{barro13} to identify compact galaxies in
observations,
$\Ms/\Re^{1.5}\!\ge\!10^{10.3} \msun \kpc^{-1.5}$.
Redshift along each track is marked by colour.
The beginning and end of each track are marked by triangles and the galaxy
name. The points of maximum gas compactness are marked by squares, 
indicating the peak of the blue-nugget phase. 
The 12 galaxies shown (out of 26) go through a compact, nugget phase
during the given redshift range, consistent with the compactness of
observed blue nuggets.
%\adr{Try to fill the whole textwidth}
}
\label{fig:re_mass_tracks}
\end{figure}

\smallskip % re_mass_tracks
To what extent do the simulated blue and red nuggets match the 
compactness of observed nuggets? 
\citet{barro13} defined the locus of compact galaxies in the mass-radius
plane by the threshold line (in log-log)  
$\Ms/\Re^{1.5}\!\ge\!10^{10.3} \msun \kpc^{-1.5}$.
\Fig{re_mass_tracks} shows evolution tracks of simulated galaxies in this
plane with respect to the observational threshold line.
One can see that each of the 12 galaxies shown  
goes through a compact nugget phase as defined by the observational threshold.
Thus, in terms of compactness, a significant fraction of the simulated nuggets 
are in the ball park of the observed nuggets. 
In other cases, especially involving low-mass galaxies, a similar wet 
compaction process leads to a compact star-forming galaxy that also deserves 
to be termed a ``blue nugget".
Thus, the simulated evolution allows us to define the BN phase as the
product of a wet-compaction process rather than by an absolute threshold for
an instantaneous compactness measure. This should enable us 
to propose a more physically motivated definition for BNs 
based on their observed properties.

%------------------------------- 4.4
\subsection{Kinematics of Compacting Galaxies}

\begin{figure} % 10 <- 8
\includegraphics[angle=0,width=0.48\textwidth]{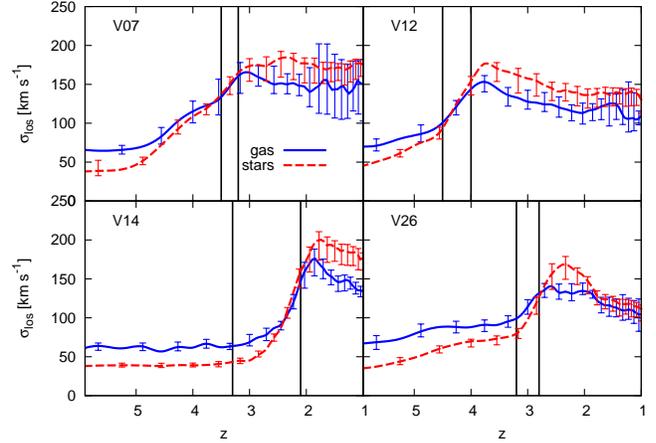}
\caption{The evolution of line-of-sight velocity dispersion,
through a beam of diameter 8 kpc,
averaged over random directions with the standard deviation shown,
for gas (blue) and for stars (red).
Vertical lines mark the onset of gas compaction in the central $1\kpc$ and
the time of maximum gas density inside this volume.
The velocity rises steeply during the compaction phase from $\sim 50\kms$ to
$150-200\kms$, and then levels off at $\sim 150\kms$, roughly the
circular velocity of the given potential well.
}
\label{fig:los}
\end{figure}

\begin{figure} % 11 <- 9
\includegraphics[angle=0,width=0.48\textwidth]
{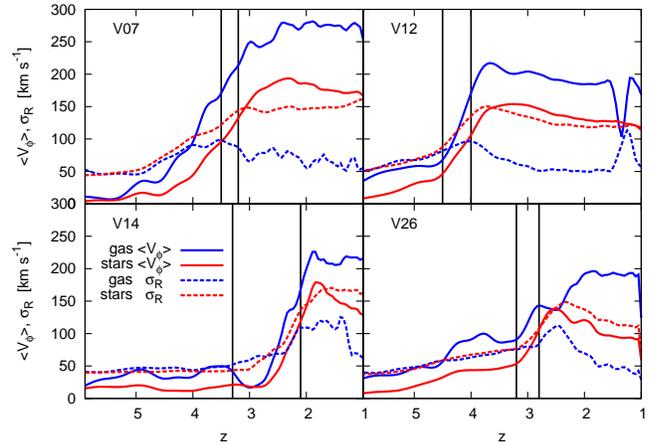}
\caption{The evolution of rotation velocity and radial velocity dispersion,
mass-weighted averaged within the central 4 kpc, for gas and for stars. 
The velocities were averaged across three consecutive snapshots. 
Vertical lines mark the onset of gas compaction in the central $1\kpc$ and
the time of maximum gas density inside this volume.
In the compact phase, the stellar rotation velocity and velocity dispersion
tend to become comparable, while the gas tends to be rotation dominated.
}
\label{fig:v_and_sigma}
\end{figure}

The process of compaction outlined so far is expected to be
associated with a drastic change in the
morphology and the kinematics of the galaxy. When the galaxy is not
resolved observationally,
an interesting observable is the line-width, for gas or stars,
which may reflect either rotation or velocity dispersion or both.
As a proxy for line-width, we compute $\slos$,
the mass-weighted velocity dispersion
along a given line of sight, through a cylindrical beam of diameter 8 kpc
about the galaxy centre (corresponding to $\sim\!1$").
\Fig{los} shows the average and standard deviation of $\slos$ over 64 random
directions for the gas and for the stars in the four example massive
galaxies examined in \fig{ssfr_sigma_ex}.
The gas and stellar $\slos$ roughly evolve together.
In the early diffuse phase, the galaxies all have low values of
$\slos \sim 50\kms$, slowly increasing with time in V12 and V26.
During the compaction phase, $\slos$ rises steeply to peak values of $\sim
150-200\kms$. 
Following the point of maximum gas compactness, the $\slos$ settles
to a level of $\sim 150\kms$ that remains roughly constant during the
quenching phase. This velocity is in the ballpark of the circular velocity in
the given potential well, namely the halo virial velocity.
The scatter between different lines of sight is rather small prior to
compaction, and is only $\sim \pm 25\%$ after the compaction.
These results are consistent with the observed line-widths for galaxies
in the different distinct phases \citep{dokkum09,barro14b,nelson14}.
We note that in the post-compaction phase the linewidth of the stars is 
on average $\sim\!30\%$ higher than that of the gas, and it can be almost
a factor of two higher in some cases. This is an indication for rotational
support in the gas, with a significantly higher rotation-to-dispersion ratio
than for the stars, as explained next.

\smallskip
The line-of-sight velocity dispersion originates both from systematic 
rotation and three-dimensional velocity dispersion.
In the simulations, we can follow separately the evolution of these 
components of the velocity field, both for gas and for stars,
as shown in \fig{v_and_sigma}.
The rotation velocity is the tangential
velocity component $V_\phi$ in cylindrical coordinates aligned with the galaxy
angular-momentum vector,
mass-weighted averaged over a cylinder of radius $4\kpc$ and height 
$\pm 2\kpc$.
The radial velocity dispersion $\sigma_r$ about the average rotation velocity 
is computed within the same cylindrical volume.
We find for the stars that the velocity dispersion dominates over the 
rotation velocity during the compaction and BN phase, but they become
comparable to each other in the subsequent compact state during the quenching
phase; the rotation becomes larger for V07 and V12, but the dispersion 
remains larger in V14 and V26. 
Thus, for the stars, $V_\phi/\sigma_r\!\sim\!1$.
We recall that the model of DB14 for wet compaction in VDI indeed predicted 
that the star-forming compact galaxies should develop a high velocity
dispersion, $V_\phi/\sigma_r\!\leq\!2$.  
On the other hand, the post-compaction gas component is dominated by rotation, 
with a post-compaction ratio of $V_\phi/\sigma_r\!\sim\!2\!-\!5$.
%
%The large fluctuations in $v_\phi$ seen in \fig{v_and_sigma}
%could be due to difficulties in defining the
%center of rotation during clump mergers near the galaxy centre, 
%sometimes showing as low rotation velocity or even counter-rotation.

\smallskip

The similarity between $\slos$ for gas and stars seen in \fig{los} 
reflects the validity of radial equilibrium, Jeans equilibrium for the stars
and hydrostatic equilibrium for the gas,
\be
V_{\rm c}^2 \sim V_\phi^2 + \alpha\, \sigma_r^2 \, .
\label{eq:jeans}
\ee
Here $V_{\rm c}^2 \simeq GM/R$ characterizes the same potential for the gas
and the stars; $V_{\rm c}$ would have been the rotation speed had the disc been
cold, with negligible dispersion.
The value of the parameter $\alpha$ depends on the
density profile of the relevant component, $\alpha=\dd\ln\rho/\dd\ln r$. 
For an isothermal sphere
or for an exponential disk at the exponential radius, $\alpha\!\simeq\!2$
\citep{burkert10,bt08}.
This is valid for the stars and the gas separately, with a different
rotation-to-dispersion ratio.

\smallskip
Along a line-of-sight (los) with inclination $i$,
the los velocity dispersion (or line width) can be written as
\be
\sigma_{\rm los}^2 = \beta\, (\sin\!i\, V_\phi)^2 + \sigma_i^2 \, .
\label{eq:los}
\ee
The factor $\beta$ represents the projection of the rotation velocity along
the line of sight. For a transparent cylindrical disk edge on
$\beta = 2/\pi \simeq 0.64$, which serves as an upper limit.
The dispersion $\sigma_i$ is the one-dimensional velocity dispersion along
the line of sight.
In the face-on view $\sigma_i=\sigma_z$,
and in the edge-on view $\sigma_i$ is an average of $\sigma_r$ and
$\sigma_\phi$.
As the simplest case one can assume isotropy, $\sigma_i=\sigma_r$ for every $i$.

\smallskip

Denote, separately for gas and for stars,
$\gamma \equiv \frac{V_\phi}{\sigma_r}$.
Combining \equ{jeans} and \equ{los} for a given $\gamma$ we obtain
\be
\frac{\sigma_{\rm los}^2}{V_{\rm c}^2} =
\frac{1+\beta\,\gamma^2\,(\sin\!i)^2}{\gamma^2+\alpha} \, .
\label{eq:slos}
\ee
In the face-on view, only the first term in the numerator contributes, 
with $\gamma$ representing $\sigma_z$.
The second term, representing rotation, adds a maximum contribution in the
edge-on view.
We can deduce from \equ{slos} that in the edge-on view one expects the gas and
stellar $\sigma_{\rm los}$ to be comparable as long as
$\beta\!\sim\!\alpha^{-1}$. 
In the face-on view,  
\be
\frac{\sigma_{\rm los,stars}}{\sigma_{\rm los,gas}}
=\frac{(\gamma_{\rm gas}^2+\alpha)^{1/2}}{(\gamma_{\rm stars}^2+\alpha)^{1/2}} 
\, ,
\label{eq:face}
\ee
which can obtain a value in the range 1-2 depending on how high
$\gamma_{\rm gas}$ is.
The trend seen in \fig{los} in the post-compaction phase, with the gas
$\sigma_{\rm los}$ somewhat lower than the stellar $\sigma_{\rm los}$,
is thus an indication
for a rotation-dominated gas disk in this phase, consistent with the images
showing extended gas rings in \figs{images_07} to \ref{fig:images_26}, and with
the inside-out quenching indicated by comparing the two right panels in
\figs{ssfr_sigma_ex} and \ref{fig:ssfr_sigma_low}.
A similar feature of a somewhat lower $\sigma_{\rm los}$ for gas versus stars
is observed in a $z\!\sim\!1.7$ green nugget \citep{barro15_kin}.

\begin{figure*} % 12 <- 10 sSFR vs Sigma, all galaxies 
\centering
\subfigure{\includegraphics[width=0.49\textwidth]
{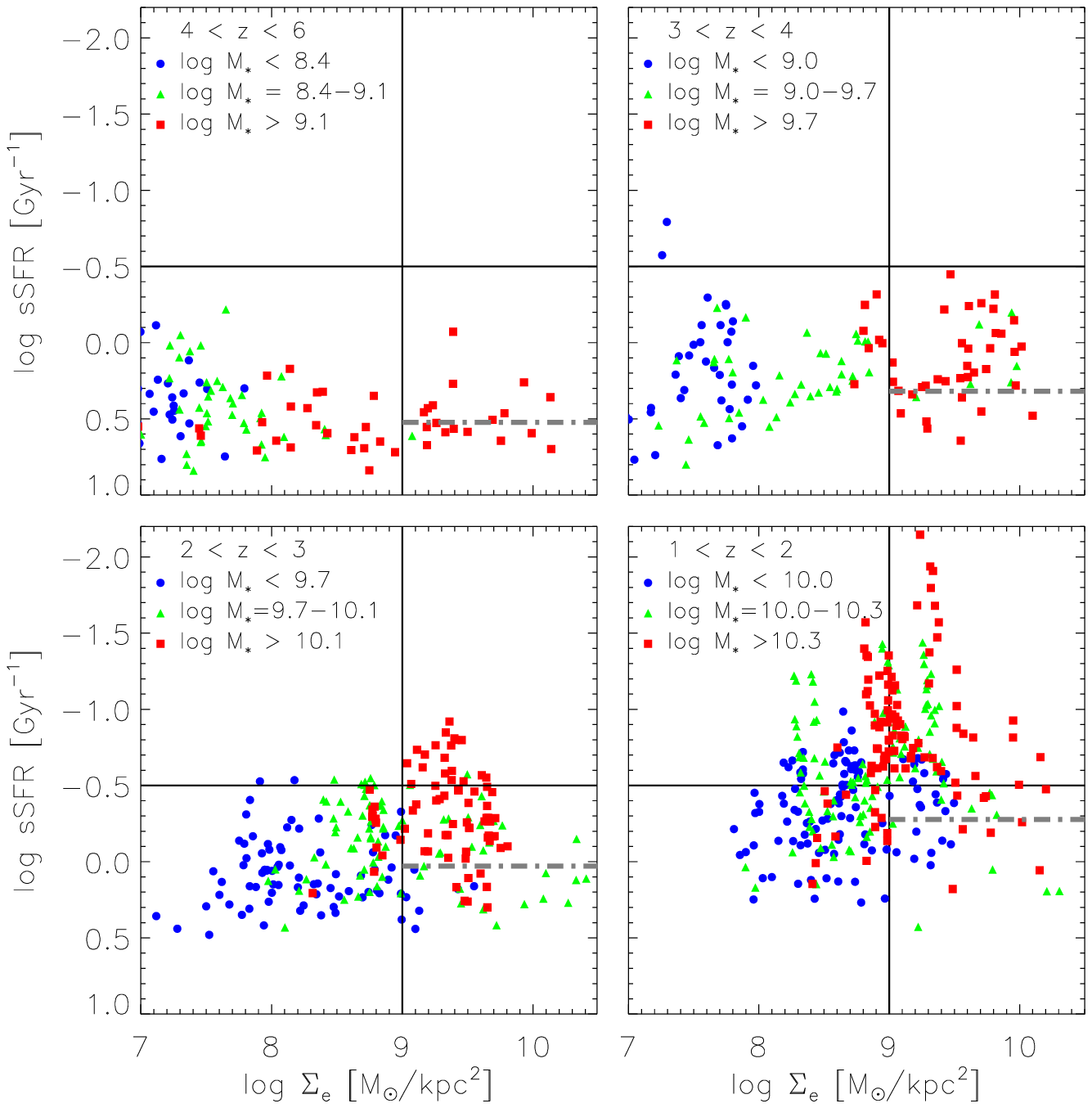}}
\subfigure{\includegraphics[width=0.49\textwidth]
{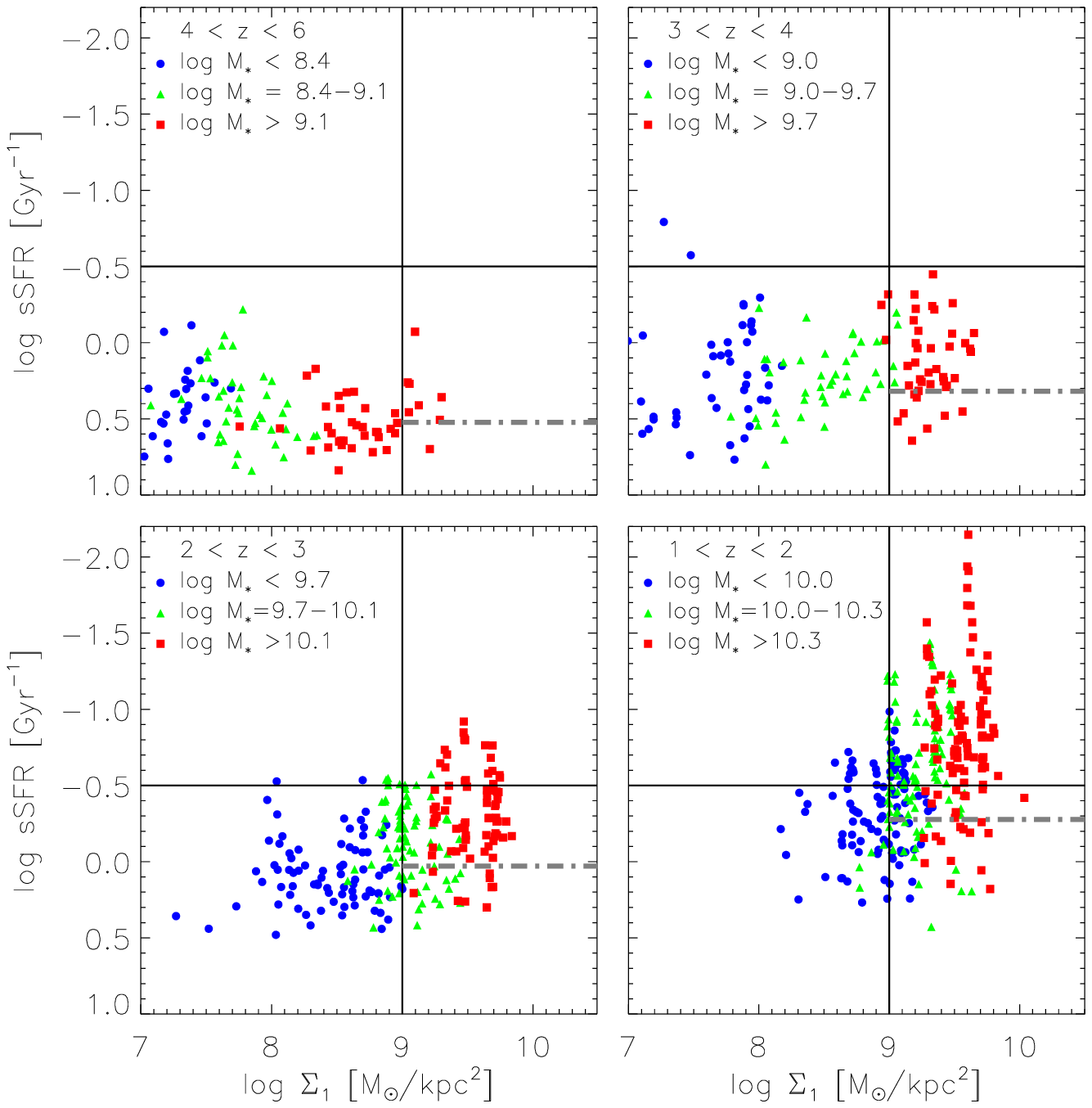}}
\caption{
Time evolution of the simulated galaxy sample in the plane of sSFR
and compactness.
The compactness is measured by the stellar surface density either within
the effective radius ($\Se$, left) or within 1 kpc ($\Skpc$, right).
The snapshots are divided into 4 redshift bins.
In each redshift bin, the sample is divided by mass to 3 subsamples
with a third of the galaxies in each (lowest mass blue, highest mass red).
The solid lines crudely distinguish between diffuse and compact galaxies 
(vertical line) and between SFGs and quenched galaxies (horizontal line).
The horizontal dot-dashed lines mark the ridge of the MS 
according to \equ{sSFR}, evaluated at the median redshift and mass of the 
massive galaxies in the given redshift bin (the red points).
The evolution is from diffuse to compact SFGs (``blue" nuggets),
and then to compact quenched galaxies (red nuggets).
The more massive galaxies evolve earlier.
% median_mass = 4.0e+09   1.5e+10   3.4e+10   6.1e+10
% median_z = 4.9, 3.5, 2.3, 1.4
}
\label{fig:ssfr_sigma_all}
\end{figure*}

\begin{figure*} % 13 new 
\centering
\includegraphics[angle=90,width=\linewidth]{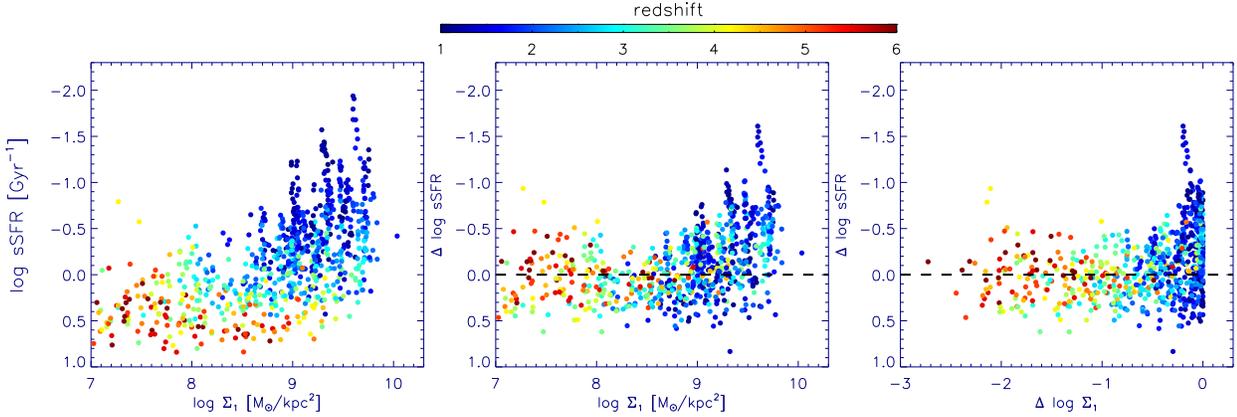}
\caption{A universal track of evolution in the sSFR-$\Skpc$ plane.
All snapshots of all galaxies, each represented by a point, are put together 
with the redshift marked by colour.
{\bf Left:} The raw data from \fig{ssfr_sigma_all}.
{\bf Middle:} The sSFR is scaled, showing the deviation from the ridge of the
main sequence as defined in \equ{sSFR}.
{\bf Right:} In addition, the values of $\Skpc$ for each galaxy are scaled to
match all other galaxies at the same $\Skpcm$. 
The galaxies evolve along a universal L-shape track with a small scatter, 
first along a horizontal branch corresponding to the star-forming main sequence,
and then, after the blue-nugget phase, along a vertical, quenching branch.
The scatter about the horizontal branch is $\pm 0.25$ dex,
and the scatter about the vertical branch is 0.24 dex and 0.08 dex 
in the middle and right panels respectively.
%\adr{Adi says:
%left, $S1<10^9$, 0.40 dex.
%middle, $S1<10^9$, 0.26 dex.
%right, log delta $S1 < -0.5$, 0.25 dex.}
%\adr{Quote scatter in the vertical branch}
}
\label{fig:L-shape}
\end{figure*}

%%%%%%%%%%%%%%%%%5   
\section{Properties of the Evolving Sample}
\label{sec:sample}

We now turn to the whole sample of galaxies as it evolves in time,
and address relevant galaxy properties and correlations between them,
which may shed light on the processes of compaction and quenching.
This allows comparisons to observations, where one should recall that we
are following here an evolving sample, where the masses systematically
grow in time, while the observed samples may be selected according
to different criteria, e.g., at a fixed mass.
The study of the evolving sample also allows comparisons to the
model predictions by DB14.

%------------------
\subsection{sSFR versus Compactness}
\label{sec:ssfr_sigma}

\Fig{ssfr_sigma_all} shows how the galaxies populate the $\Sigma$-sSFR diagrams
in 4 redshift bins between $z=6$ and $z=1$. All the outputted snapshots for all
the simulated galaxies are shown.
This diagram is a proxy for the similar diagrams commonly used to present
observational results \citep[e.g.][]{barro13}, where the vertical
axis is a measure
of sSFR (increasing from top to bottom) and the horizontal
axis is a measure of central
compactness. Like in \figs{ssfr_sigma_ex} and \ref{fig:ssfr_sigma_low}, 
here we refer to two such measures, $\Skpc$ and $\Se$.
We distinguish between diffuse and compact galaxies at 
$\Sigma=10^9\msun\kpc^{-2}$ 
and between star-forming galaxies and quenched galaxies at
sSFR$= 0.3 \Gyr^{-1}$,
thus dividing each figure into 4 quadrants.
The distribution of our simulated galaxies in the $\Sigma$-sSFR plane
at the different redshift bins qualitatively resembles the
observational results based on the CANDELS survey at $z=1.4-3$
\citep{barro13,barro14a,barro14b}.

\smallskip % compaction
We see that at $z>4$ all the galaxies are star-forming galaxies (SFG), 
with sSFR$\geq 1 \Gyr^{-1}$, and most of them are diffuse, namely they
populate the lower-left quadrant. 
$\Se$ is typically larger than $\Skpc$ because $\Re<1\kpc$.
By $z=3$, more of the SFGs have undergone compaction, populating
the lower-right quadrant, while some galaxies have already started
their quenching process to lower sSFR and can be found near the
``green valley", which we quite arbitrarily identify with 
sSFR$\sim 0.3\Gyr^{-1}$.
The values of $\Se$ reach higher values than $\Skpc$ since $\Re$ is still
smaller than 1 kpc.
At $z=2-3$ most of the galaxies in our evolving sample are 
in the blue-nugget quadrant, star forming and compact, 
while several of the compact galaxies have already crossed the green valley 
in their quenching process.

\smallskip % quenching 4
Finally, at $z=1-2$, a large fraction of the galaxies have quenched, most 
to compact red nuggets.
We note that there is no one single value of $\Sigma$ where quenching occurs
--- different galaxies in our sample quench at different densities, spanning
a range of $\sim 0.8$ dex in $\Skpc$ and $\sim 1.3$ dex in $\Se$. 
When using $\Skpc$ as the measure of compactness, the top-left quadrant 
remains empty, as there is almost no quenching directly from the diffuse 
stage and no significant de-compaction during the quenching phase.
This is not the case for $\Se$, which at late times, when $\Re$ is growing 
above 1 kpc, refers to a larger and growing volume and therefore to lower 
and decreasing densities.
Recall that the quenching in our simulations may be incomplete because of the
potential absence of additional sources of feedback, such as AGN feedback,
so galaxies that have reached sSFR values significantly below the green 
valley, say sSFR$<0.16 \Gyr^{-1}$, may be considered practically quenched.

\begin{figure*} % 14 <-- 11 a,b distributions SFGs compact
\centering
\subfigure{\includegraphics[width=0.45\textwidth]
{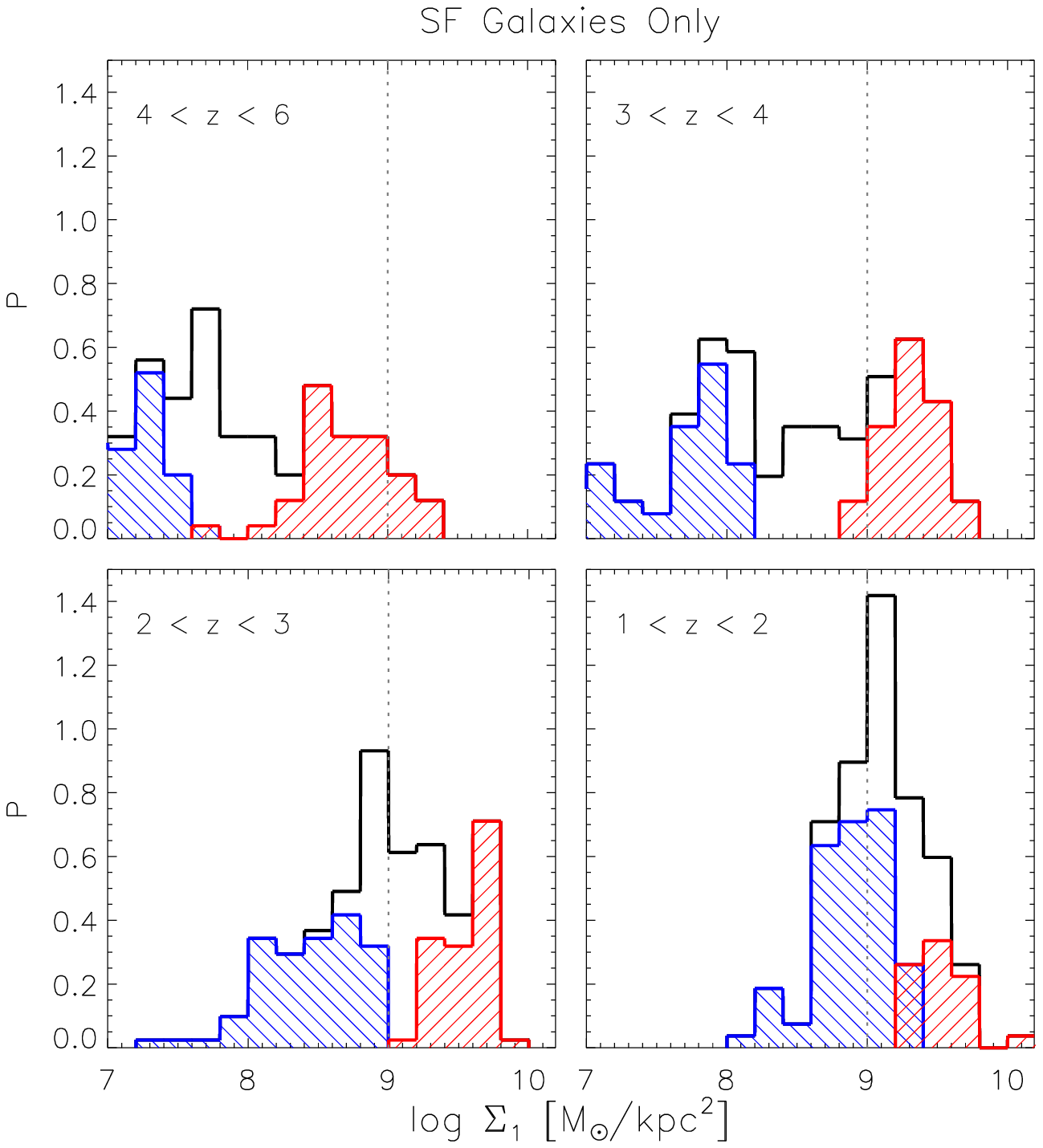}}
\subfigure{\includegraphics[width=0.45\textwidth]
{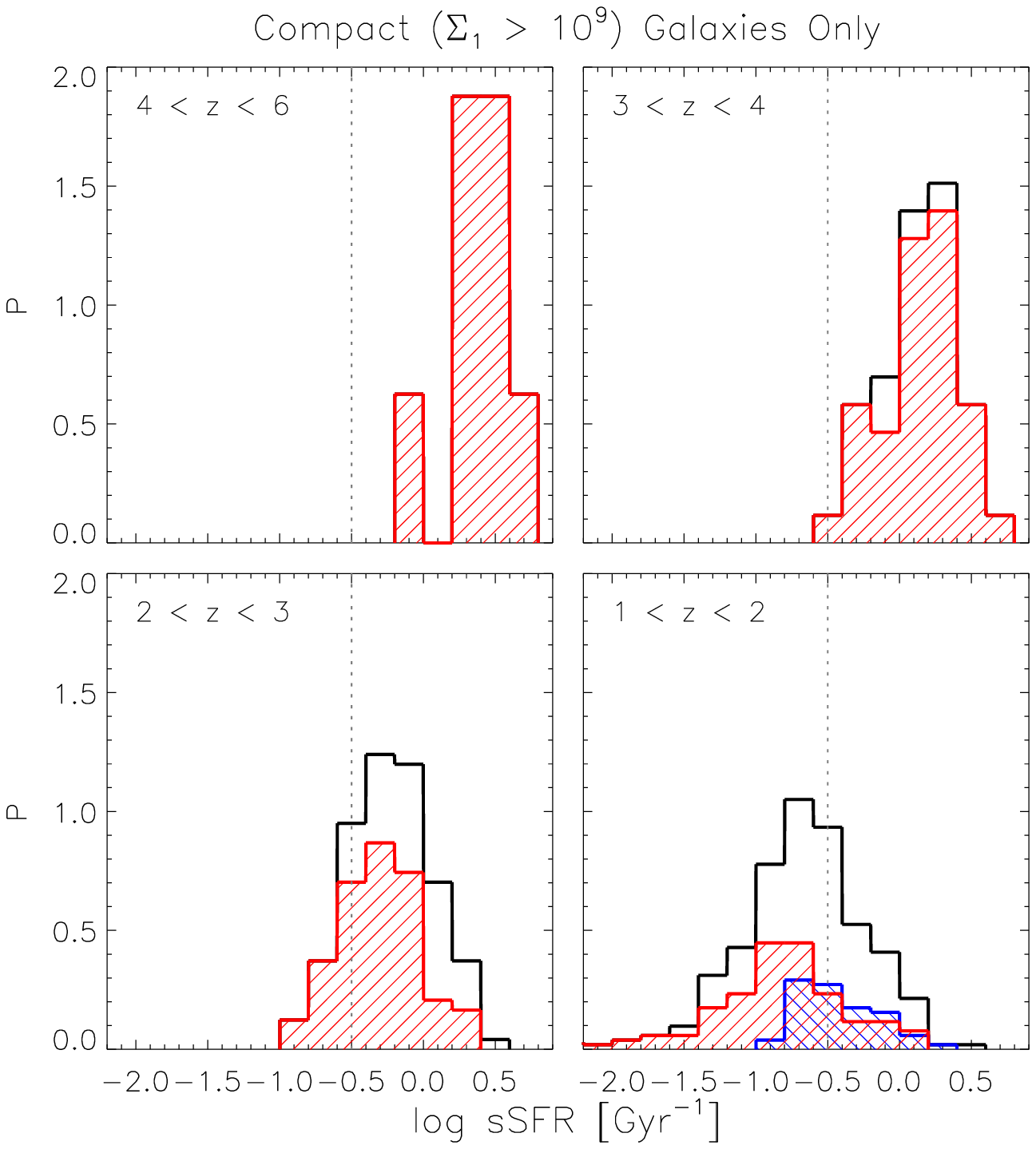}}
\caption{
Probability distributions
following \fig{ssfr_sigma_all}, at the same redshift bins and mass bins
(blue and red corresponding to the lowest and highest thirds by mass
respectively).
{\bf Left:} Distribution of $\Skpc$
for star-forming galaxies (sSFR$>\!0.3\Gyr^{-1}$).
{\bf Right:} Distribution of
sSFR for compact galaxies ($\Skpc >10^9 \msun\kpc^{-2}$).
A bimodality is 
indicated
in the distribution of $\Skpc$ at $z \geq 3$.
This distribution is gradually shifting in time towards higher $\Skpc$.
Galaxies in the higher mass bin compactify earlier, and quench earlier,
while galaxies of the lower mass bin compactify only after $z\!\sim\!2$.
The sSFR distribution of the compact galaxies is shifting towards lower
values at $z<3$.  % total normalized to unity, and the mass bins to 1/3 each. 
}
\label{fig:quad_histo}
\end{figure*}

\begin{figure*} % 15 <-- 12 <- 13+14
\centering
\subfigure{\includegraphics[width=0.41\textwidth]{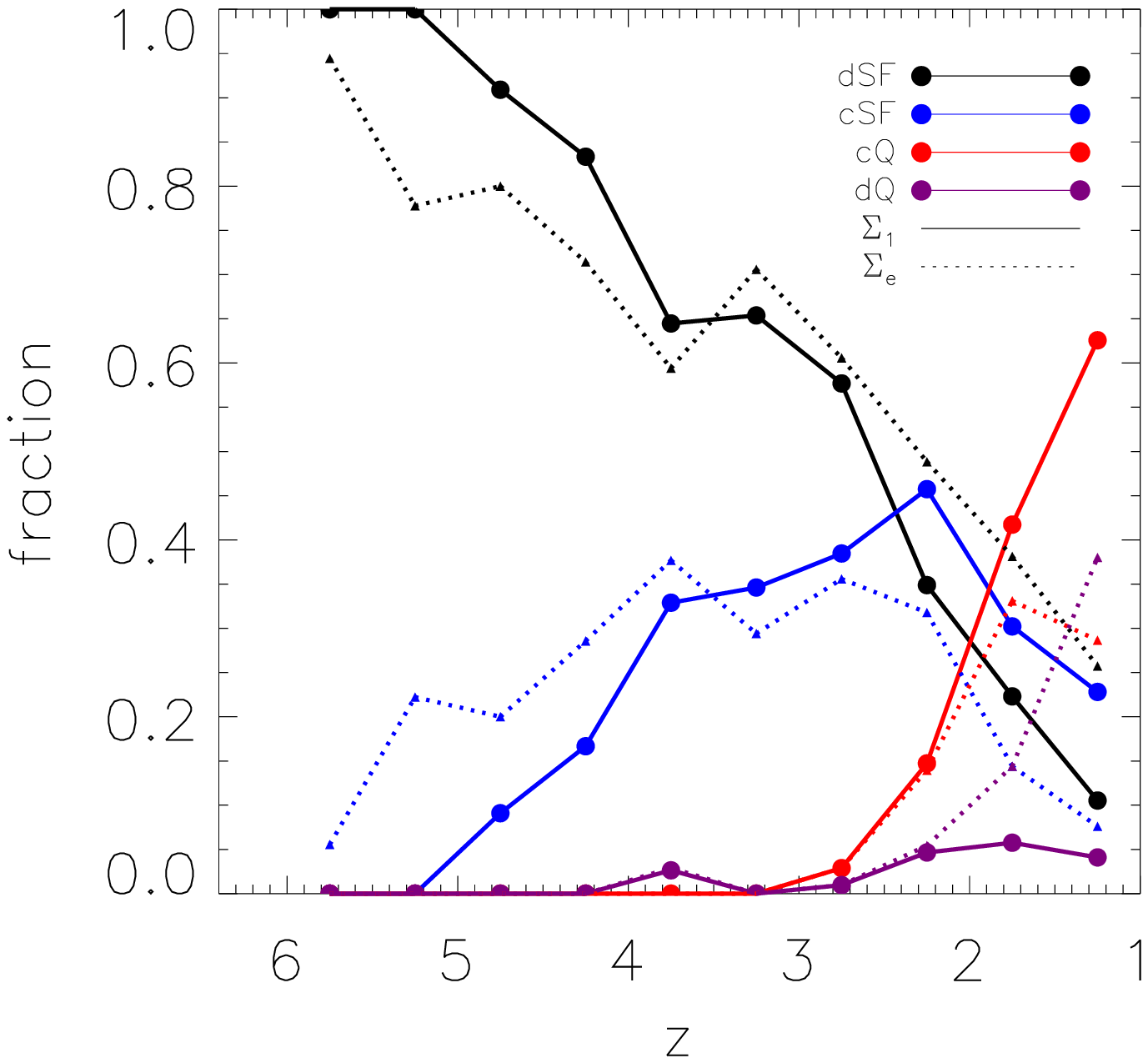}}
\subfigure{\includegraphics[width=0.41\textwidth]{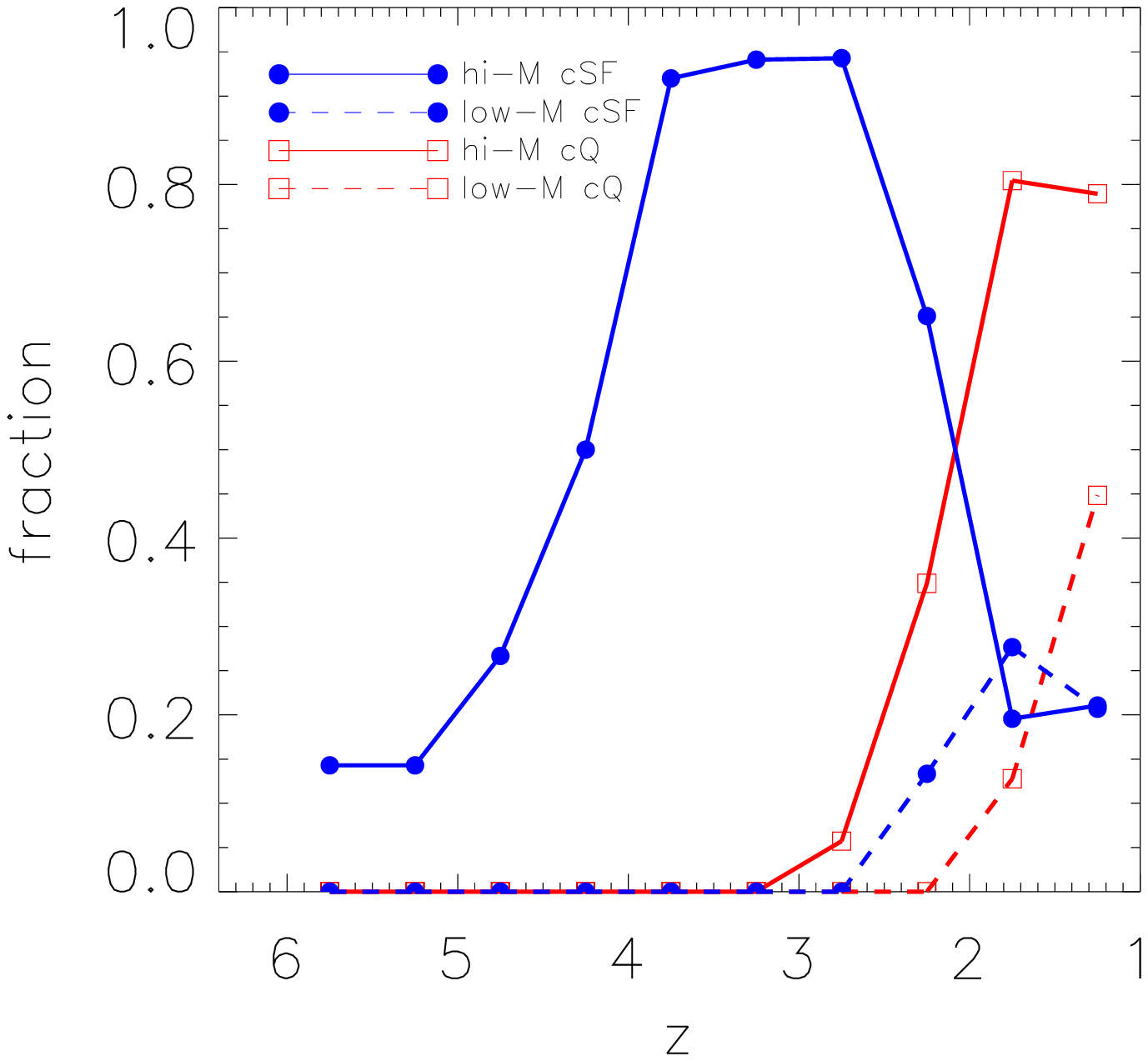}}
\caption{Time evolution of fractions of different components,
following \fig{ssfr_sigma_all}.
{\bf Left:}
Fraction of galaxies in each quadrant, with
the compactness defined either by $\Skpc$ (solid) or by $\Se$ (dotted).
The fraction of diffuse SFGs (black) is declining in time,
at the expense of the gradual growth of the compact fraction (blue plus red).
The fraction of compact SFGs (blue) is increasing in the range $z=5-2.5$,
and the fraction of compact quenched galaxies (red) is increasing after $z=3$.
The fraction of galaxies that quench to the diffuse quadrant
(purple) is negligible when defined by $\Skpc$, but not so with respect to
$\Se$.
{\bf Right:}
For each of the high-mass (solid lines) and low-mass (dashed lines) bins,
the fractions of compact-SFG and of compact-quenched.
The more massive galaxies become compact-SFG earlier and quench earlier.
%\adr{Consider comparing to observations in the text.}
%For each mass bin, the fractions in the 4 quadrants should sum
%%up to unity, but we show only the two quadrants of main interest.
}
\label{fig:quad_frac}
\end{figure*}

\smallskip % L-shape
The distribution of points in \fig{ssfr_sigma_all}, which reflect  
evolution tracks as in \figs{ssfr_sigma_ex} and \ref{fig:ssfr_sigma_low},
seem to trace a characteristic
L-shape, with the star-forming galaxies compactifying along a horizontal track
from left to right, and then quenching along a vertical track. To see this more
clearly, we stack all snapshots of all galaxies in \fig{L-shape}.
The left panel shows the raw quantities.
In the middle panel, the sSFR in each snapshot is scaled by the sSFR of the
ridge of the MS at that stellar mass and redshift.
This ridge can be fitted by the expression
\be
{\rm sSFR_{\rm MS}} = 
s\, \Gyr^{-1}\, (1+z)^\mu \left( \frac{\Ms}{10^{10}\msun} \right) ^{\beta} \, ,
\label{eq:sSFR}
\ee
with $\mu=2.5$ and $\beta=0.14$.
This approximation has been derived for $z\!>\!1$ from
the cosmological specific accretion rate (sAR) into haloes,
plus the notion that the sSFR roughly equals the sAR,
both predicted analytically and measured in simulations
\citep{neistein08b,dekel13,dm14}.

\smallskip
In order to determine the value of the free normalization parameter $s$ for the
current simulated galaxies, we select a subsample of star-forming galaxies that
reside in the main sequence in two alternative ways; 
either as all snapshots in the redshift range $z\!=\!3\!-\!6$,
or as all snapshots where $\Skpc\!\leq\!10^9\msun\kpc^{-2}$.
The scaled quantity of interest is
\be
\Delta \log {\rm sSFR} = \log {\rm sSFR} - \log {\rm sSFR_{\rm MS}} \, .
\ee
When determining the best-fit value of $s$, we eliminate the outliers  
outside the 16\% percentiles in both sides (in order to focus on the
$\pm 1\sigma$ range). 
We obtain a normalization of $s=0.0446\pm 0.0015$ and 
$0.0476\pm 0.0015\Gyr^{-1}$ 
for the selection based on $z$ and on $\Skpc$ respectively,
so we adopt here $s=0.046\Gyr$. 
The scaled quantity $\Delta \log {\rm sSFR}$ with this value of $s$
is shown along the vertical axis of the middle and left panels of 
\fig{L-shape}.
We see that with this scaling the star-forming branch became rather horizontal,
with a small scatter of $\pm 0.25$ dex for the
$\Skpc\!\leq\!10^9\msun\kpc^{-2}$ sample 
(and $\pm 0.24$ dex for the $z=3-6$ sample).

The scatter of the vertical branch in the left panel, 
for $\log {\rm sSFR} < -0.5$, is 0.35 dex.
After scaling the sSFR in the middle panel, for $\Delta\log {\rm sSFR} <-0.5$, 
it becomes 0.24 dex.
The scatter of the vertical branch is narrowed further by brute force in the 
right panel, where $\Skpc$ for each galaxy
is scaled such that all the galaxies match at the same $\Skpcm$.
The quantity shown in the horizontal axis of the right panel is thus
\be
\Delta \Skpc = \log \Skpc - \log \Skpcm \, .
\ee
The resultant scatter is 0.08 dex.

\smallskip % MS ridge in Fig. 12
By comparing in \fig{ssfr_sigma_all} the sSFR of the massive galaxies (marked
red) with the ridge of the MS from \equ{sSFR} (dashed line), 
one can see that in the cases that are considered quenched by our fixed sSFR
threshold criterion, the decline rate of sSFR is faster 
than the overall decline rate of the ridge of the MS,
indicating that this is indeed a real quenching process
(see also \fig{ssfr_mass} in \se{corr} below).

\smallskip % high masses evolve earlier
The galaxies in each redshift bin of \fig{ssfr_sigma_all} were divided into 
three mass bins of equal numbers in each, marked by different 
colors.\footnote{This division by the mass in every redshift bin is  
similar but not identical to implementing mass cuts at the highest redshift 
bin and following the galaxies in each of the three initial mass bins 
as they grow in mass with time.}
We immediately notice that the more massive galaxies evolve earlier --
they compactify earlier and quench earlier. 
They start at higher central densities, and compactify to higher $\Skpc$
values at which they start and pursue their decisive quenching process to 
low sSFR values.
The galaxies of lower masses evolve later through a similar pattern, with the
main difference being that after compaction
they reach smaller maximum $\Skpc$ values, 
and sometimes attempt to quench at relatively low central densities. 
In terms of $\Se$, many of the low-mass galaxies tend to quench into the 
top-left quadrant of \fig{ssfr_sigma_all}.  
While these galaxies may not be above the $\Sigma$ threshold
commonly used to define ``nuggets", they are not qualitatively different,
as they also underwent compaction-triggered quenching.

\smallskip
The evolution of the distributions of galaxies in the quadrants of 
\fig{ssfr_sigma_all} is quantified in \figs{quad_histo} and \ref{fig:quad_frac}.
\Fig{quad_histo} shows in the same redshift bins the probability distributions
of
(a) $\Skpc$ (left) 
in the two SFG phases (lower quadrants of \fig{ssfr_sigma_all}), 
and (b) sSFR (right) in the two compact phases (right quadrants of
\fig{ssfr_sigma_all}).
The $\Skpc$ distribution gradually evolves to larger densities,
from being diffuse-dominated at $z>4$ to
compact-dominated at $z<3$, as observed \citep{barro13,barro14a}. 
The SFGs show a bimodality in $\Skpc$ at $z>3$, as predicted by DB14.
The reason is that once the pre-compaction galaxy is gas-rich enough, 
with a gas surface density above a ``wetness" threshold, a quick
compaction occurs before most of the gas turns into stars 
and $\Skpc$ becomes higher, evacuating the gap near 
$\Skpc \sim 10^8-10^9 \msun\kpc^{-2}$.
We notice, again, that the more massive galaxies compactify earlier and to
higher densities.
The sSFR distribution shifts to lower values starting at $z \sim 3$.  

\begin{figure*} % 16 <-- 13 
\centering
\subfigure{\includegraphics[width=0.37\textwidth]{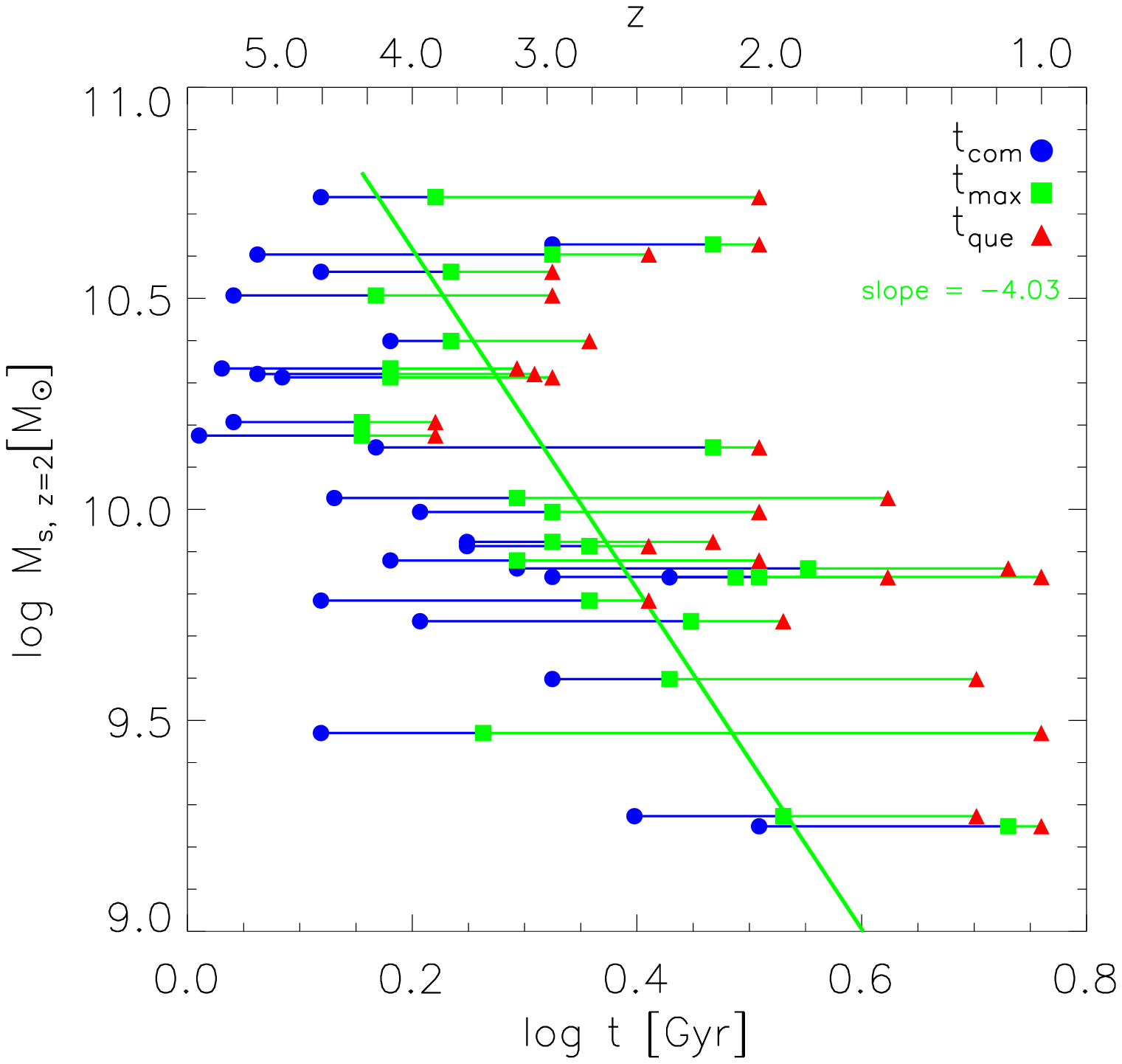}}
\subfigure{\includegraphics[width=0.37\textwidth]{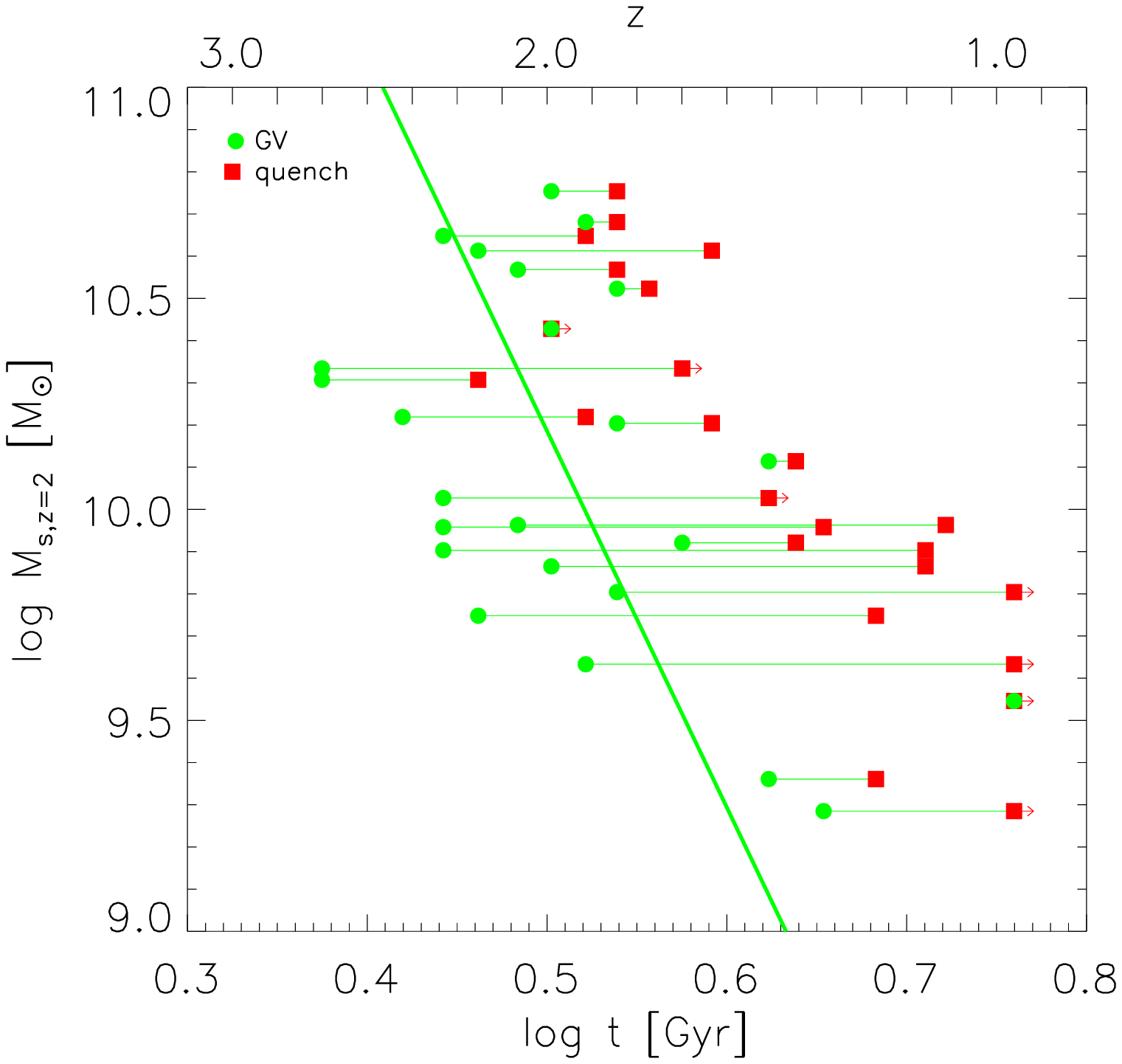}}
\subfigure{\includegraphics[width=0.37\textwidth]{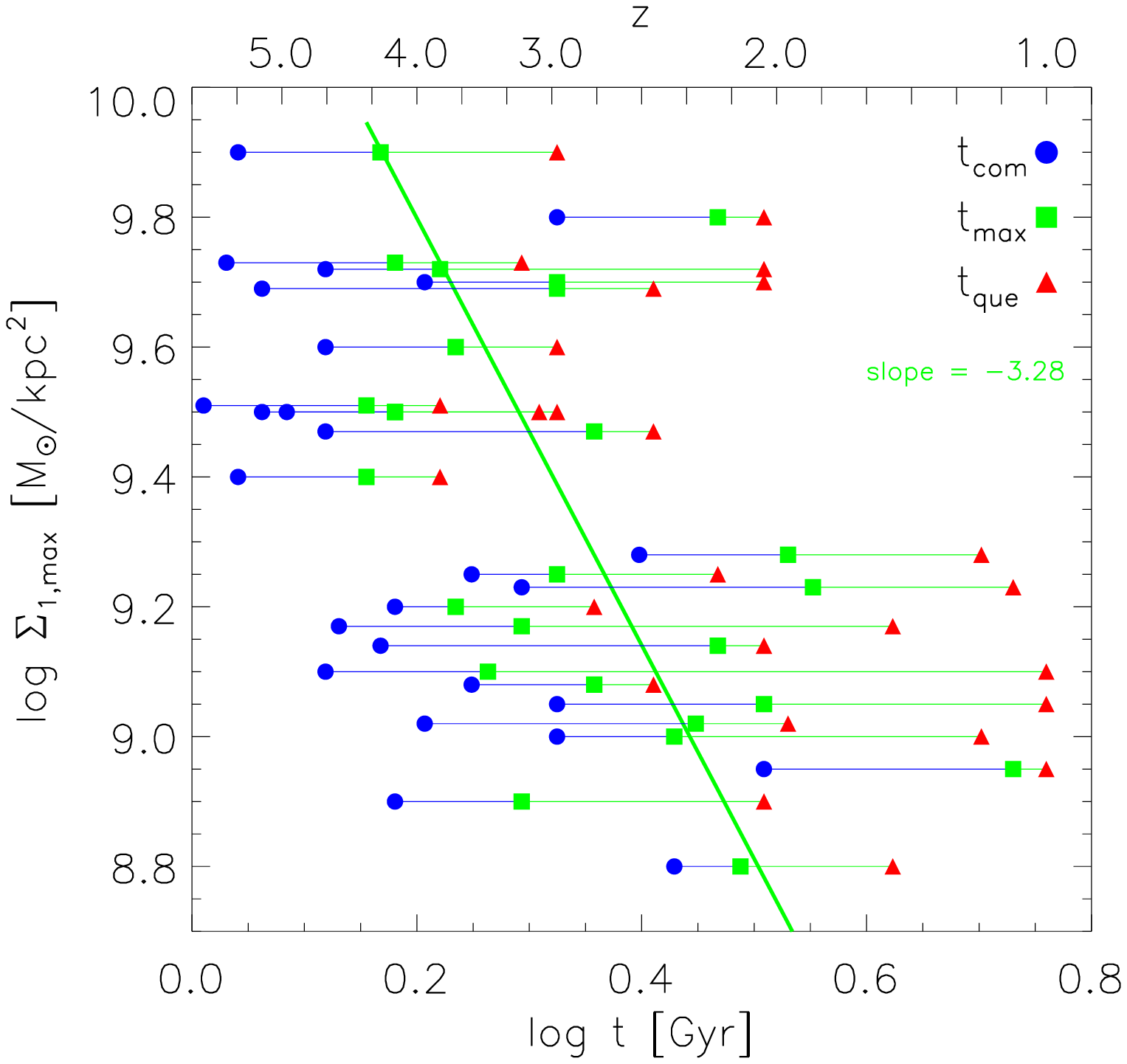}}
\subfigure{\includegraphics[width=0.37\textwidth]{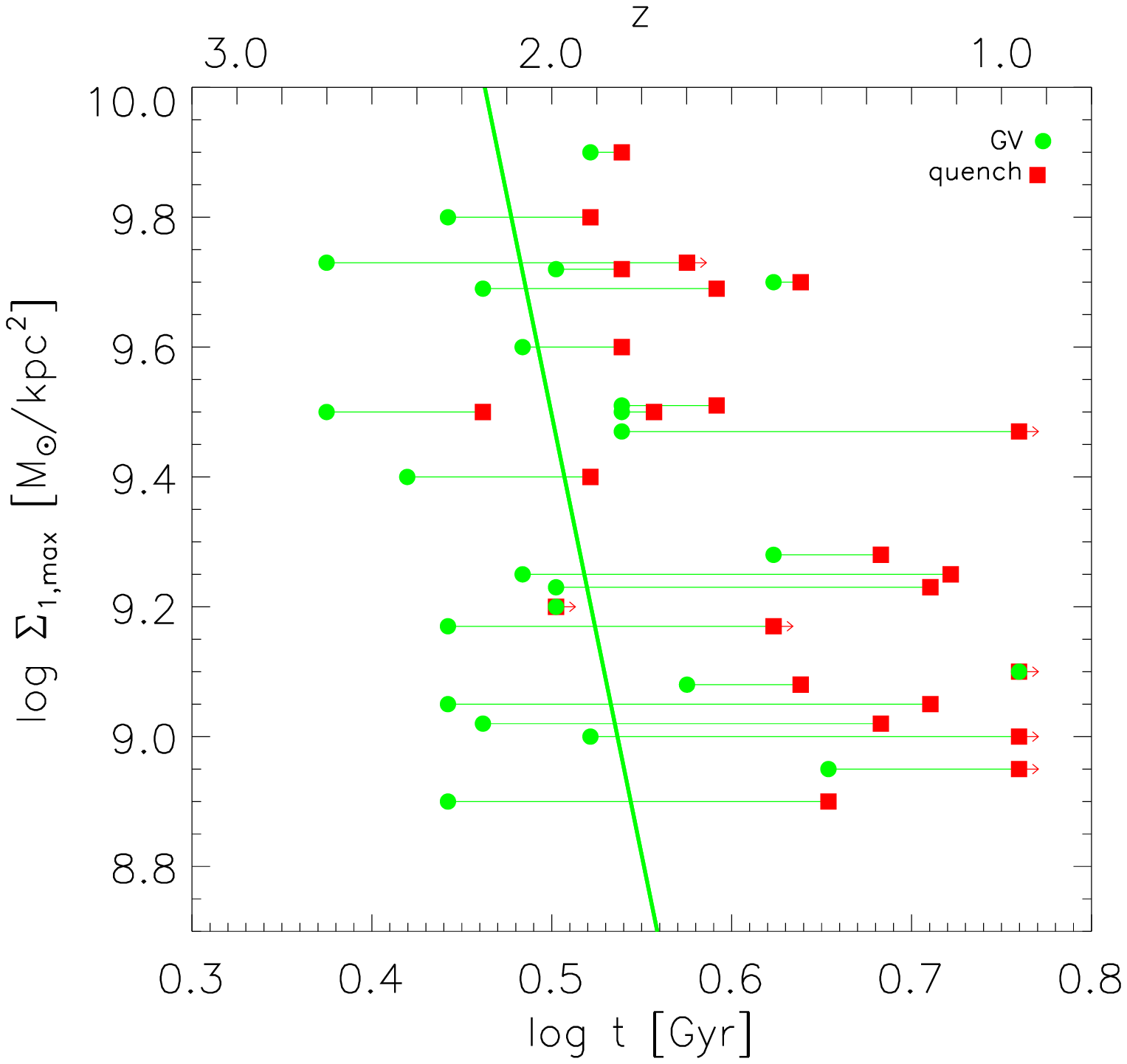}}
\caption{
Characteristic times for compaction and quenching within the inner
1 kpc versus the stellar mass at $z=2$ (top)
and versus maximum stellar surface density (bottom)
{\bf Left:}
Based on the gas density, the three times refer to
(a) the onset of gas compaction, $t_{\rm com}$ (blue circles),
(b) the maximum gas density where the compaction ends and the quenching starts,
$t_{\rm max}$ (green squares),
and (c) the quenching time $t_{\rm que}$ (red triangles), when the gas density
has dropped by a factor of 10 from its maximum.
{\bf Right:}
Based on sSFR, the two times refer to
(a) the first crossing of the green valley at sSFR$=0.3\Gyr^{-1}$,
and (b) the successful quenching beyond sSFR$=0.16\Gyr^{-1}$.
Arrows mark lower limits to the quenching time, in the seven cases that have 
not successfully quenched by the final snapshot of the simulation.
}
\label{fig:times}
\end{figure*}

\smallskip
\Fig{quad_frac} summarizes the time evolution of the fractions of galaxies in 
the different quadrants of \fig{ssfr_sigma_all}.
We see in the left panel 
that the fraction of diffuse SFGs (black) is gradually declining in time.
This is compensated by a growth in the fraction of compact galaxies, both 
star-forming and quenched (blue plus red). 
The fraction of compact SFGs (blue) is increasing in
the range $z=5-2.5$ and decreasing after $z=2.5$,
while the fraction of compact quenched galaxies (red) is negligible prior to
$z=3$ and is increasing steeply after $z=3$.
When compactness is measured by $\Se$, the fraction of compact SFGs is rather
flat between $z=4.5$ and 2.5, because $\Re$ is systematically growing in that
period.
The fraction of galaxies that quench to the diffuse quadrant
(purple) is negligible when defined by $\Skpc$, but not so with respect to
$\Se$.
The right panel of \fig{quad_frac} distinguishes between the highest and 
lowest mass bins.
For each mass bin, it shows the evolution of the fractions of compact-SFG and 
compact-quenched galaxies.
It clearly demonstrates that the more massive galaxies become compact-SFG
earlier and quench earlier. 

%\smallskip
%\adr{Adi: xxx Consider adding a figure (13) like figure 10 in 4 z bins
%for sSFR vs $\Ms$ (namely the main-sequence plane).
%Is this useful?  Postpone tio Taccheella+

%--------------------
\subsection{Characteristic Times}
\label{sec:times}

\smallskip % times
\Fig{times} shows the characteristic times (and corresponding 
redshifts\footnote{An approximate translation of time in Gyr to redshift, 
valid in the Einstein-deSitter regime, $z\!>\!1$, 
is provided by $1+z\!=\!(t/17.5\Gyr)^{-2/3}$.})
for compaction and quenching for our simulated galaxies.
In the left panels, the times are based on the gas density within the
central 1 kpc (e.g. the blue line in the second-from-right panels of
\figs{ssfr_sigma_ex} and \ref{fig:ssfr_sigma_low}).
The three times refer to   
(a) the onset of gas compaction, $t_{\rm com}$ (blue circles),
where the gas density growth rate steepens abruptly,
(b) the peak of gas density, $t_{\rm max}$ (green squares),
where the gas compaction ends and the quenching starts, 
and (c) the quenching time $t_{\rm que}$ (red triangles), when the gas density
has dropped by a factor of 10 from its maximum value. 
In the right panels, the times are based on sSFR (e.g. the left
panels of \figs{ssfr_sigma_ex} and \ref{fig:ssfr_sigma_low}). 
The two times refer to 
(a) the first crossing of the green valley (green circles), 
defined at sSFR$=0.3\Gyr^{-1}$, and viewed as the first quenching attempt,
and (b) the successful quenching (red squares), where the sSFR drops below  
$0.16\Gyr^{-1}$.
Two galaxies are omitted from the right panels (V34 and V13) because they
have not reached the green valley by their last snapshots ($z=1.86$ and 
$z=1.5$).
Squares with arrows mark lower limits to the successful quenching time, in 
cases that have not successfully quenched by the final snapshot of the 
simulation.
Some of the galaxies have several periods of compaction in the redshift range
studied, but here we pick only one of these compaction events, 
tending to identify the one where the central gas density reaches the highest
peak value, and favoring the latest compaction event prior to the last 
snapshot of the simulation (which is typically $z=1$).

\smallskip
The onset of compaction occurs in the range $t\!=\!1\!-\!2.7\Gyr$ 
($z\!=\!5.7\!-\!2.5$).  
The compaction typically takes $0.5-1\Gyr$. 
The end of compaction and beginning of quenching typically
happens in the range $t\!=\!1.4\!-\!3.5\Gyr$ ($z\!=\!4.4\!-\!1.9$).
The first crossing of the green valley occurs in most cases
in the range $t\!=\!2.4\!-\!4.5\Gyr$ ($z\!=\!2.8\!-\!1.5$),
and successful quenching is achieved after $t\!=\!2.8$ ($z\!=\!2.4$)
and possibly only after $t\!\sim\! 6\Gyr$ ($z\!\sim\!1$).
The quenching process can take anywhere between half a Gyr to several Gyrs.
As a rule of thumb, the typical duration of the comapction and the quenching 
events is roughly a constant fraction of the Hubble time, $(0.3\!-\!0.4)\,t$.

\smallskip
In \fig{times}, 
the galaxies are ranked along the vertical axis either by their stellar mass at 
$z\!=\!2$ (top) 
or by their maximum central stellar surface density $\Skpcm$ (bottom), 
which characterizes the central density of that galaxy during the whole
quenching process.
%\adr{Adi, in figs 2 and 3 it seems that the actual max of Sigma1 is reached in
%most cases well after the turning point of sSFR. Is the max of Sigma1 really a
%good indicator of BN? How do we deal with this?}
We see a clear trend of the characteristic times with galaxy mass and with 
$\Skpcm$, showing that more massive galaxies evolve earlier and to higher
maximum central densities. 
A similar trend exists with respect to halo mass (not shown here).
The {\it most massive} galaxies in our sample
typically start compaction at $t\!=\!1\!-\!1.5\Gyr$ (namely by $z\!=\!4$),
and compactify within less than $0.5\Gyr$. 
They then immediately start quenching, and most of them successfully 
quench by $t\!=\!3.7$ ($z\!=\!1.8$). 
The {\it least massive} in our sample start compaction at  
$t\!=\!1.5\!-\!2.7\Gyr$ ($z\!=\!4\!-\!2.5$), and compactify within 
$0.5\!-\!1\Gyr$, sometimes over a longer period.
The quenching process of the least massive galaxies can take between 
1 to 3 Gyr.
This is consistent with the less-massive examples shown in 
\fig{ssfr_sigma_low}, where the quenching tends to be 
indecisive. In these galaxies the sSFR fluctuaties down and back up, 
representing several quenching attempts, each followed by a recurrent 
compaction and star-formation episode.
(This will be discussed in the context of halo quenching in \se{mass}.)

\begin{figure*} % 17 <- 14   $\lambda$ and sSFR vs $\Skpcm$
\centering
\subfigure{\includegraphics[width=0.38\textwidth]{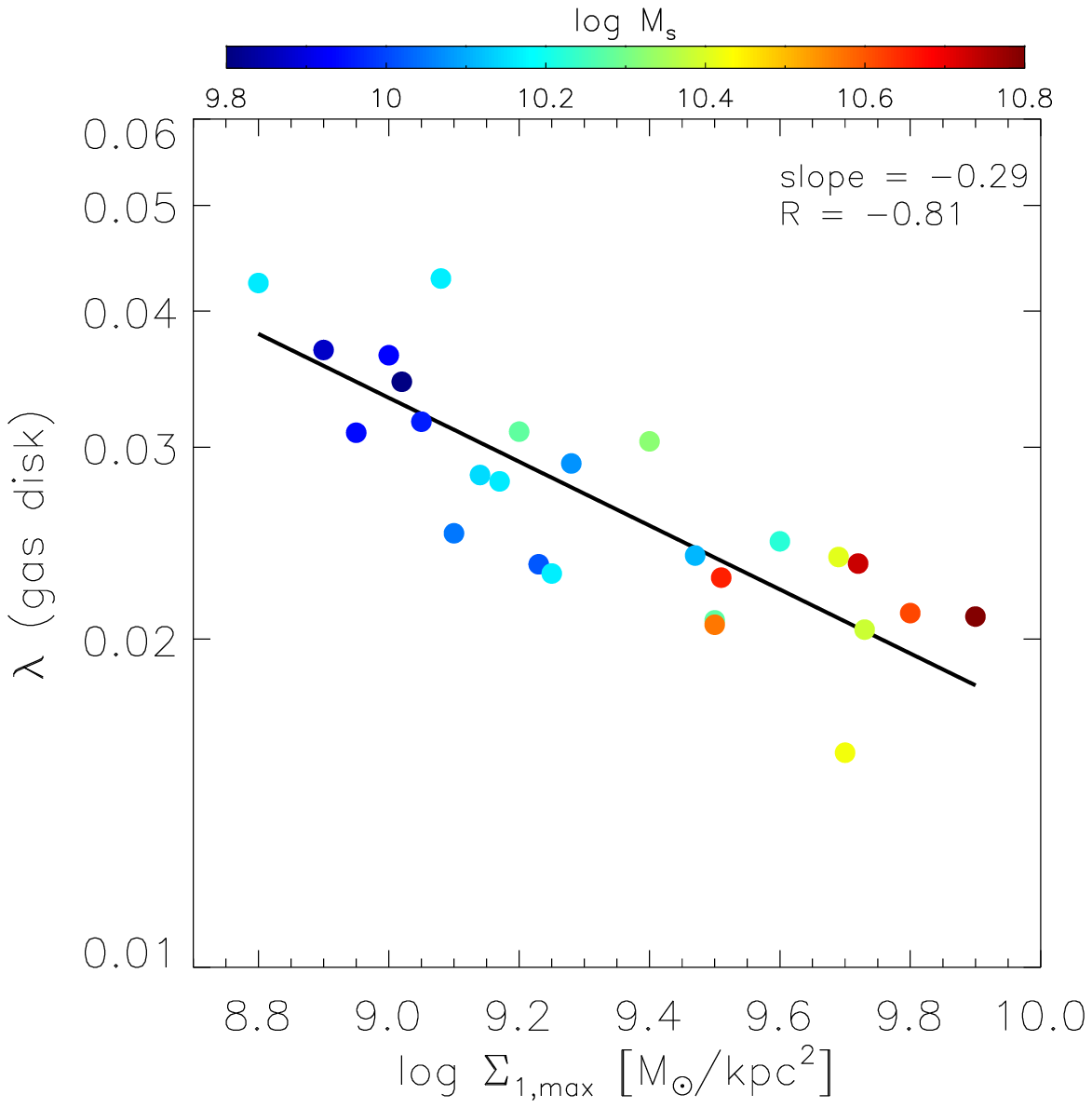}}
\subfigure{\includegraphics[width=0.38\textwidth]{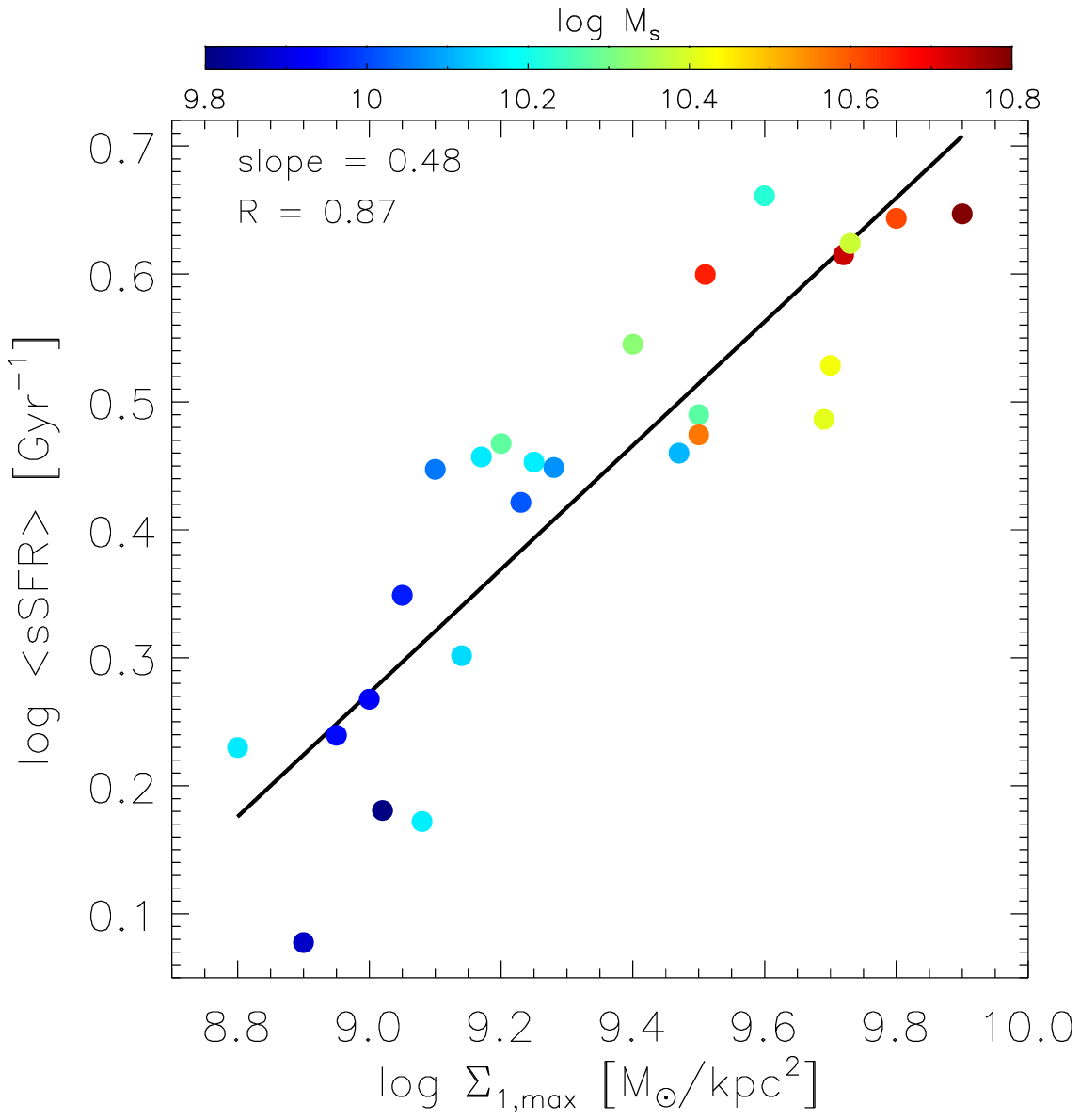}}
\caption{
Properties of galaxies in the pre-compaction phase 
against the maximum value of stellar surface density $\Skpcm$ that is
reached after compaction.
The color refers to stellar mass at the time when $\Skpcm$ is reached.
{\bf Left:}
The spin parameter $\lambda$ of the cold gas (T$<10^5K$) within the disc
radius.
{\bf Right:}
The average sSFR.
There is an anti-correlation between $\lambda$ and $\Skpcm$ and between
$\lambda$ and $\Ms$, and there is a correlation 
between sSFR and $\Skpcm$ and between sSFR and $\Ms$, 
as predicted by the model of DB14.
}
\label{fig:lambda_ssfr_sigma_max}
\end{figure*}

\smallskip
The fact that similar trends are seen as a function of mass and of maximum
central density is a manifestation of the strong correlation between the two
during the SFG phase (to be discussed in \se{mass}).
Also worth noting in the bottom panels of \fig{times} is that the quenching
does not occur at a very specific value of $\Skpc$ but rather in a range of
values, between $10^9$ and $10^{10}\msun\kpc^{-2}$ for the given sample,
though this range can possibly be considered as not very broad.
This corresponds to a comparable range of values for $\Ms$ at quenching 
(\se{mass}).

%---------------------

\subsection{Pre-compaction Spin and sSFR}

% lambda_sigma_max
The model for wet contraction by DB14 suggested an anti-correlation between the
compactness of blue nuggets and the pre-compaction spin parameter of their 
gas disc.
The idea is that galaxies with a low initial spin start with a high surface 
density gas disc, which implies a high wetness parameter.  
This means
that the gas is driven into the centre of the galaxy before it turns into 
stars, leading to a high central density in these galaxies. 
This is tested in
\fig{lambda_ssfr_sigma_max}, which shows, for all of the galaxies in our sample,
the pre-compaction spin parameter $\lambda$ of the cold gas (T$<10^5K$) 
within the disc radius,  
against the maximum value of stellar surface density within 1 kpc, $\Skpcm$.  
The spin parameter is taken to be the average over the three consecutive 
output times prior to the onset of gas compaction. The latter is identified 
visually 
as the last sharp upturn in the slope of the growth curve for the gas mass 
inside $1\kpc$ as a function of time 
(see \figs{ssfr_sigma_ex} and \ref{fig:ssfr_sigma_low}), 
referring to the latest compaction
event that leads to the maximum stellar density $\Skpcm$. 
Details on how the spin parameter is computed in a given snapshot
are provided in \citet{danovich14}.
We indeed see a significant anti-correlation between the two quantities,
with a log slope $-0.29$ and a correlation coefficient $R=-0.81$. 
Also shown, marked by colour, is the stellar mass at the time when 
$\Skpcm$ is obtained. It shows an anti-correlation between $\lambda$ and $\Ms$, 
with a similar slope. This is consistent with the tight linear scaling of 
$\Skpc$ with $\Ms$ for all the simulated galaxies at all redshifts, 
as discussed below in \se{corr}.

\smallskip

The model of DB14 predicts the mass and redshift dependence of $\lambda$ and 
$\Skpcm$, in their equation 15 (with equations 18 and 22) and equation 29, 
respectively.
From the bottom-left panel of our \fig{times}, with $\Ms \propto \Skpc$, 
one can deduce a relation between the redshift and stellar mass 
at $\Skpcm$, approximately $1+z \propto \Ms^{0.22}$. 
Inserting this in the equations of DB14, using the scaling relations 
(their equations 18 and 20) in the high-mass regime 
($\Mv\!\sim\!10^{11.5-12.5}\msun$), one obtains at $\Skpcm$ the scaling
$\lambda_{w=1} \propto \Skpcm^{-0.54} \propto \Ms^{-0.67}$. 
This is qualitatively consistent with the trends seen in 
\fig{lambda_ssfr_sigma_max}, where the differences in slopes reflects the 
crudeness of the DB14 toy model and measurement uncertainties in the 
simulations.

\smallskip % ssfr_sigma_max
The DB14 model also implies a correlation between the central density
at maximum compaction and the pre-quenching sSFR.
The idea is that a higher sSFR is associated with a higher gas surface density,
which is associated with a higher value of $w$, and thus a wetter
compaction, leading to a higher $\Skpcm$. 
For testing this prediction,
\fig{lambda_ssfr_sigma_max} displays the average sSFR during the pre-compaction
phase (as defined above) versus $\Skpcm$. 
It shows a correlation between the two quantities, as expected, with a
log slope $0.48$ and a correlation coefficient $R=0.87$.
As already noted based on \fig{times}, the higher values of $\Skpcm$, 
and therefore the higher values of sSFR, 
tend to be associated with compaction at an earlier redshift.
This is consistent with having a higher gas density at earlier epochs,
both because the Universe was denser and because the gas fraction was 
systematically higher.
 
\smallskip % a range of track characteristic. fast and slow modes
We can thus characterize the evolutionary tracks of galaxies during compaction 
and quenching by a number of correlated properties: 
the wetness of the pre-compaction phase, 
the duration of the shrinkage, the SFR in this phase,   
the central surface density at the peak of compaction, 
and the efficiency of the quenching process.
The high-mass, high-$\Sm$, high-sSFR tracks tend to be associated with 
high-redshift compaction (\figs{ssfr_sigma_all}, \ref{fig:quad_histo} 
and \ref{fig:times})
and/or with low pre-compaction spin parameter (\fig{lambda_ssfr_sigma_max}).
The low-mass, low-$\Sm$, low-sSFR tracks tend to occur at low redshift 
(\fig{times}) and/or in high-spin galaxies (\fig{lambda_ssfr_sigma_max}). 
These properties of the tracks are also correlated with the rate of
evolution along the tracks, where the high-$\Sm$ tracks of massive galaxies
represent a {\it fast mode} of compaction and quenching, 
while the lower-$\Sm$ tracks of lower-mass galaxies represent a 
{\it slower mode} of evolution through similar stages at less compact 
configurations (\fig{times}). 

\begin{figure*} % 18 <- 15
\centering
\subfigure{\includegraphics[angle=-90,width=0.48\textwidth]
{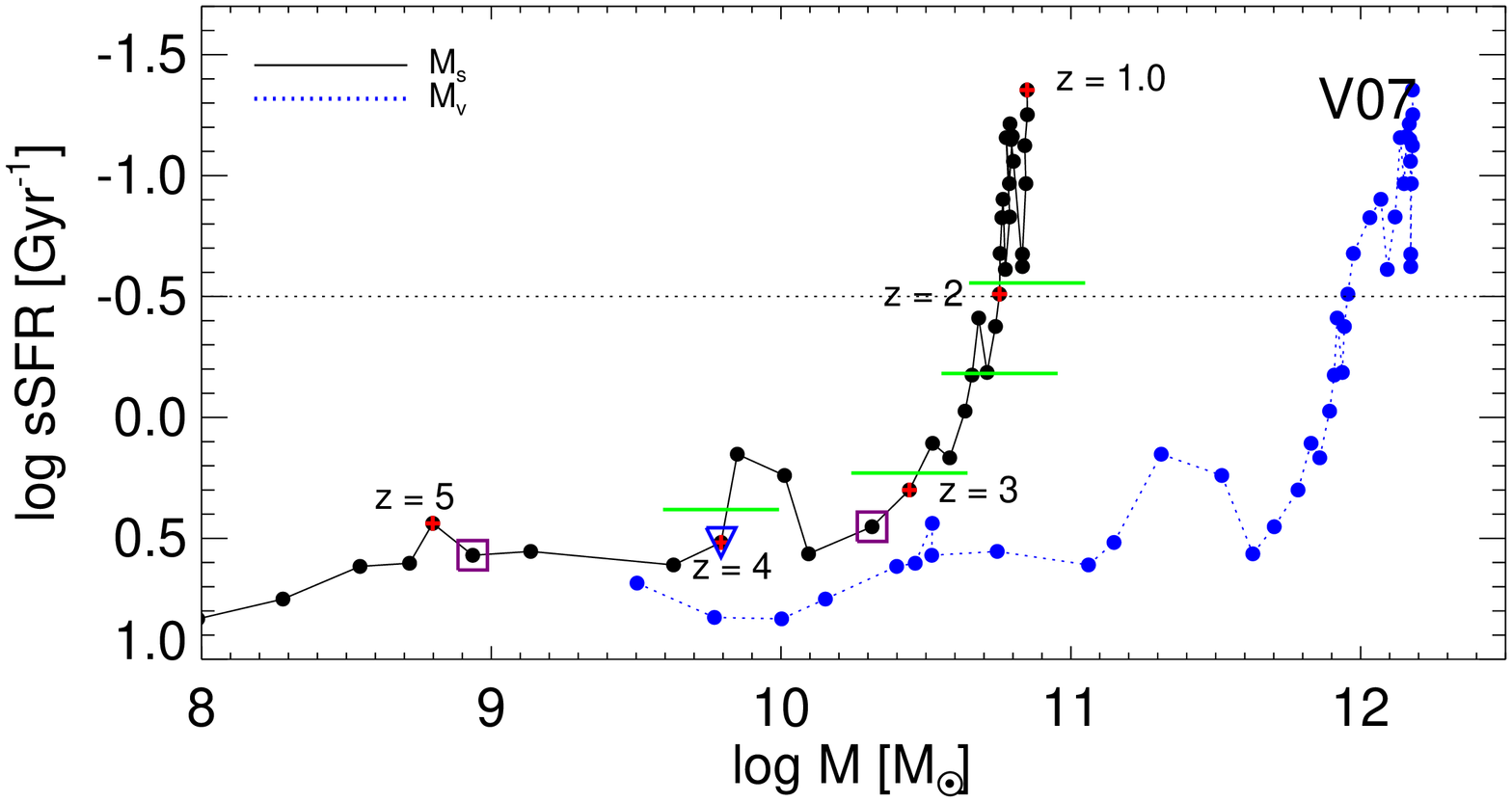}}
\subfigure{\includegraphics[angle=-90,width=0.48\textwidth]
{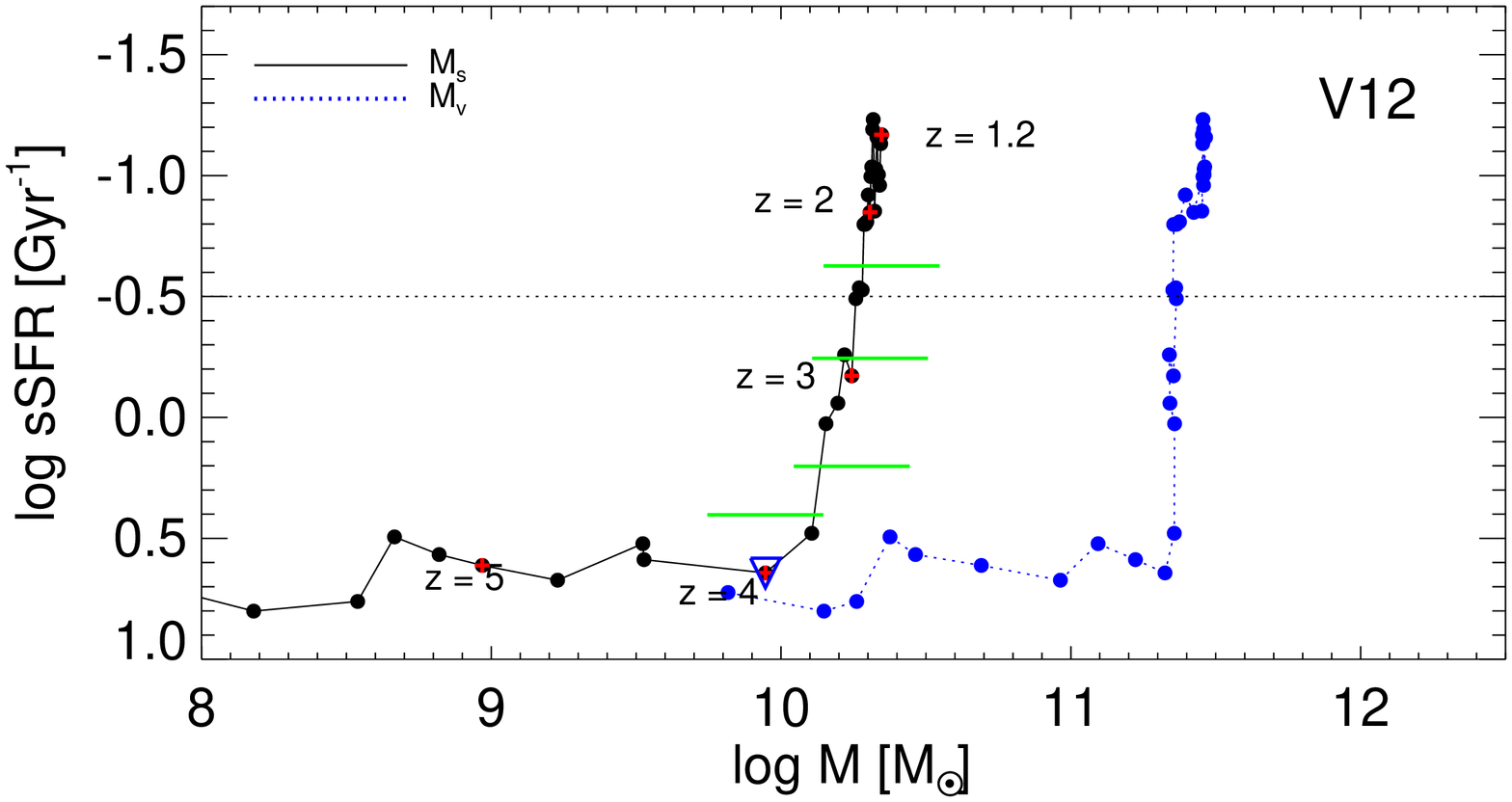}}
\\
\subfigure{\includegraphics[angle=-90,width=0.48\textwidth]
{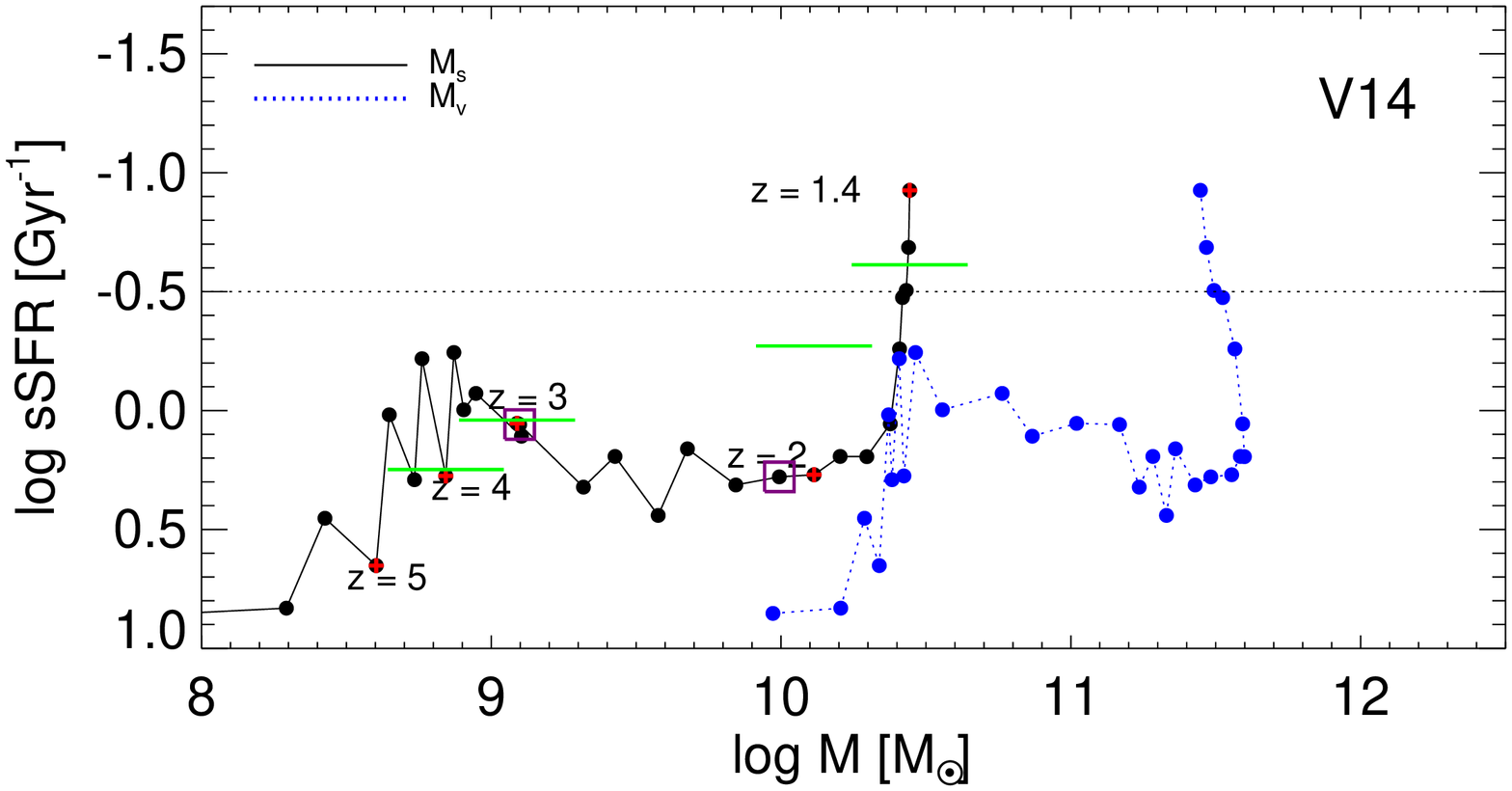}}
\subfigure{\includegraphics[angle=-90,width=0.48\textwidth]
{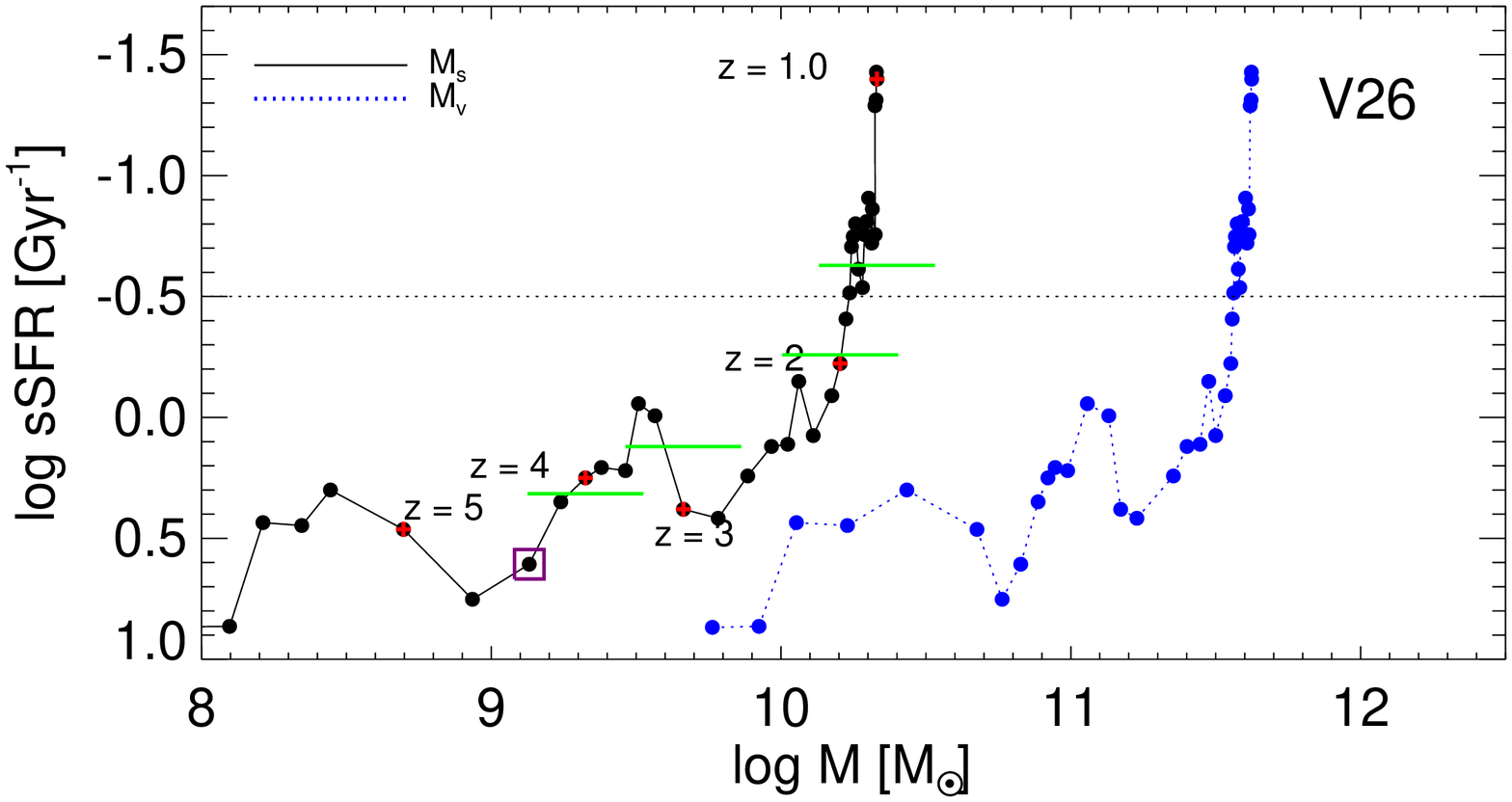}}
\\
\vskip 5mm
\subfigure{\includegraphics[angle=-90,width=0.48\textwidth]
{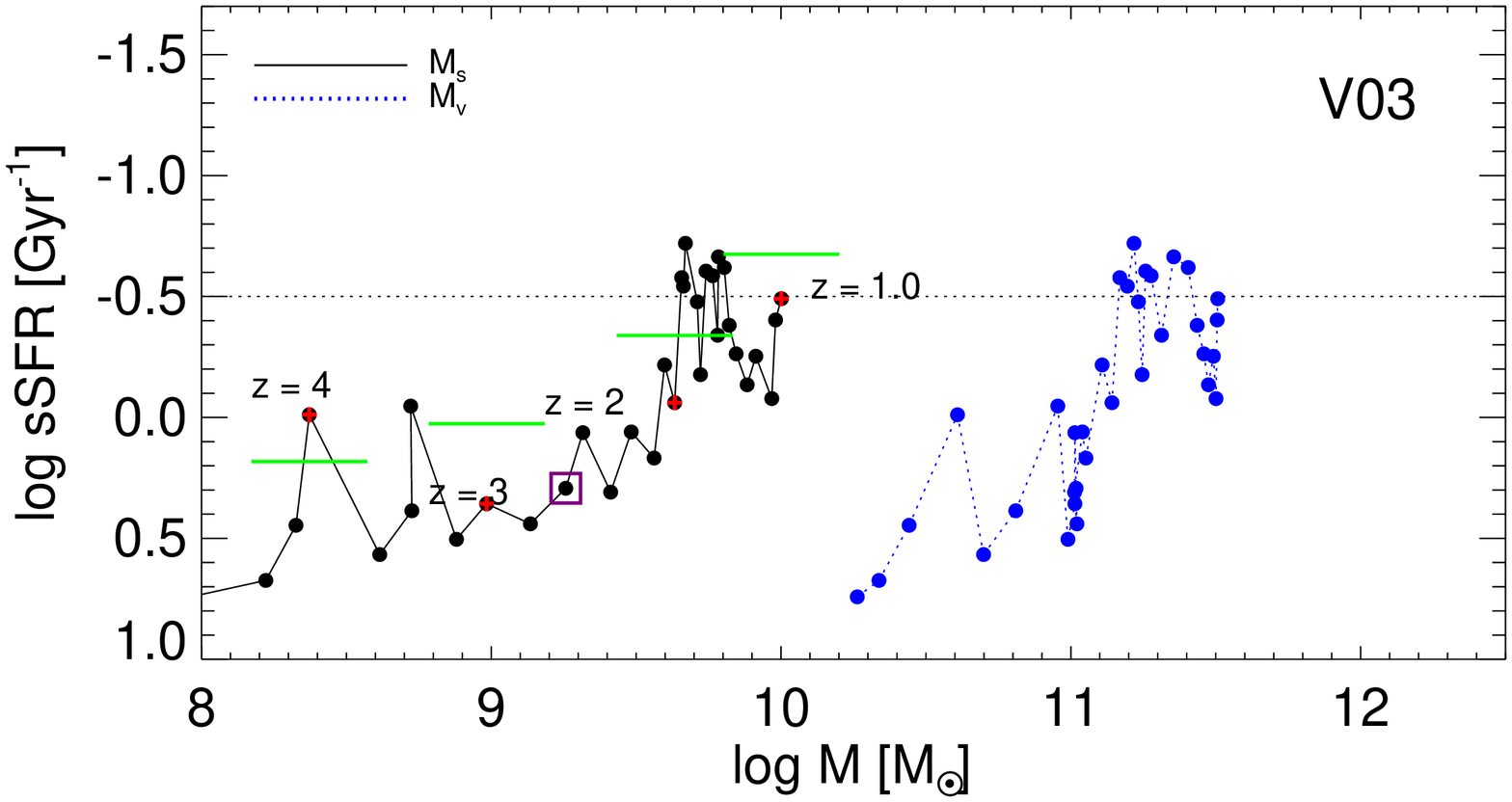}}
\subfigure{\includegraphics[angle=-90,width=0.48\textwidth]
{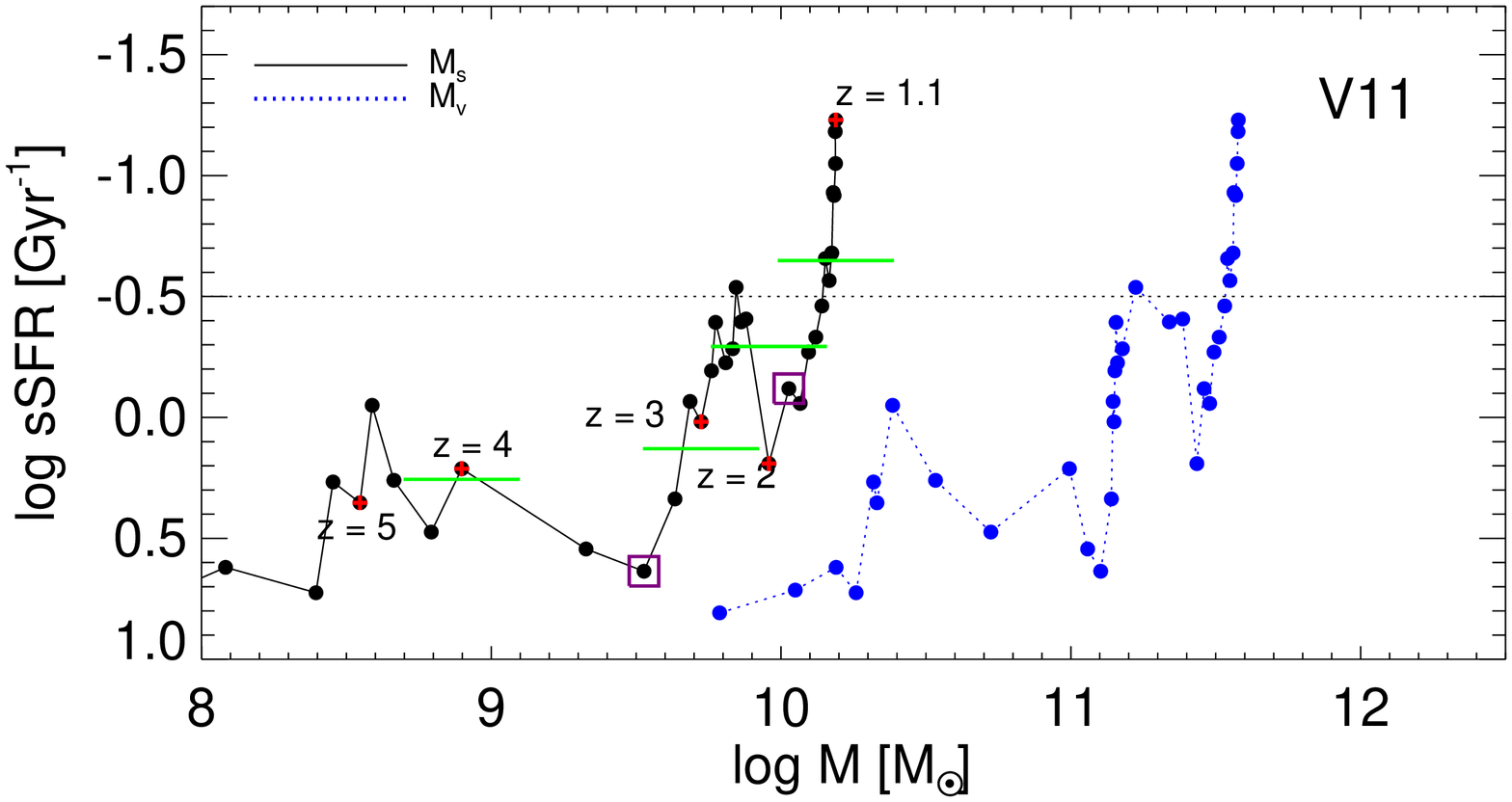}}
\\
\subfigure{\includegraphics[angle=-90,width=0.48\textwidth]
{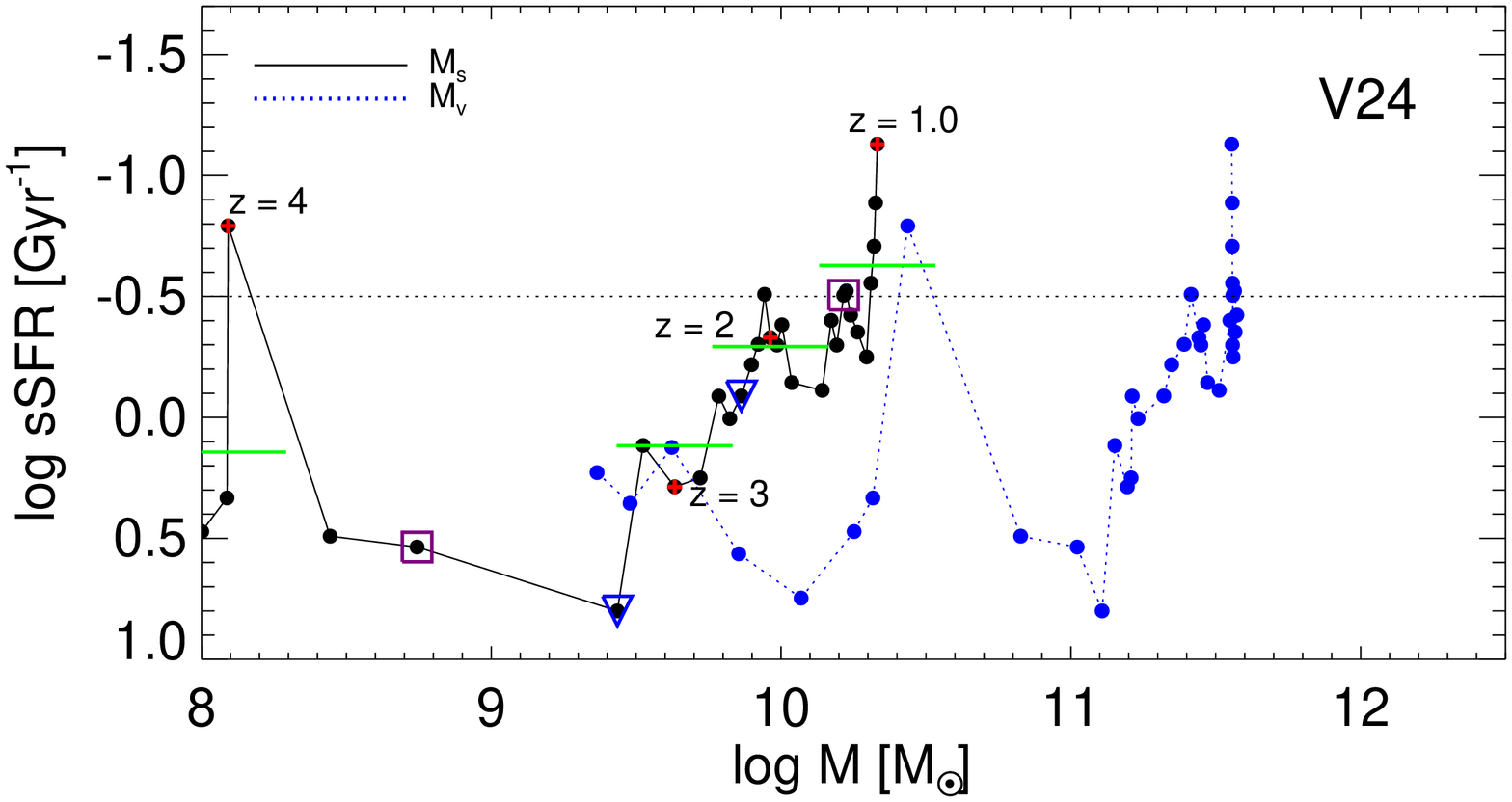}}
\subfigure{\includegraphics[angle=-90,width=0.48\textwidth]
{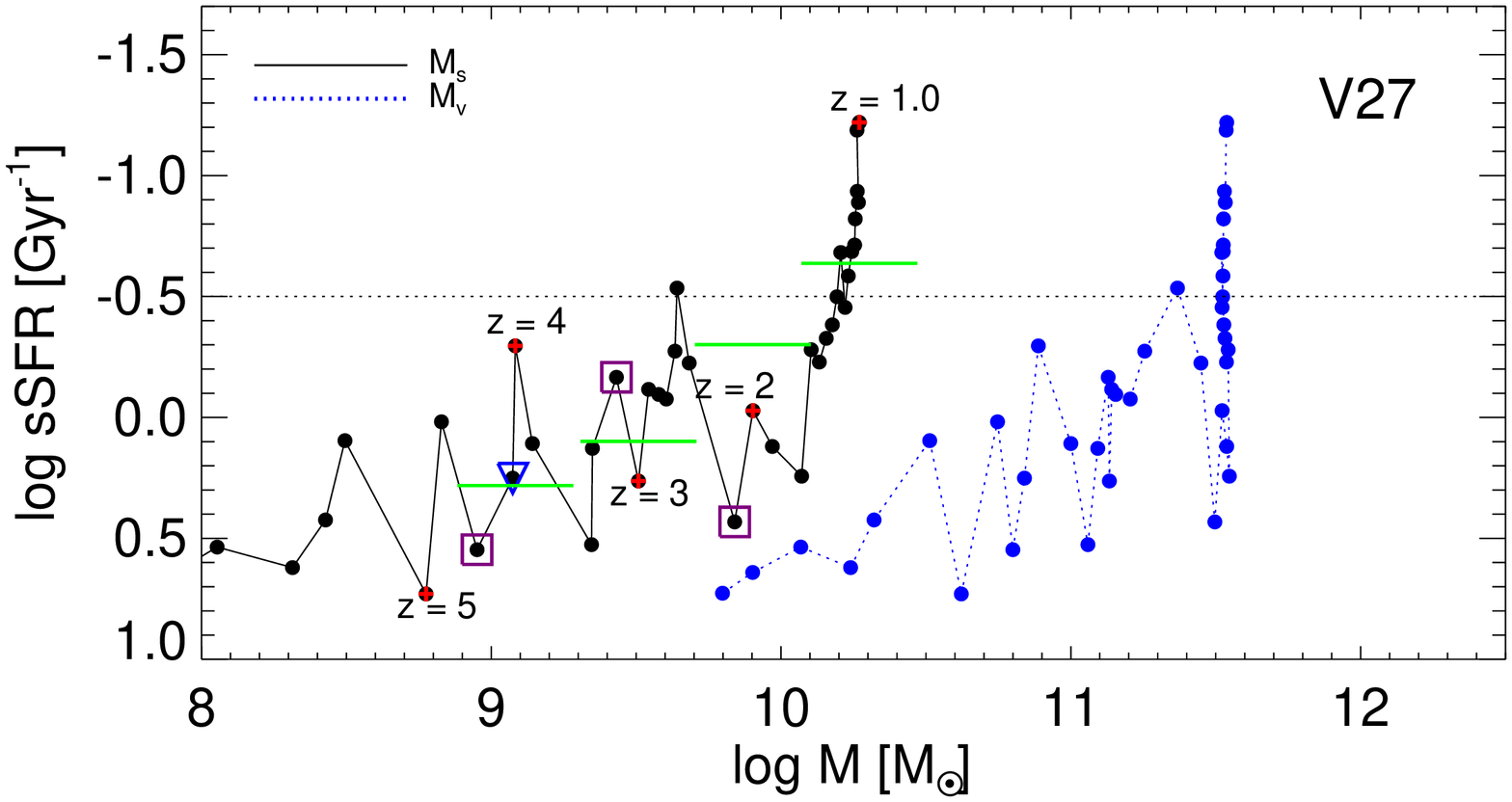}}
\caption{Evolution tracks in sSFR versus stellar mass (black solid lines)
and halo mass (blue dotted lines), for the eight examples shown in previous
figures, the high-mass examples
of \fig{ssfr_sigma_ex} (top four panels) and the low-mass examples
of \fig{ssfr_sigma_low} (bottom four panels).
The ridge of the MS is marked by green horizontal bars
at $z=5,4,3,2,1$, 
as evaluated by \equ{sSFR} for the mass of the galaxy at that
redshift.
The positive deviation of the sSFR from the MS ridge is
maximal during the blue-nugget phase at the onset of the first quenching
attempt.
The sSFR of galaxies that are identified as quenched by the fixed sSFR 
threshold declines faster than the MS ridge, indicating that the galaxy is 
indeed in the process of quenching.
}
\label{fig:ssfr_mass}
\end{figure*}

\begin{figure*} % 19 <- 16 
\centering
\subfigure{\includegraphics[width=0.49\linewidth]{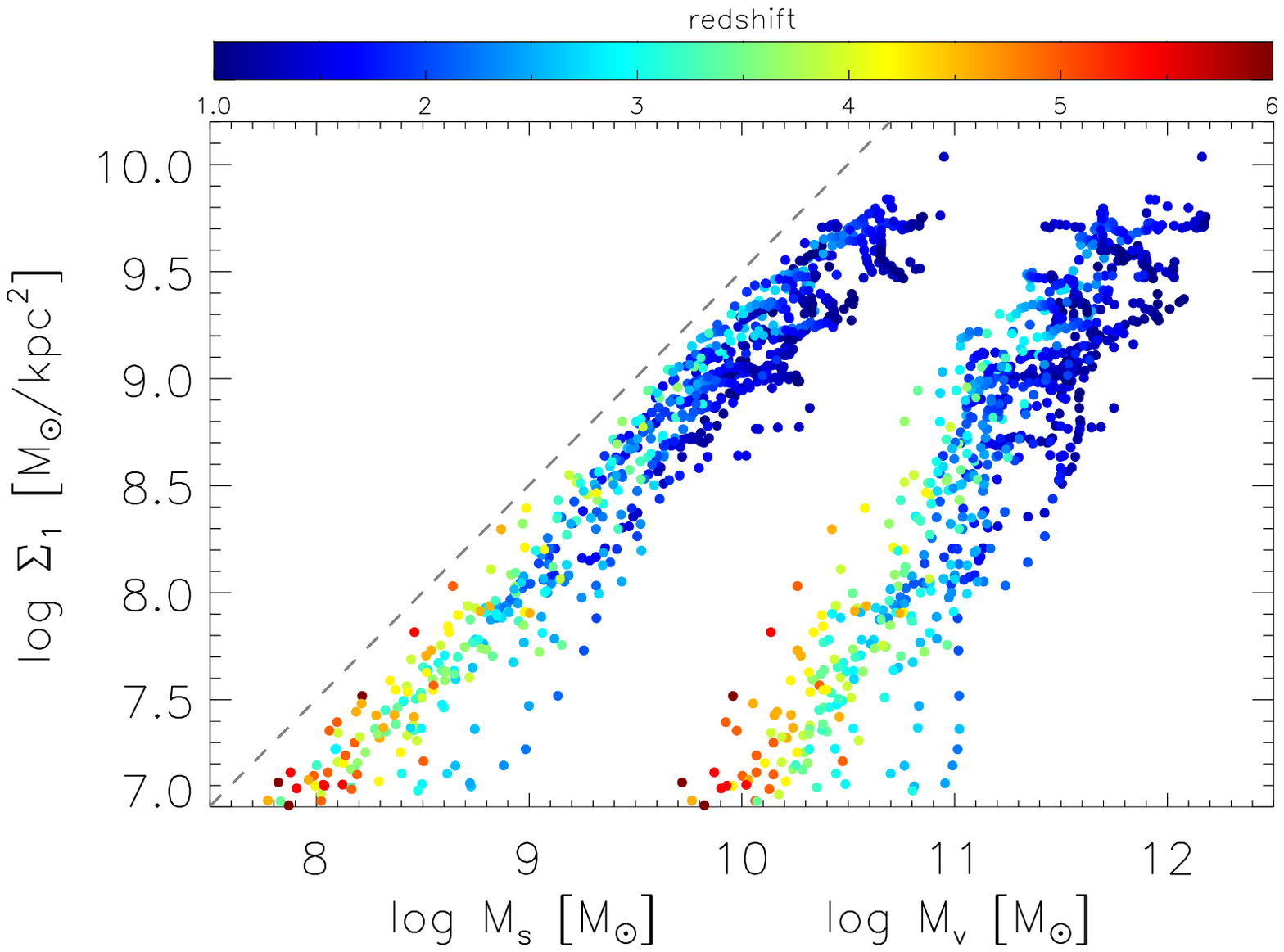}}
\subfigure{\includegraphics[width=0.49\linewidth]{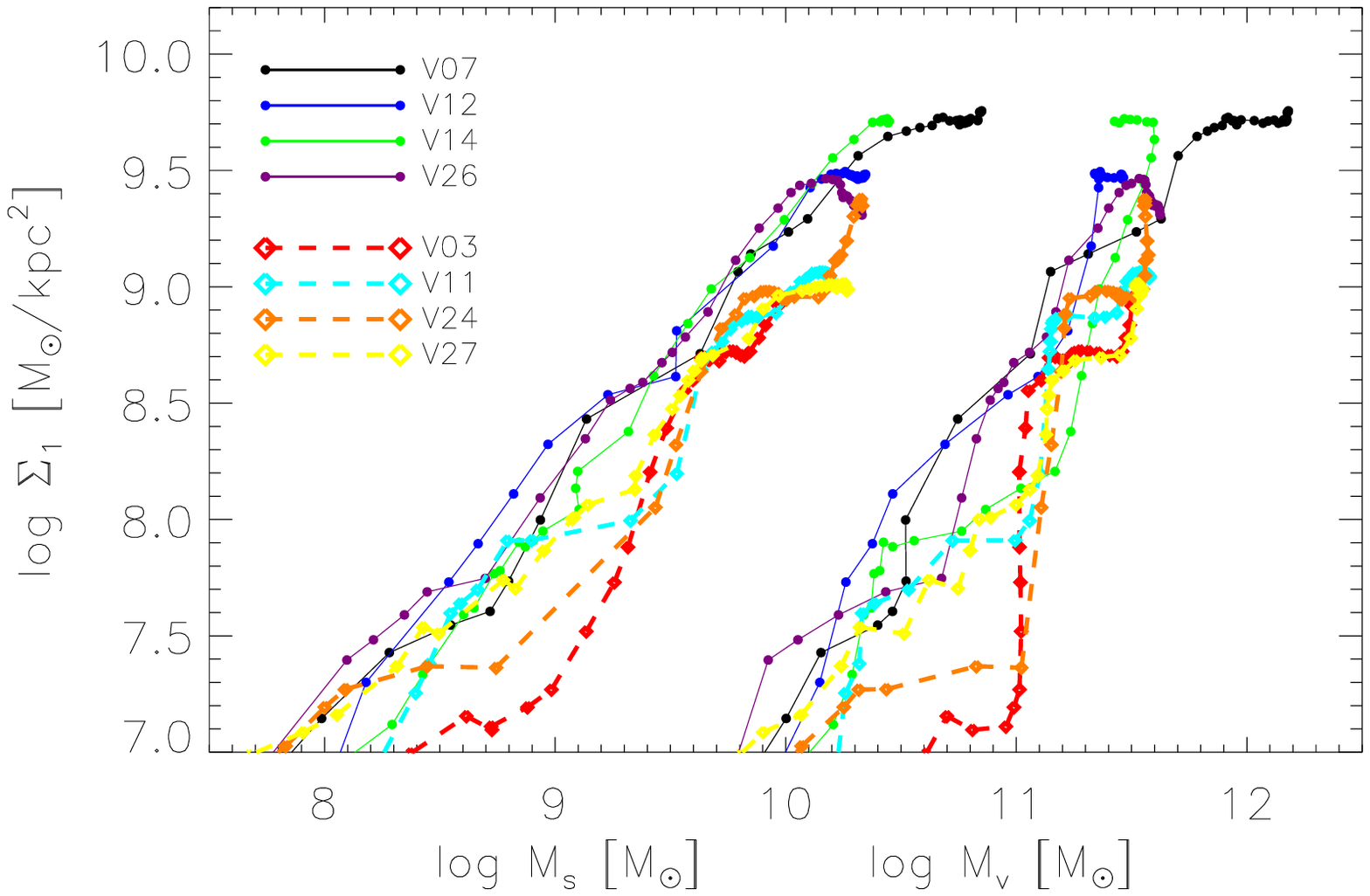}}
\caption{
Central stellar surface density versus galaxy stellar mass and halo mass.
{\bf Left:} All snapshots together.
{\bf Right:} Evolution tracks of the eight example galaxies shown in previous
figures.
The central surface density
$\Skpc$ is strongly correlated with mass.
The galaxies evolve along the correlation line in the SFG phases,
and branch out to a short horizontal segment (little mass growth)
during the long quenching phase (many densly packed snapshots),
at smaller $\Skpc$ for lower-mass galaxies. 
%\adr{Adi: Try to show SF vs quenched in the left diagram, 
%perhaps by using open symbols for the quenched, or any other way.}
}
\label{fig:sigma_mass}
\end{figure*}

%%%%%%%%%%%%%%%%%%%%%%%%%%%%%%%%%
\section{Stellar Mass and Halo Mass}
\label{sec:mass}

\subsection{Correlations Between Stellar and Halo Mass}
\label{sec:corr}

In the analysis so far, we have demonstrated correlations between the 
characteristic evolution of galaxies through compaction and quenching 
and two galaxy properties -- the total stellar mass $\Ms$ and the central 
surface density $\Skpc$, or $\Se$. We now examine the correlation
between these two variables, as well as with a third quantity -- 
the total halo virial mass $\Mv$. 
The quantities $\Ms$ and $\Skpc$ are expected to be strongly correlated 
in a trivial way as
long at $\Re$ is smaller than or comparable to $1\kpc$,
where $\Skpc \sim \Ms/\pi\kpc^{-2}$. This is indeed the 
case for our galaxies at the high redshift range where the compaction typically
occurs. Somewhat larger deviations may be expected for massive galaxies at
lower redshifts. 
On the other hand, $\Ms$ and $\Mv$ are naturally correlated.
So one expects the three variables to be correlated rather tightly.
This makes it hard to identify the actual physical source for quenching,
and in particular to address the potentially different roles played by
the (internal) central density and the (external) halo mass in the quenching 
process.

\smallskip %15
\Fig{ssfr_mass} shows evolution tracks in sSFR versus stellar mass and versus
halo mass for the eight example galaxies, the four more-massive galaxies of
\fig{ssfr_sigma_ex} in the top two rows, and the four less-massive galaxies of
\fig{ssfr_sigma_low} in the bottom two rows.
One can see that the characteristic L shape of the evolution tracks in
the sSFR-$\Skpc$ plane is translated to tracks of a similar shape in the
sSFR-$\Ms$ plane and in the sSFR-$\Mv$ plane, 
emphasizing in particular the sharp onset of quenching
corresponding to the peak of central
gas density and SFR in the blue-nugget phase.

\smallskip % MS main sequence
Recall that the sSFR-$\Ms$ plane at a given redshift is where one commonly
identifies the basic galaxy bimodality into a main-sequence of star-forming
galaxies (MS) and a red-sequence of quenched galaxies. Thus, the tracks shown
in \fig{ssfr_mass} tell how individual galaxies evolve in this plane (shown
upside down), 
while the zero-point of the MS ridge is gradually shifting 
toward smaller sSFR values at later times according to \equ{sSFR}.
This evolving ridge is marked by green horizontal bars in \fig{ssfr_mass}
at $z=5,4,3,2,1$.
One can see that the galaxies tend to lie near the ridge at high
redshift. At the blue-nugget phase, near the onset of the first major 
quenching attempt, the positive deviation of sSFR from the ridge tends to 
be at a maximum.
After quenching across the green valley (sSFR$\sim\!0.3\Gyr^{-1}$),
the sSFR tends to be below the ridge line, eventually dropping to much lower 
values of sSFR as the quenching proceeds to the red-nugget phase. 
We note that the the decline rate of sSFR is indeed faster than that of 
the MS, confirming the interpretation of a quenching process
away from the MS ridge and toward the red sequence.  
In galaxy V27 the galaxy fluctuates above and below the main-sequence 
ridge several times, in a sequence of compaction-driven SFR episodes
followed by quenching attempts, before the final quenching occurs.
This demonstrates a tight association of the evolution through 
compaction and quenching events and the deviation of the sSFR 
from the MS ridge. 
It may explain the gradients of galaxy properties
across the main sequence, as well as the small scatter about its ridge
\citep[explored in detail in][]{tacchella15_ms}.

\begin{figure} % 20 <- 17 
\includegraphics[width=0.40\textwidth]{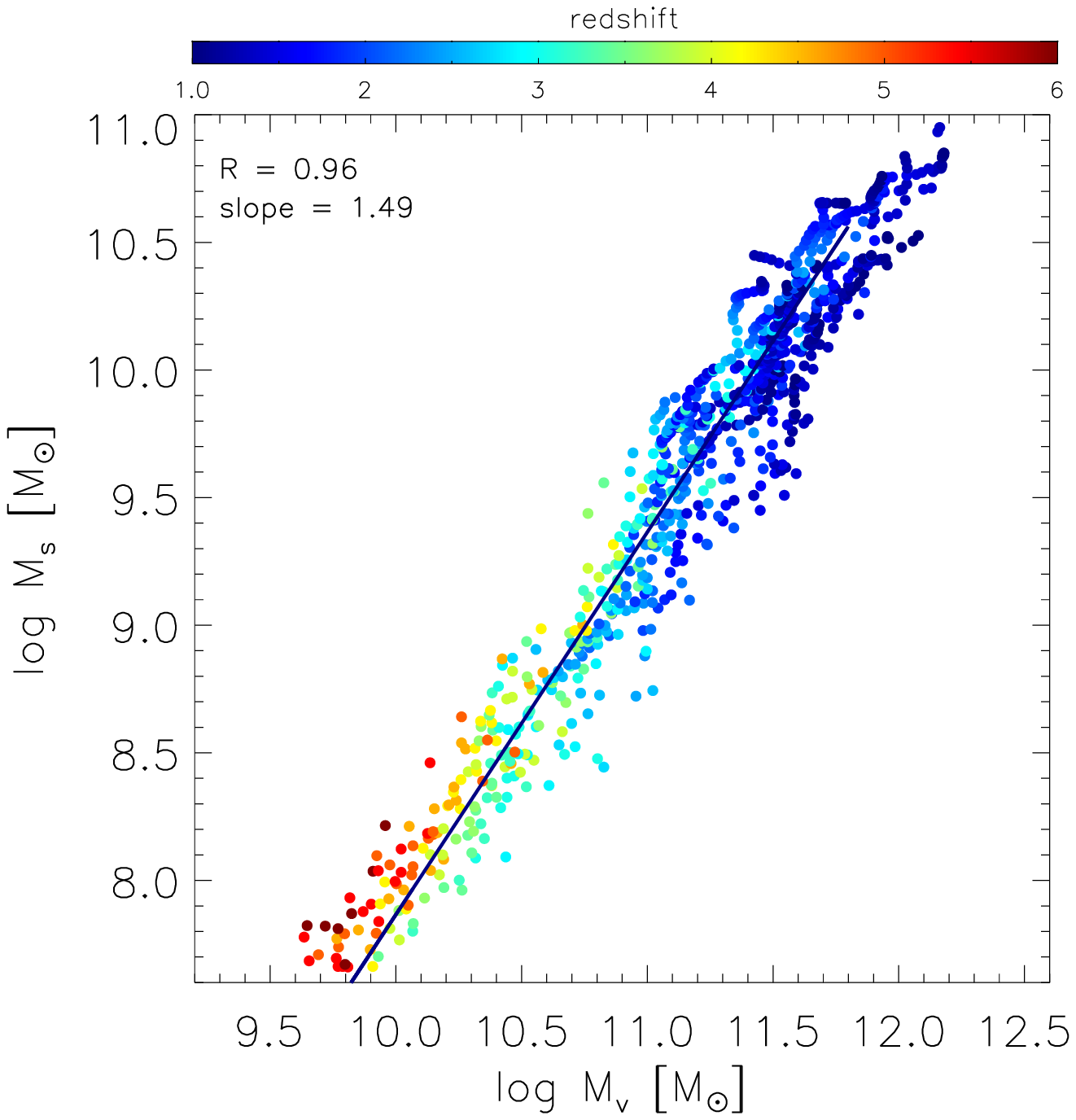}
\caption{
Stellar mass versus halo virial mass for all galaxies and snapshots.
Points are coloured by redshift.
There is a tight correlation, with the linear regression
(for $\Mv\!<\!10^{11.8}\msun$) shown,
and with the slope and correlation coefficient quoted.
The individual galaxy tracks evolve along the line, and then bend over
and flatten as the $\Ms$ growth is suppressed during the quenching process.
}
\label{fig:mstar_mhalo}
\end{figure}

\smallskip %16, 17  Skpc vs Ms
The correlation between $\Skpc$, $\Ms$, and $\Mv$ are directly addressed in
\figs{sigma_mass} and \ref{fig:mstar_mhalo}. 
The left panel of \fig{sigma_mass}, which shows all snapshots,
demonstrates that
$\Skpc$ and $\Ms$ are strongly correlated about the line $\Skpc \propto \Ms$,
where the upper envelope is defined by $\Skpc \leq \Ms/\pi \kpc^{-2}$.
The evolution tracks of our eight examples, shown in the right panel of
\fig{sigma_mass}, indicate that these variables grow together in time as the
galaxy evolves along the SFG phases. 
This is true in many cases also during the compaction phase, as $\Ms$ is
growing due to the high SFR and the continuing accretion. 
In some cases the track steepens as $\Skpc$ is growing faster than $\Ms$.
Then, at the end of compaction, the evolution track flattens, as $\Ms$ keeps
growing while $\Skpc$ becomes rather constant and remains so during the
quenching phase. In this phase the growth of $\Ms$ is slow, so the horizontal
track segment is short.
We note that the tracks of the less massive galaxies 
(when ranked at a given redshift) tend to flatten off at a lower $\Ms$ and 
a lower $\Skpcm$, as indicated in \fig{times}.

\smallskip % SF vs quenched
Back to the scatter diagram on the left of \fig{sigma_mass}, 
the quenched galaxies naturally populate the upper-right part of the 
distribution of points. The slope of the $\Skpc-\Ms$ relation for the quenched
galaxies is only somewhat flatter than that of the SFGs, 
and its zero point is slowly shifting with
time toward larger masses, qualitatively similar to observations 
\citep{fang13,tacchella15_science}.  
The star-forming galaxies show a somewhat steeper relation, with a larger 
scatter in the lower-left part of the scatter diagram
\citep[see also][]{ceverino14_e}.

\smallskip %17 Ms - Mv correlation
The examples of \fig{ssfr_mass} indicate that the stellar mass and halo mass
grow together, as expected.
The correlation between stellar mass and halo mass is addressed directly
in \fig{mstar_mhalo}. In the range of masses spanned by our sample in the high
redshift range, all the way up to $\Mv=10^{11.8}\msun$,
the data is well fit by a power law, $\Ms \propto \Mv^{1.5}$, with a
correlation coefficient $R=0.96$.
This power law is consistent with the correlation
deduced from observations through abundance matching
\citep{moster10,behroozi13}. 
At larger masses there is a hint for flattening, as expected
from virial shock heating in halos more massive than $\sim\!10^{11.5-12}\msun$
\citep{db06}.

%----------
\subsection{Halo Quenching}

We know from theory that at late times a halo has to be above a threshold
mass, in the range $10^{11.5-12}\msun$,
in order to support a stable virial shock and sustain a hot 
circum-galactic medium (CGM) that can suppress cold gas supply to the galaxy 
and thus maintain long-term quenching \citep{db06}.
After establishing the role played by central density in triggering 
the quenching process, we now attempt to identify clues for the role
played by a hot halo, through the halo mass, in allowing and maintaining the
quenching. 

\begin{figure*} % 21 <- 18 
\centering
\subfigure{\includegraphics[width=0.45\textwidth]{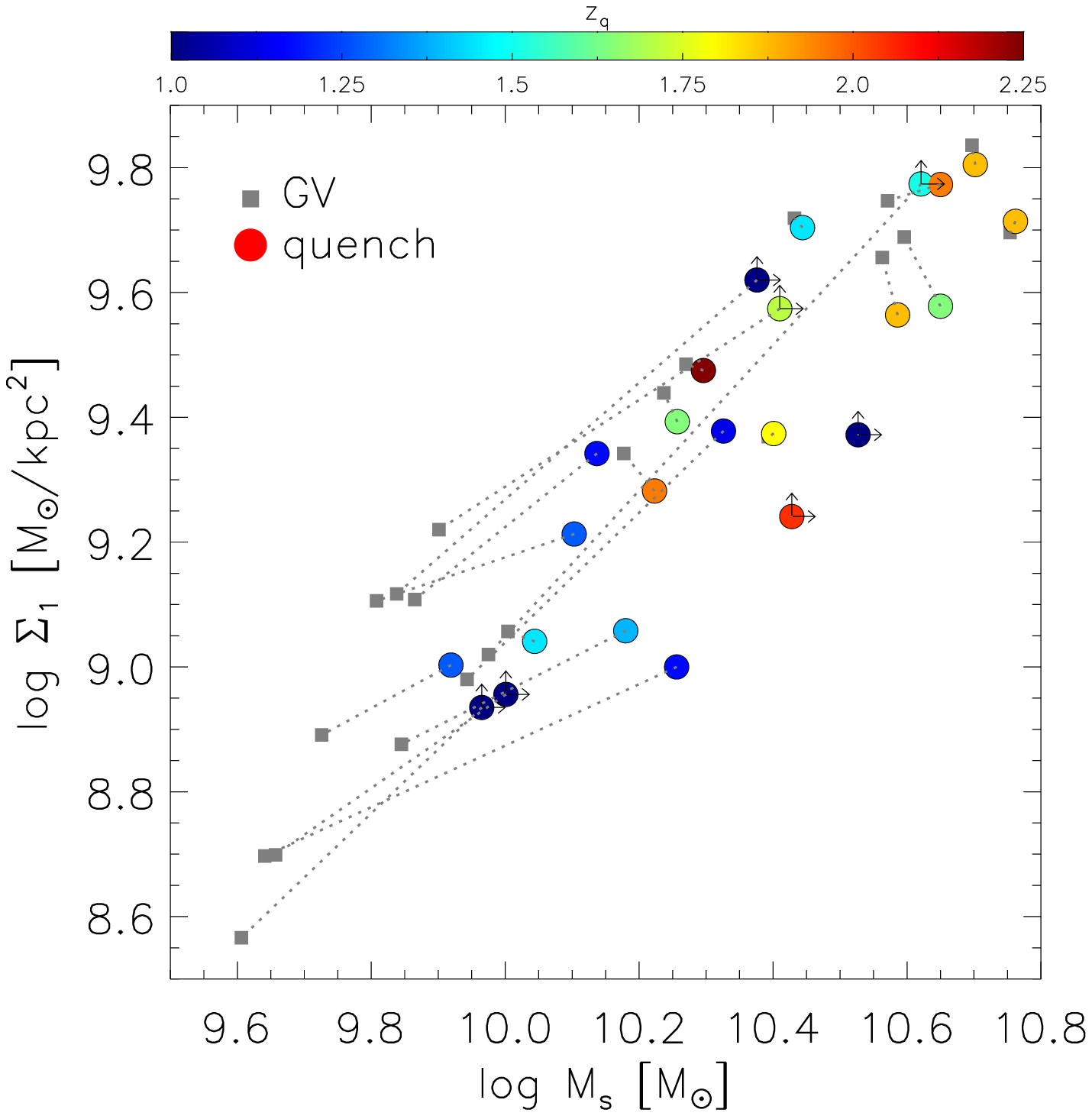}}
\subfigure{\includegraphics[width=0.45\textwidth]{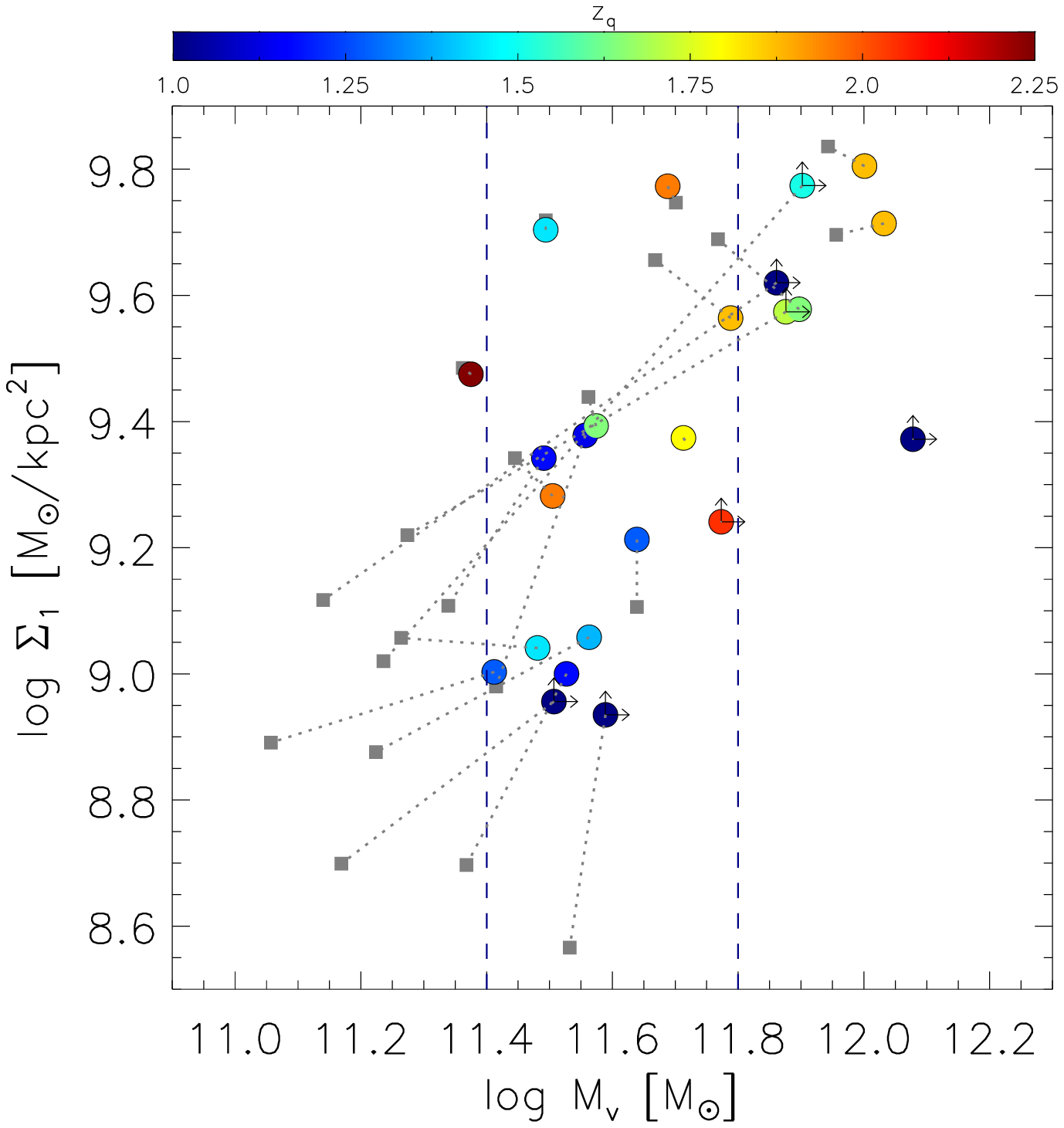}}
\caption{
The values at quenching of central surface density, $\Skpc$, versus mass,
$\Ms$ (left) and $\Mv$ (right).
The 24 grey squares refer to the first green-valley crossing, where
sSFR$<\!0.3\Gyr^{-1}$ for the first time.
The 17 circles with no arrows refer to galaxies that successfully quenched,
reaching sSFR$<0.16\Gyr^{-1}$.
The 7 circles with arrows refer to galaxies that did not successfully quench
by the final snapshot, and thus provide a lower limit to the mass and $\Skpc$
at possible later quenching.
The color marks the redshift at quenching (or an upper limit for it).
Each pair of points that belong to the same galaxy is connected by a dashed
line.
Succesful quenching occurs when the halo mass is above a threshold,
$\Mv \geq 10^{11.4}\msun$, and about 70\% of the galaxies quench when
the halo is in the narrow range $\Mv = 10^{11.4}-10^{11.8}\msun$.
At smaller halo masses,
the partial quenching attempt triggered by moderate compaction fails.
We see no sharp lower limit for $\Skpc$ or $\Ms$ at final quenching.
Galaxies that quench earlier tend to do it at a higher $\Skpcm$
and a higher $\Ms$, but only at a marginally higher $\Mv$.
}
\label{fig:mv_sigma_quench}
\end{figure*}

\smallskip
\Fig{mv_sigma_quench} %21
shows the central density ($\Skpc$) and masses ($\Ms$ and $\Mv$) 
of each galaxy at two times that characterize the quenching process.
One time (marked by grey squares) refers to the first crossing of the green 
valley, defined by sSFR$=0.3\Gyr^{-1}$, which is interpreted as a quenching 
attempt that may or may not proceed to significantly lower values of sSFR.
The second time (filled circles) refers to the crossing of 
sSFR$=0.16\Gyr^{-1}$, which is interpreted as successful quenching. 
For galaxies that have not successfully quenched by the final snapshot 
($z \sim 1$ in most cases) we interpret the last crossing of the green valley
as a lower limit for the mass and central density at the time of
possible later quenching (solid circles with arrows).
Only two of the 26 galaxies in our sample (V34 and V13)
have not crossed the green valley by their last snapshot 
($z=1.86$ and $z=1.5$), and
are therefore omitted from this figure.
Seventeen galaxies have successfully quenched by the final snapshot,
while 7 galaxies provide only lower limits to possible quenching.

\smallskip % Fig 21 most important
The general correlation seen in \fig{mv_sigma_quench}
between density and mass at quenching is not 
surprising given that this correlation holds during the evolution of each
galaxy, as seen in \figs{sigma_mass} and \ref{fig:mstar_mhalo}.  
We learn from \fig{mv_sigma_quench} that quenching attempts (grey squares) may 
occur for different galaxies at different densities and masses,
spanning a relatively broad range, without picking out a special scale.
However, successful quenching seems to occur when the parameters span 
relatively narrow ranges, about 0.9 dex, 0.9 dex and 0.7 dex for 
$\Skpc$, $\Ms$ and $\Mv$ respectively. 
The log of the corresponding medians (in the units used here)
and the $1\sigma$ scatter about them are roughly
$9.5 \substack{+0.3 \\ -0.3}$,
$10.4 \substack{+0.3 \\ -0.2}$, and
$11.6 \substack{+0.3 \\ -0.1}$.

\smallskip % threshold and narrow range for halo mass
Most interestingly, there seems to be a threshold for successful quenching at
$\Mv \geq 10^{11.4}\msun$.
If a galaxy attempts to quench while its halo is below this threshold,
its quenching attempt temporarily fails, waiting for the halo mass
to grow above the threshold before it can successfully quench.
Consistent with this threshold, the halo mass at successful quenching is 
typically in a relatively narrow range above the threshold.
If we consider the lower limits as detections, 
we find that about 70\% of the galaxies quench when the halo mass is in the 
narrow range $\Mv = 10^{11.4}-10^{11.8}\msun$. 

\smallskip % theory halo threshold
This is consistent with the basic theoretical understanding of halo quenching,
where being above a threshold mass of $\sim\!10^{11.5-12}\msun$ is a necessary 
condition for the halo to sustain a stable virial shock. 
Such a shock is typically 
triggered by a minor merger event that happens after the halo grew to above 
the threshold \citep{db06}.
A more direct test of halo quenching can be provided by measuring the halo gas
temperature and the absence of penetrating cold inflows, which is deferred to a
future study.

\smallskip % induced range for stellar quantities
The corresponding ``thresholds" for $\Ms$ and $\Skpc$ at quenching are softer, 
and the ranges of values above their minima are somewhat broader, 
0.6 dex for 76\% of the galaxies compared to the 0.4 dex for $\Mv$. 
This is consistent with the theory-motivated assertion that $\Mv$ is more 
strongly associated with the cause of quenching, and the narrow but
somewhat broader ranges of values for the stellar quantities at quenching
are induced by their correlations with $\Mv$.

\smallskip
The colours of the circles in \fig{mv_sigma_quench} mark the redshift at
successful quenching (or an upper limit for it).
This shows that galaxies that quench earlier tend to do so at a higher $\Skpcm$
and correspondingly at a higher stellar mass, confirming the 
quenching downsizing discussed above in \se{sample} and below in \se{origin}. 
A similar general trend is also seen for the halo mass, but here the trend is 
weaker, indicating a tendency to quench near a similar characteristic halo 
mass at the different redshifts.

%%%%%%%%%%%%%%%%%%%%%%7
\section{Discussion: Origin of compaction \& Quenching}
\label{sec:origin}

%-------------------- 7.1
\subsection{Onset of Compaction}
\label{sec:onset_comp}
%Intense Streams, Mergers and Non-linear VDI}

% in the examples
As seen in previous sections, the compaction is wet, 
in the sense that it is driven by gas 
contraction at a rate that is faster than the SFR.
This wet compaction could possibly be associated with violent disc instability
(DB14), and it is likely triggered by an intense episode of gas inflow
through gas-rich mergers or smoother gas streams. We discuss the possible
origin of compaction next.

\subsubsection{Mergers}

\smallskip % mergers in the sims
The possible correlation between merger events and the onset of compaction 
and/or quenching can be crudely explored in our simulations
using the green curves in the right 
panels of \fig{ssfr_sigma_ex} and \fig{ssfr_sigma_low}, which
show the mass growth, in the inner 1 kpc and in the whole galaxy,
due to ex-situ stars that formed outside the galaxy.
The addition of a large stellar mass over a single output timestep, 
of a typical duration of $150\Myr$, is an indication for a merger event. 
In many cases there are indications for mergers prior to 
compaction (e.g. V11, V12, V14). 
In other cases there is no evidence for a
significant merger prior to compaction, possibly indicating that the 
compaction is associated with VDI (e.g. V26). In most cases we can see
a sequence of mergers prior to compaction but they do not seem to coincide with 
the onset of quenching (e.g., the first snapshot shown for V12 in 
\fig{images_12}).
The eight galaxies shown in \figs{ssfr_sigma_ex} and \ref{fig:ssfr_sigma_low},
like the other galaxies in the sample, give the 
impression that in many cases mergers are important in triggering the 
compaction, and are possibly also involved in the quenching process.   
These mergers can be major in some cases, namely with a mass ratios higher
than 1:3, but in most cases they are minor mergers with a mass ratio
between 1:10 and 1:3.

\subsubsection{Stimulated VDI}

\smallskip % nonlinear VDI
Our developing picture is that the VDI that is very abundant in high-redshift
disc galaxies is different from the secular instability common in 
late-redshift discs like the Milky Way. At high redshift, where the gas
fraction is higher, the perturbations, including the giant clumps, are large
and with non-negligible masses compared to the whole disc, 
so they have important dynamical effects that operate on timescales comparable 
to the disc dynamical timescale. This refers, for example, to the formation of 
in-situ clumps and to their migration to the 
disc centre. Furthermore, the simulations reveal that the Toomre $Q$
parameter in most of the disc outside the clumps is 
significantly larger than unity, while the disc still produces new clumps
(Inoue et al., in preparation). 
This indicates a non-linear unstable state that is stimulated by
non-linear perturbations. These perturbations may be associated with the 
intense instreaming including minor mergers, that at those epochs are rather 
frequent with respect to the disc orbital time. 
The emerging wisdom is that VDI and mergers are not
two distinct phenomena but rather closely associated with each other. 
In a gas-rich disc, they seem to work in concert in driving the intense inflow 
within the disc, thus compactifying the system into a blue nugget.
This non-linear disc instability is under investigation (Inoue et al., in
preparation).

\subsubsection{Counter-rotating streams}

\smallskip % counter-rotating  instreaming
The intense gas streams could naturally drive compaction if they are 
counter-rotating with respect to the existing gas disc. 
Our simulations demonstrate that this phenomenon is common in high-redshift 
galaxies. \citet{danovich14}
found that on average about $30\%$ of the instreaming gas mass 
in the inner halo %at $(0.1-0.3)\Rv$
% At 1Rv the mass fraction of counter rotating infalling gas: ~0.44 \pm 0.13
% At 0.1-0.3Rv: 0.27 \pm 0.19
is counter-rotating with respect to the net instreaming angular momentum (AM)
in the same volume, 
carrying a negative contribution with an amplitude that is on average  
$43\%$ of the positive component, namely $75\%$ of the net total inflowing AM.

\smallskip % disc flips
In order to evaluate the effect of the counter-rotating streams on the disc, 
\citet{danovich14}
measured $\Jneg$, the AM component in the direction anti-parallel to the disc 
AM, for the instreaming gas at radial distances and velocities that will 
bring it to the disc in less than an outer-disc orbital time.                   
This time is typically $\torb \sim 250 \Myr$ in our galaxies at $z \sim 2$,
involving most of the instreaming gas at $0.1-0.5\Rv$.
For the galaxies at $z=2-3$,  
the median of $\Jneg/\Jd$ (where $\Jd$ is the disc AM)
is about 10\%, but about 20\% of the galaxies have $\Jneg/\Jd \sim 1$. 
Indeed, in a period of $250\Myr$, about $15-20\%$ of the discs flip their 
orientations, namely change their spin direction by more than $90^\circ$.
We note that this is suspiciously similar to the fraction of galaxies that 
start their compaction event in a similar period.

\smallskip % lovelace etc.
The interaction of a counter-rotating stream with the rotating disc
leads to a cancellation of angular momentum that can drive a significant
shrinkage (to be studied elsewhere).
This phenomenon has been studied in the context of proto-stellar discs.
\citet{quach15} performed an analytic calculation based on the work of
\citet{lovelace09}, discussing the Kelvin-Helmholtz instability between a
rotating inner disc and a counter-rotating outer disc.
\citet{dyda15} performed idealized simulations of such a two-component disc
as well as a disc and a counter-rotating stream, and found in both cases
a significant shrinkage of the disc (their Fig.~11) or part of it (Fig.~2). 

\subsubsection{Low-AM recycled gas}

Most of the gas that inflows into galaxies at $z\sim 1-3$ 
is likely to be recycled gas that has been ejected from the galaxy earlier. 
This has been argued based on the failure of models based on cosmological 
accretion alone to match the high sSFR observed in SFGs at these redshifts 
\citep{dm14}, and it has been directly measured in the simulations  
used in the current paper (DeGraf et al., in preparation).
The gas is ejected preferentially from the dense central regions that are
typically of low AM. The interaction between the low-AM returning gas and
the high-AM fresh instreaming gas is another natural way to generate shrinkage.
This has been addressed by \citet{elmegreen14}, who studied shrinking
galaxy discs by fountain-driven accretion from the halo. 

\subsubsection{Tidal compression}

Tidal interactions are commonly perceived as causing stripping,
but in certain circumstances they could actually cause compaction.
In particular, when a satellite galaxy is orbiting in a core of a
host halo, where the density profile is flatter than $r^{-1}$,
the tidal forces from all directions are pushing inwards, 
toward the satellite centre \citep{ddh03}.
Thus, a galaxy passing through a cluster core is likely to compactify subject
to tidal compression.

\citet{renaud14} measured the compressive tides during a simulated major 
merger, and found them to become strong during the merger.
They learned that the compressive tides boost the compressive, irrotational
mode of the gas turbulence well above its usual equipartition with the 
rotational, solenoidal mode. This compressive mode can induce local collapse. 
Indeed, it has been argued analytically that compressive tides reduce the 
Jeans mass, which makes clump formation more efficient \citep{jog13,jog14}.
One may therefore deduce that compressive tides may stimulate violent disc 
instability with giant clumps, and suspect that they could also induce a 
central compaction of the galaxy core.

\subsubsection{Origin of mass dependence}

One wonders what could be the reason for 
more massive galaxies to compactify at a higher redshift.   
The total specific accretion rate into haloes is only a weak function of mass, 
$\Mdot/M \propto M^{0.14}$, which could be derived from the slope of the
fluctuation power spectrum on galactic scales 
\citep{neistein06,neistein08b,dekel13}. 
The associated merger rate, for mergers with a mass ratio above
any threshold, is a fixed fraction of this overall accretion rate
\citep{neistein08b}.
The resultant specific accretion rate of baryons onto the galaxy at the halo
centre
is similar \citep{dekel13}, and the associated galaxy merger rate is expected
to crudely follow. Therefore, 
unless the fraction of mergers is smaller in less massive
haloes due to more efficient pre-merger disruption of satellites, 
we do not expect a mass dependence of compaction time to arise from a 
mass dependence in the accretion or merger rate.

\smallskip
According to DB14 (eq. 29), at a given high redshift, for haloes in the mass 
range simulated here, the critical effective surface density for wet 
compaction, $\Sigma_{e,w=1}$,
is expected to be slowly increasing with mass, at a rate comparable
to the natural dependence of surface density on mass,  
$\Sigma \propto M^{1/3}$, induced by the virial relations 
\citep[e.g.][eq. 15]{dekel13}.
This, also, does not introduce a strong mass dependence in the compaction
efficiency at a given redshift.
On the other hand, at a given mass, $\Sigma_{e,w=1}$ is expected to increase 
with time as $\propto (1+z)^{-3/2}$, while the natural evolution of the
effective surface density in galaxies of a given mass is $\propto (1+z)^2$,
which makes it easier to compactify at higher redshift but for any galaxy mass.
We thus do not see a compelling theoretical preference for a late compaction
in low-mass galaxies. 

\smallskip  % mass dependence of quenching comes from halo
An inspection of the histories of our lower-mass simulated galaxies indeed 
indicates that the apparently late compaction of lower-mass galaxies
refers only to the {\it last} and most dramatic compaction event that 
leads to successful quenching. 
These galaxies may undergo earlier compaction events that do not end up in full
quenching.
Our tentative conclusion (as argued in DB14) is that the downsizing refers to
the quenching process, and is possibly associated with the effect of a
threshold halo mass for successful quenching, as discussed in \se{mass}.

%--------------------- 7.2
\subsection{Onset of Quenching at Maximum Compactness}

% compaction stops because of AM?

% Mdot
The simulations reveal that the point of maximum gas 
compactness defines the onset of quenching, in terms of gas depletion 
in the central regions and the associated decline in SFR.
It would be interesting to find out what makes the gas compaction stop at a
given configuration, and what causes the onset of quenching at the same time.
Clues for the mechanisms involved are provided in the second-from-right panels
of \figs{ssfr_sigma_ex} and \ref{fig:ssfr_sigma_low}
by the curves displaying the evolution of the central masses of the different 
components and their different rates of change. 

\smallskip % gas consumption into SFR
Since the SFR gradually increases during the compaction phase and reaches
a maximum near the peak of gas compactness, the gas consumption into stars
itself is likely to be an important source of quenching.
Indeed, the gas depletion time by star formation at the peak compactness
is short, $\sim 100\Myr$. 
With an average SFR of $\sim\!10\sy$ over a typical compaction duration
of $\sim\!0.5\Gyr$, the gas mass consumed into stars is a few times 
$10^9\msun$, comparable to the peak total gas mass within the 1-kpc sphere.

\smallskip % inflows and outflows
Both the inflow rate and outflow rate through the boundary of the 1-kpc 
sphere are
expected to be correlated with the SFR, and, indeed, they seem to be comparable
to it, and thus also important in the quenching process.
During the compaction and the early phases of the quenching process, 
part of the inflow is associated with the compaction process, 
while additional inflow may be associated with recycling
of earlier outflows, which is likely to be substantial \citep[e.g.][]{dm14}.
We do not identify a large burst of outflows at the onset of quenching,
but there is a broad maximum of outflow rate (and inflow rate)
during the maximum compaction phase, with a mass-loading factor of order unity
with respect to the SFR. This is consistent with preliminary
observations of outflows from blue-nugget candidates at $z \sim 3$ 
\citep{williams14b}.    

\smallskip % quenching
The pre-compaction and especially the compaction phase are characterized by 
the gas inflow rate into the central 1 kpc being higher than the sum of the
SFR and the outflow rate, leading to an overall growth with time of the central
gas density. Immediately after the compaction peaks, where the SFR and the
outflow rate are at their peak, not much gas is left outside the central 
compact object so the inflow is dropping, shifting the balance to a situation
where the SFR plus outflow dominate over the inflow. This leads to gas
depletion from the centre, which in turn reduces the SFR and leads to
gradual quenching. The fact that the inflow, SFR, and outflow rates remain
comparable to each other implies that the inequality
{\it SFR$+$outflow$>$inflow} remains valid, which
allows the quenching process to continue till completion.  
This situation may change only if there is a new episode of external gas
supply that boosts the inflow rate into the centre.
More likely in the post-compaction phase is a slow buildup of an extended
gaseous disc about the quenched centre, driven by high-AM
accretion from the halo,
partly in a hot mode, and certain outflow from the centre.
This leads to a an extended star-forming ring surrounding a quenched red
nugget, which can be interpreted as inside-out quenching during the
post-compaction phase.
We comment, again, that star formation in this ring (as well as in the bulge)
could be impeded by AGN feedback or stronger stellar feedback had those been 
included in our simulations.

\smallskip %In VDI
\citet{db14} considered the process of VDI-driven wet compaction in the 
dark-matter dominated regime. They argued that when $w>1$, as the baryons 
contract within a fixed dark-matter halo that dominates the mass, 
the compaction leads to an increase in $\delta=\Mcold/\Mt$, 
as the given $\Mcold$ is shrinking to a radius that contains less halo mass.
Based on \equ{w}, where the wetness is $w\propto \delta^2$, 
the growth of $\delta$ makes $w$ increase, 
thus leading to a runaway compaction until it is halted by the growing 
velocity dispersion or another process.
The central quenching is helped by the suppression of 
gas supply to the inner region, which may result from the suppression of
disc instability in any disc that may remain after the shrinkage. 
The increase of the Toomre $Q$ parameter to above unity can
be caused either by the depletion of gas surface density in the disc 
\citep{forbes12,forbes14a}, 
by an increase in the turbulent velocity dispersion, or by the growing
contribution of the bulge to the potential well and therefore to the 
epicyclic frequency $\kappa$ \citep[morphological quenching,][]{martig09}.
In the late stages of compaction,
when the baryons are self-gravitating with a negligible contribution from the 
dark matter, the compaction may actually lead to a decrease in $\delta$.  
This is because the ``hot" component, made of stars, grows at the expense of 
the depleting cold-gas component.
This makes $w$ decline, which may suppress the gas inflow in the disc  
as the SFR becomes faster than the inflow rate, and thus help the central
quenching.

\smallskip % halo
The halo quenching naturally introduces a mass dependence in the quenching 
process.
A halo of mass below a few times $10^{11}\msun$ allows efficient cold gas 
inflow into the galaxy \citep{db06}, which builds a gaseous disc that 
undergoes VDI. The incoming mergers and counter-rotating streams 
trigger compaction events, each followed by a temporary quenching 
attempt driven by the high central gas density with enhanced SFR 
and the associated
temporary shortage of gas supply from the disc that has been diluted by the
shrinkage.
However, the continuing cold gas supply from the halo rebuilds a disc that
provides gas mass for a recurrent compaction event, leading to a new central
star-formation event followed by a new quenching attempt, and so on.  
Only when the halo becomes massive enough to support a stable virial shock
and thus sustain a hot CGM, and when the cosmological time is late enough
such that cold streams do not efficiently penetrate through the hot CGM, 
can the quenching attempt be completed. In such a halo, the disc that has 
been diluted by compaction cannot be replenished by fresh external gas supply.
The feedback from the star-forming centre may help triggering the hot CGM
and suppress gas infall from the halo.
The inclusion of AGN feedback may do so even more efficiently and for a
longer period. This may connect the internal quenching process with the
long-term quenching by the hot CGM in the halo.

%------------------------
\subsection{Transition to Self Gravity and Quenching}

We noticed that in the pre-compaction phase, while the galaxy forms stars
at a rather constant sSFR, the mass within $\Re$ or 1 kpc is dominated by 
the dark matter. 
Then, after compaction, the baryons become self-gravitating within the central
region, where they dominate the gravitational potential over the dark matter.
This transition may provide an alternative explanation for the onset of 
quenching at the peak of central gas density.
We already mentioned in the previous section how such a transition can help the
quenching by modifying the DB14 argument for a runaway shrinkage, and we now
discuss an alternative interpretation of the quenching in the self-gravitating
state via a bathtub toy model for galaxy evolution \citep[e.g.,][]{dm14}.

\smallskip
%In bathtub model
We apply here the bathtub model to the gas mass within the central regions.
The gas mass conservation due to inflow and star formation plus outflow
is expressed as
\be
\Mgdot = \dot{M}_{\rm in} - (1+\eta_{\rm out}) 
\frac{\Mg}{\tau_{\rm sfr}} \, .
\label{eq:bathtub}
\ee
Here $\dot{M}_{\rm in}$ is the gas mass inflow rate,
and $\eta_{\rm out}$ is the outflow mass loading factor, namely the outflow 
rate with respect to the SFR.
The SFR is assumed to be proportional to the gas mass and to occur on
a characteristic timescale $\tau_{\rm sfr}$.
This timescale is assumed to be proportional to the local free-fall time in the
star-forming cloud, which scales inversely with the square root of the local
mass density \citep[e.g.,][]{kdm12}.
 
\smallskip
In the dark-matter dominated pre-compaction regime, 
$\tau_{\rm sfr}$ at a given cosmological epoch can be assumed to be roughly
proportional to the cosmological time and independent of mass. This is because
the inner-region density is in a similar ballpark for all haloes at a given
redshift, once ignoring differences associated with variations in the
halo concentration or the halo-mass growth history \citep{wechsler02}.
In this case, as long as $\dot{M}_{\rm in}$ and 
$\tau = \tau_{\rm sfr}/(1+\eta_{\rm out})$ vary slowly in time,
the approximate solution of \equ{bathtub} is
\be
\Mg = \dot{M}_{\rm in} \tau (1- e^{-t/\tau}) \, .
\label{eq:QSS}
\ee
Here the gas mass in the inner halo is growing fast at first, 
exponentially converging to the quasi-steady-state (QSS) solution
$\Mg = \dot{M}_{\rm in} \tau$, which continues to grow slowly
with time till after $z\sim 1$ \citep[e.g.][]{dekel13}. 
This is indeed the behavior seen prior to compaction.
The convergence to the QSS solution is due to the negative sign of the SFR
term in \equ{bathtub} and its linear dependence on $\Mg$.

\smallskip
Once the baryons become self-gravitating in the late compaction and
post-compaction phases, the depletion time may not be independent of gas mass
any more. We learn from the right panels of \figs{ssfr_sigma_ex}
and \ref{fig:ssfr_sigma_low} that the depletion time $\tau_{\rm sfr}=\Mg/$SFR
reaches a minimum when $\Mg$ is at its peak, and it later grows in time as 
$\Mg$ declines, namely $\tau_{\rm sfr}$ is a decreasing function of $\Mg$.  
This is consistent with the free-fall time being determined by the gas density. 
If we assume, for example, $\tau_{\rm sfr} \propto M_{\rm gas}^{-1/2}$,
the second term in \equ{bathtub} becomes $\propto M_{\rm gas}^{3/2}$. 
In this case the SFR drain term in \equ{bathtub} dominates over the accretion
source term, which makes the solution for $M_{\rm gas}$ decrease in time 
rather than approach a QSS.  
This is consistent with the gradual decrease of $\Mg$ leading to
quenching in the self-gravitating phase.

%%%%%%%%%%%%%%%%%%%%%%%8
\section{Conclusion and Discussion}
\label{sec:conc}

Our suite of zoom-in cosmological simulations reveals 
three characteristic phases  
in the evolution of galaxies at high redshift: a diffuse phase, 
a compaction phase, and a subsequent quenching phase, which in some cases occur
more than once in the lifetime of a galaxy. 
The main features of this chain of events are as follows.
\rf\bul
In the diffuse phase the intense accretion though streams, including mergers, 
grows a gas-rich, highly perturbed, turbulent, rotating disc, 
and drives a high sSFR, often associated with violent disc instability.  
\rf\bul
The typical galaxy undergoes a phase (or phases) of
dissipative compaction into a compact, gas-rich, star-forming system, 
which resembles the observed ``blue" nuggets.
\rf\bul
The moment of maximum gas compaction marks the onset of gas depletion from
the central regions, followed by gradual central quenching into
a compact elliptical that resembles the observed red nuggets.
\rf\bul
In some cases, a post-compaction clumpy ring of gas forming stars 
develops around the quenched massive compact bulge, which appears as
inside-out quenching in this phase of the evolution. 
An extended stellar envelope may gradually grow by dry accretion.
\rf\bul

The compaction process is associated with a steep increase in both rotation and
velocity dispersion for the stars, and an even steeper increase in rotation 
velocity of the gas. These velocities remain high in the post-compaction phase,
and give rise to high and comparable gas and stellar linewidths when observed
through a broad beam.

\rf\bul
There is downsizing in the sense that more massive galaxies tend to quench 
earlier, faster, and at a higher stellar density. They quench more 
efficiently and decisively than their lower-mass counterparts.
The latter tend to oscillate about the main-sequence ridge between the
blue-nugget and green-nugget phases before they eventually fully quench.

\smallskip
Other noticeable features of these events are as follows.
\rf\bul
As a galaxy undergoes compaction it maintains a roughly constant sSFR.
When it subsequently quench, the sSFR decreases rapidly while 
the density within the inner $1\kpc$ stays roughly constant 
(and the effective surface density may slowly decrease), 
consistent with the way observed galaxies populate the sSFR-$\Sigma$ plane.
\rf\bul
The compaction is commonly triggered by an intense inflow episode.
This may include a major merger, multiple minor mergers, counter-rotating
streams, recycled gas, or tidal-compression,
and is sometimes associated with stimulated violent disc instability.
\rf\bul
As predicted by DB14,
the compaction process is wet, with the inflow rate higher than the SFR.
The maximum surface density reached after compaction
is indeed anti-correlated with cosmological 
time and with the initial spin parameter, and it is correlated with the sSFR. 
The simulated population of star-forming galaxies at $z\geq 3$ shows 
a bimodality in central surface density, being either diffuse or compact, as
predicted.
\rf\bul
The central quenching in the current simulations is driven by gas depletion
from star formation and 
stellar/supernovae feedback, while the gas inflow to the centre is 
(at leaset temporarily) suppressed.
\rf\bul
Quenching occurs after the baryons become self-gravitating 
in the central kpc, which may be explained in the context of VDI-driven
compaction or via a bathtub toy model for the evolution of the central
region of the galaxy.
\rf\bul
A halo mass above a threshold of $10^{11.4}\msun$ is required for successful
quenching, with most galaxies quenching in the narrow range 
$\Mv=10^{11.4-11.8}\msun$.
This indicates that the halo mass has an important role in allowing significant
quenching and maintaining it, while the central density provides the trigger
for quenching. 
The stellar mass and central surface density at final quenching 
span a somewhat broader but also rather narrow range of values, 
presumably reflecting their tight correlations with the halo mass.

\smallskip % comparison to observations.
The simulation results, which confirm the toy-model predictions of DB14,
are qualitatively consistent with observational findings concerning 
high-redshift star-forming and quenched galaxy
populations, the compact ``blue" nuggets and red nuggets, 
both at $z<1$ \citep{cheung12,kauffmann12,fang13}
and at $z=1-4$
\citep{barro13,kaviraj13a,kaviraj13b,lee13,williams14a,williams14b,bruce14,
barro14a,barro14b}.

\smallskip % linewidth
DB14 predicted using a toy model for VDI-driven wet compaction
that the ``blue" nuggets should be 
dispersion-dominated, rotating, star-forming systems, 
with $\sigma/V\!\geq\!0.5$.
Indeed, we find in our simulations that the compaction is associated with a
steep rise in both rotation velocity and velocity dispersion (for the stars).
The line-of-sight mass-weighted velocity dispersion
tends to rise during the compaction from $\sim\!50\kms$ to $100\!-\!200\kms$,
and to remain high (or slightly decreasing) during the subsequent 
quenching to a red nugget. 
This is consistent with the observed line-widths for galaxies
in the three distinct phases \citep{barro14b}. 
As pointed out by DB14, the ``blue" nuggets could potentially be associated 
with observed dispersion-dominated systems \citep{law12,newman13}. 
We found that the ``blue" nuggets could either proceed or precede a VDI clumpy
phase.
Low-mass, low-metallicity, dispersion-dominated galaxies more likely
represent the violent pre-VDI stages of galaxy buildup driven by intense 
instreaming, while the post-compaction dispersion-dominated ``blue" nuggets 
are expected to include more massive and more metal rich systems.

\smallskip % rings
Our simulations, and the scenario of DB14, predict that compact galaxies,
following the ``blue"-nugget phase, may be surrounded by an extended disc or
ring of low-density gas and young stars. This ring is gravitationally unstable
showing clumps and is forming stars at a gradually declining rate. 
%\adr{Adi: xxx Can we quantify this?}
There are already possible observational detections
of such cases \citep{elmegreen_d05,bournaud11b,genzel14a}.
The high-$z$ post-compaction quenching is thus from the inside out.
We address this in detail in a companion paper \citep{tacchella15_prof}.
It is consistent with preliminary observational indications for inside-out
quenching based on the rising sSFR profiles of a sample of galaxies at 
$z\sim 2.2$
\citep{tacchella15_science}.

\smallskip % bulge quenching
Our work reveals how the dissipative shrinkage of a galaxy into a compact
bulge leads to quenching. The process is driven by efficient gas consumption 
into stars and the associated feedback-driven gas removal by outflows,
while the inflow to the center is suppressed, first because of the 
disappearance of the extended gas disc, and then by the shutdown of fresh 
gas accretion from the halo. 
In parallel, the disc instability, which drives star formation and inflow 
within the disc, is suppressed by the growth of bulge mass
\citep[morphological quenching][]{martig09,martig13}, the increase of turbulent
velocity dispersion, and the decline of gas surface density.
AGN feedback, which is likely to be correlated with the high SFR in the 
``blue"-nugget phase \citep[e.g.,][]{barro13}
but is not included in the current simulations, 
may help boosting the quenching process associated with the internal
compactness. 

\smallskip % halo quenching
The maintenance of long-term quenching requires a shutdown of fresh gas supply,
which is naturally provided by virial shock heating of the CGM in the halo,
occurring in haloes more massive than a threshold mass that can vary about 
a mass scale $\lsim\!10^{12}\msun$ and at sufficiently low redshift where cold
streams fail to penetrate deep into the galaxy \citep{db06}.
This is a slower quenching mode that is potentially dominant at lower
redshifts,
commonly after the quenching has been triggered by the internal bulge quenching
processes discussed above \citep{woo13,db14,woo14}.
The lower-mass galaxies in our simulations start as discs with moderate
wetness, and they contract later, more slowly and to a lower density.
Their compaction events trigger unsuccessful short-term quenching attempts 
as long as the halo mass is below the threshold for virial shock heating. 
The less massive galaxies manage to
quench significantly only when the halo becomes massive enough to sustain a hot
CGM that suppresses the cold gas supply.  

\smallskip % de-compaction by dry mergers.
Once a compact red nugget forms at a high redshift, 
it is likely to be subject to further accretion (including mergers), 
which becomes more and more star-dominated with time
\citep{oser10,porter14a,porter14b},
and thus lead to a dry growth. 
In some cases, the accretion is slow, leaving the red nugget naked, with no
stellar envelope, until $z \sim 1$.
In cases of substantial dry growth, an extended stellar envelope develops 
around the red nugget, and the galaxy becomes a rotating spheroid compatible 
with the typical observed ellipticals at low redshift 
\citep[e.g.,][]{dokkum14}. 
The stellar core itself typically remains dense within the inner kpc,
while the density within the effective radius can decrease as the effective
radius grows. 
In many cases, the overall stellar density profile seems to evolve 
self-similarly,
crudely matching the De Vaucouleurs' profile \citep{tacchella15_prof}.

\smallskip % limitations of the simulations -- which results are robust?
As a word of caution, 
one should recall that the reliability of cosmological simulations is 
still limited by the uncertain treatment of star formation and especially
feedback. This commonly leads to too-early formation of stars at redshifts 
above 3, leaving the galaxies with a gas fraction somewhat
lower than observed at $z\!\sim\!2$. The simulations used here, with lower SFR
efficiency and added radiative feedback compared to earlier simulations in our
series, do better than the earlier versions,
but the simulated gas fractions in massive galaxies at $z\!\sim\!2$ may still 
be lower than deduced from observations \citep{daddi10,tacconi13}, 
and the stellar-to-virial mass ratio may still be higher than deduced from 
observations \citep{moster10,behroozi13}, both by factors of $\sim\!2$
\citep{moody14}.
If true, the compaction events may tend to occur in the simulations
at somewhat earlier epochs than in the real Universe.

Furthermore, because of the potentially insufficient feedback and in 
particular the absence of AGN feedback in the current simulations, 
the quenching is sometimes incomplete. Nevertheless, the simulations do 
appear to capture the key features of the
compaction and quenching processes and thus provide at least a qualitative  
description of these major events in the lifetime of a galaxy, 
pointing to their frequent abundance at $z\!>\!2$, and to the vital role they
play in the formation of compact massive galaxies, star forming and quenched
ellipticals.

\bigskip % MS main sequence

Based on \figs{ssfr_sigma_all}, \ref{fig:L-shape} and \ref{fig:ssfr_mass}, 
we propose that the evolution through the phases of 
compaction and quenching may be responsible for variations of galaxy 
properties across the MS at a given stellar mass  
and may be the key for explaining the small scatter about the 
main-sequence ridge 
\citep[addressed in detail in a companion paper][]{tacchella15_ms}.
Observations, simulations and analytic theory indicate that
the average sSFR on the main-sequence at any redshift is a weak  
function of stellar mass and its zero point is declining with time
as sSFR$\propto (1+z)^{2.5}$ 
\citep[e.g.][]{noeske07a,noeske07b,whitaker12,dekel13,dm14}.
We see in the eight example galaxies shown in \fig{ssfr_mass} 
that each galaxy starts near the main-sequence 
ridge with a constant or slightly declining sSFR. 
During the pre-compaction phase and especially during the compaction phase 
the positive deviation of the sSFR from the ridge is rising, 
reaching a maximum at the ``blue"-nugget phase. 
From then on the sSFR of the quenching galaxy
drops faster than $\prop (1+z)^{2.5}$, thus bringing the galaxy to a negative
deviation from the ridge before it reaches the green-valley phase.   
We thus associate the galaxies that lie ``above" the main-sequence with 
the ``blue"-nugget phase of evolution, and the galaxies that lie ``below" 
it with the post-compaction green-valley phase, on the way to quenching.
The simulations indicate that the ``blue"-nugget phase is associated with a
minimum in the depletion time and a maximum in the gas fraction during the 
short-term history of the galaxy. This predicts gradients 
in these quantities orthogonal to the main sequence, as observed
\citep{genzel14c}. 
Furthermore, the onset of quenching at the peak of central gas density,
based on the balance between the rates of inflow and the sum of SFR and
outflow, as argued above, may explain the upper bound for the sSFR deviation 
from the main-sequence ridge. In turn, the lower bound can possibly be
explained by a temporary recovery of the inflow rate that triggers a new
episode of compaction and rising sSFR, till the final rapid drop of the sSFR 
toward the red sequence.

\smallskip
We conclude that compaction and subsequent quenching, assisted by a hot massive 
halo, combine to a chain of events of major importance in the history of 
massive galaxies. In particular,
the high-redshift ``blue" nuggets represent a key phase in galaxy evolution.

%%%%%%%%%%%%%
\section*{Acknowledgments}

We acknowledge stimulating discussions with Andi Burkert,
Marcella Carollo, Reinhard Genzel, David Koo, Simon Lilly 
and Sandro Tacchella.
This work was supported by ISF grant 24/12,
by GIF grant G-1052-104.7/2009, by a DIP grant,
by the I-CORE Program of the PBC, by ISF grant 1829/12,
and by NSF grants AST-1010033 and AST-1405962.
SI and CD acknowledge Lady Davis Fellowships at HU.
DC acknowledges support from MINECO grant AYA2012-31101,
MICINN grant AYA-2009-13875-C03-02, and a Juan-de-la-Cierva fellowship.
The simulations were performed at NERSC, LBNL, NASA Ames, and at the astro
cluster at HU.

%%%%%%%%%%%%%%%%%%%%%%%%%%%%%%%%%%%%%%%%%%%%%
\bibliographystyle{mn2e}
\bibliography{nuggets}
%%%%%%%%%%%%%%%%%%%%%%%%%%%%%%%%%%%%%%%%%%%%%

\label{lastpage}
\end{document}

%%%%%%%%%%%%%%%%%%%%%%%%%%%%%%%%%%%%%%%%%%%%%%%%%%%%%%%%%%%%%%%%%%%%%%%%%
%%%%%%%%%%%%%%%%%%%%%%%%%%%%%%%%%%%%%%%%%%%%%%%%%%%%%%%%%%%%%%%%%%%%%%%%%%